\documentclass[aps, prd, superscriptaddress, preprintnumbers, floatfix, longbibliography,10pt,twocolumn]{revtex4-2}

\usepackage{ graphicx}
\usepackage{  amsfonts,amsmath,amssymb,amstext}
\usepackage{  float,wrapfig}
\usepackage{  subfigure, psfrag}
\usepackage{  dsfont}
\usepackage{  color}
\usepackage{  comment}

\usepackage[colorlinks=true,urlcolor=blue,citecolor=blue,linkcolor=blue]{hyperref}
\usepackage{  bm}
\usepackage{  mathtools}

\graphicspath{{Figures/}} 

\usepackage[colorlinks=true,urlcolor=blue,citecolor=blue,linkcolor=blue]{hyperref}
\hypersetup{breaklinks=true}

\newcommand{\be}{\begin{equation}}
\newcommand{\ee}{\end{equation}}
\newcommand{\ba}{\begin{eqnarray}}
\newcommand{\ea}{\end{eqnarray}}

\definecolor{purple}{rgb}{0.8,0,0.6}
\definecolor{darkgreen}{rgb}{0.00,0.6,0.00}

\definecolor{Blue}{rgb}{0.,0.75,0.85}

\extrafloats{100}

\begin{document}

\title{Viscoelastic tensor and hydrodynamics of altermagnets}
\date{August 27, 2025}

\author{A.~A.~Herasymchuk}
\email{arsengerasymchuk@gmail.com}
\affiliation{Department of Physics, Taras Shevchenko National Kyiv University, Kyiv, 01601, Ukraine}
\affiliation{Bogolyubov Institute for Theoretical Physics, Kyiv, 03143, Ukraine}

\author{E. V. Gorbar}
\email{gorbar@knu.ua}
\affiliation{Department of Physics, Taras Shevchenko National Kyiv University, Kyiv, 01601, Ukraine}
\affiliation{Bogolyubov Institute for Theoretical Physics, Kyiv, 03143, Ukraine}

\author{P.~O.~Sukhachov}
\email{pavlo.sukhachov@missouri.edu}
\affiliation{Department of Physics and Astronomy, University of Missouri, Columbia, Missouri, 65211, USA}
\affiliation{MU Materials Science \& Engineering Institute,
University of Missouri, Columbia, Missouri, 65211, USA}
\affiliation{Center for Quantum Spintronics, Department of Physics, Norwegian University of Science and Technology, NO-7491 Trondheim, Norway}

\begin{abstract}
We calculate the viscoelasticity tensor for altermagnets and formulate the corresponding hydrodynamic equations. The anisotropy of altermagnetic Fermi surfaces allows for additional terms in the viscoelasticity tensor and is manifested in transport properties, including electron and spin flows in a channel and nonlocal responses. In the channel geometry, the altermagnetic spin splitting leads to nontrivial spin density and spin current. Like the electric current, the spin current acquires a Poiseuille profile for no-slip boundary conditions. In nonlocal responses, the altermagnetic anisotropy affects current streamlines and electric potential distributions in the viscous regime. Our results provide signatures of the hydrodynamic transport regime in altermagnets, potentially facilitating its experimental studies and discovery.
\end{abstract}

\maketitle

\section{Introduction}
\label{sec:Introduction}

The search for novel transport regimes is an important direction in condensed matter physics. In addition to the well-known Ohmic and ballistic electron transport regimes, a hydrodynamic regime has attracted significant theoretical and experimental attention. Being first proposed by R.~Gurzhi~\cite{Gurzhi:1963, Gurzhi:1968}, electron hydrodynamics is characterized by dominant electron-electron scattering and is manifested in several effects such as the Poiseuille profile of electric current, formation of vortices of electron fluid, Gurzhi effect, etc. Electron hydrodynamics is reviewed in Refs.~\cite{Lucas-Fong:rev-2017, Narozhny:rev-2019, Narozhny:rev-2022, Fritz-Scaffidi-HydrodynamicElectronicTransport-2024}.

Electron hydrodynamics was observed in GaAs heterostructures~\cite{Molenkamp-Jong:1994, Jong-Molenkamp:1995} and, recently, in graphene~\cite{Crossno-Fong:2016, Ghahari-Kim:2016, Krishna-Falkovich:2017, Berdyugin-Bandurin:2018, Bandurin-Falkovich:2018, Ku-Walsworth:2019, Sulpizio-Ilani:2019, Samaddar-Morgenstern:2021, Kumar-Ilani:2021, Jenkins-BleszynskiJayich-ImagingBreakdownOhmic-2022, Krebs-Brar-ImagingBreakingElectrostatic-2023}. Graphene is a particularly suitable material for the realization of electron hydrodynamics because it is clean, mechanically stiff, and has an easily tunable Fermi energy. Therefore, crucial for hydrodynamics electron-electron collisions become dominant for a wide range of temperatures. This range of temperatures is often called the hydrodynamic window. Other two-dimensional (2D) materials might also support electron hydrodynamics since the hydrodynamic window is generically favorable there. Signatures of electron hydrodynamics were also reported in three-dimensional (3D) materials such as the Weyl semimetal WP$_2$~\cite{Gooth-Felser:2018, Jaoui-Behnia-DepartureWiedemannFranz-2018}.

A common assumption made in analyzing hydrodynamic transport is the isotropic approximation for the electron band dispersion. Although it is an excellent approximation in graphene with its isotropically dispersing Dirac cones at low energy, it may not be a good approximation in other materials. The role of anisotropy of the band dispersion in electron hydrodynamics was considered in Refs.~\cite{Link-Schmalian:2017, Rao-Bradlyn:2019, Offertaler-Bradlyn:2019, Cook-Lucas:2019, Pena-Benitez-Surowka:2019, Varnavides-Narang:2020, Robredo-Bradlyn:2021, Huang-Lucas:2022, Herasymchuk-Sukhachov-ViscoelasticResponseAnisotropic-2024}. It was found that anisotropy is imprinted in the viscosity tensor, which contains additional components compared with its counterpart in isotropic systems, and can be manifested in transport, collective modes, etc.

Recently, materials with anisotropic spin-polarized dispersion relation has attracted significant attention. These materials, dubbed altermagnets~\cite{Smejkal-Sinova:2020, Smejkal-Jungwirth:2022, Smejkal-Jungwirth-ConventionalFerromagnetismAntiferromagnetism-2022}, are characterized by a combined symmetry including lattice rotations and spin reversal; while inversion symmetry is preserved, the time-reversal symmetry (TRS) is broken in altermagnets. This combined spin-lattice symmetry leads to even-parity spin-polarized Fermi surfaces with $d$-, $g$-, or $i$-wave symmetry~\cite{Noda-Nakamura-MomentumdependentBandSpin-2016, Smejkal-Sinova:2020, Hayami-Kusunose-MomentumDependentSpinSplitting-2019, Ahn-Kunes:2019, Yuan-Zunger:2020, Yuan-Zunger-PredictionLowZCollinear-2021, Ma-Liu-MultifunctionalAntiferromagneticMaterials-2021, Smejkal-Jungwirth:2022, Smejkal-Jungwirth-ConventionalFerromagnetismAntiferromagnetism-2022, Mazin:2022}.
Due to their large nonrelativistic spin splitting, altermagnets allow for several transport phenomena, including the spin-splitter effect~\cite{Gonzalez-Hernandez-Zelezny:2021, Karube-Nitta:2022, Bai-Song:2022, Bose-Ralph:2022}, giant tunneling magnetoresistance~\cite{Smejkal-Jungwirth-GiantTunnelingMagnetoresistance-2022, Samanta-Tsymbal-TunnelingMagnetoresistanceMagnetic-2023, Cui-Yang-GiantSpinHallTunneling-2023}, the magnetoelectric effect~\cite{Oike-Peters-NonlinearMagnetoelectricEffect-2024, Smejkal-AltermagneticMultiferroicsAltermagnetoelectric-2024, Sun-Cheng-RobustMagnetoelectricCoupling-2024}, and the Hall drag effect~\cite{Lin-Xie-CoulombDragAltermagnets-2024}, which occur due to the Coulomb drag in multilayer altermagnets. Transport phenomena in altermagnet-superconductor heterostructures are also rich. For example, altermagnetic spin-splitting is manifested in Andreev reflection~\cite{Sun-Linder:2023, Papaj:2023, Beenakker-Vakhtel:2023, Nagae-Ikegaya-SpinpolarizedSpecularAndreev-2024, Das-Soori-CrossedAndreevReflection-2024}, $0-\pi$ oscillations in the DC Josephson effect~\cite{Ouassou-Linder:2023, Zhang-Neupert-FinitemomentumCooperPairing-2024, Sun-Trauzettel-TunableSecondHarmornic-2025}, the magneto-electric effect~\cite{Giil-Brataas:2024hat, Zyuzin-MagnetoelectricEffectSuperconductors-2024, Hu-Song-NonlinearSuperconductingMagnetoelectric-2024, Kokkeler-Bergeret-QuantumTransportTheory-2024}, the thermoelectric effect~\cite{Sukhachov-Linder-ThermoelectricEffectAltermagnetSuperconductor-2024}, etc.

Material candidates, such as MnF$_2$, ultrathin films of RuO$_2$, Mn$_5$Si$_3$, MnTe, V$_2$Se$_2$O, V$_2$Te$_2$O, CrO, CrSb, and CoNb$_4$Se$_8$ have been suggested, see Refs.~\cite{Smejkal-Jungwirth:2022, Bai-Yao-AltermagnetismExploringNew-2024} for more comprehensive list of candidates. Signatures of spin-split electron bands were experimentally observed in Refs.~\cite{Krempasky-Jungwirth:2024, Osumi-Sato-ObservationGiantBand-2024, Reimers-Jourdan:2023, Zeng-Liu-ObservationSpinSplitting-2024, Ding-Shen-LargeBandsplittingWave-2024, Jeong-Jalan-AltermagneticPolarMetallic-2024, Betancourt2024, Weber-Schneider-AllOpticalExcitation-2024, Li-Brink-TopologicalWeylAltermagnetism-2024, Regmi-Ghimire-AltermagnetismLayeredIntercalated-2024, Dale-Griffin-NonrelativisticSpinSplitting-2024, Zhang-Chen-CrystalsymmetrypairedSpinValley-2025, Jiang-Qian-MetallicRoomtemperatureDwave-2025, Sakhya-Neupane-ElectronicStructureLayered-2025} by using spectroscopic and transport measurements, see also Ref.~\cite{Song-Pan-AltermagnetsNewClass-2025} and the references therein. However, the field of altermagnetism is still in its infancy and more different experimental probes are needed to narrow down the list of potential candidates.

In this work, we analyze the signatures of electron hydrodynamics in altermagnetic metals paying special attention to the role of anisotropy in these materials. By using a low-energy effective model of a 2D $d$-wave altermagnet, we calculate the viscoelasticity tensor and derive the hydrodynamic equations. It is found that the anisotropic band structure of altermagnets is imprinted in the viscoelasticity tensor, which acquires additional components determined by the altermagnetic spin-splitting. This anisotropy, in turn, affects the flow of electron fluid in channel geometries and can be probed via nonlocal responses. Depending on the relative orientation of the channel and altermagnetic crystal axes, the altermagnetic spin splitting can allow for nontrivial spin density and spin current. Similar to the electric current, the spin current along the channel acquires a Poiseuille profile for no-slip boundary conditions. In nonlocal responses, the altermagnetic anisotropy affects current streamlines and electric potential distributions. Weak SOC does not qualitatively affect the electron flows, but results in the Hall part of the viscoelasticity tensor and anomalous Hall effect (AHE) component of the current. The proposed results establish the signatures of electron hydrodynamics in altermagnets and suggest potential ways to investigate the interplay of altermagnetism and hydrodynamics in experimentally relevant settings.

The paper is organized as follows. In Sec.~\ref{sec:Model-Kubo}, we describe the low-energy model of altermagnets and calculate the viscoelasticity tensor in the Kubo approach. In Sec.~\ref{sec:Hydrodynamics}, we derive the hydrodynamic equations in the kinetic framework. Electron hydrodynamic flow in channel and nonlocal geometries of altermagnets is explored in Secs.~\ref{sec:Flow} and \ref{sec:nonlocal-responses}, respectively. The effects of spin-orbital coupling (SOC) are discussed in Sec.~\ref{sec:SOC}. The results are summarized in Sec.~\ref{sec:summary}. Technical details are presented in Appendixes~\ref{sec:App-Kubo}, \ref{sec:App-Spectral-function}, and \ref{sec:App-Hydro}. Throughout this paper, we use $\hbar=k_{\rm B}=1$.

\section{Kubo approach for viscoelasticity tensor}
\label{sec:Model-Kubo}

\subsection{Model of altermagnets}
\label{sec:Model}

We start by defining the effective low-energy model of altermagnets. For the sake of definiteness, we focus on $d$-wave altermagnets. The generalization to the case of other symmetries is straightforward, albeit leads to more cumbersome expressions, see, e.g., Appendix~\ref{sec:App-Kubo-giwave}.

The Hamiltonian of electron quasiparticles in a 2D $d$-wave altermagnet reads as
\begin{equation}
\label{eq:hamiltonian}
H=t_0 \left( k_x^2+ k_y^2 \right) + \left[ t_{1} \left( k_y^2 - k_x^2 \right) +2 t_{2} k_x k_y \right] \sigma_z - \hat{\mu}.
\end{equation}
Here, $\mathbf{k}$ is the momentum vector, $\sigma_{z}$ is the Pauli matrix in the spin space, $\hat{\mu}=\text{diag} \left( \mu_{+}, \mu_{-} \right)$, where $\mu_{\lambda}$ is the chemical potential for the spin projection $\lambda=\pm$~\footnote{The spin-dependent effective chemical potential can be achieved by applying a magnetic field via the Zeeman term. In 2D systems, orbital effects can be ignored if the field is in the plane of the material.}, parameter $t_0$ is related to the effective mass of electron quasiparticles $m$ ($t_0=1/(2m)$), and parameters $t_1$ and $t_2$ determine the orientation and strength of the altermagnetic spin splitting. To obtain finite Fermi surfaces, we assume that the parameters satisfy the following inequality: $t_0^2>t_1^2+t_2^2$.

\begin{figure}[h]
\includegraphics[width=.45\textwidth]{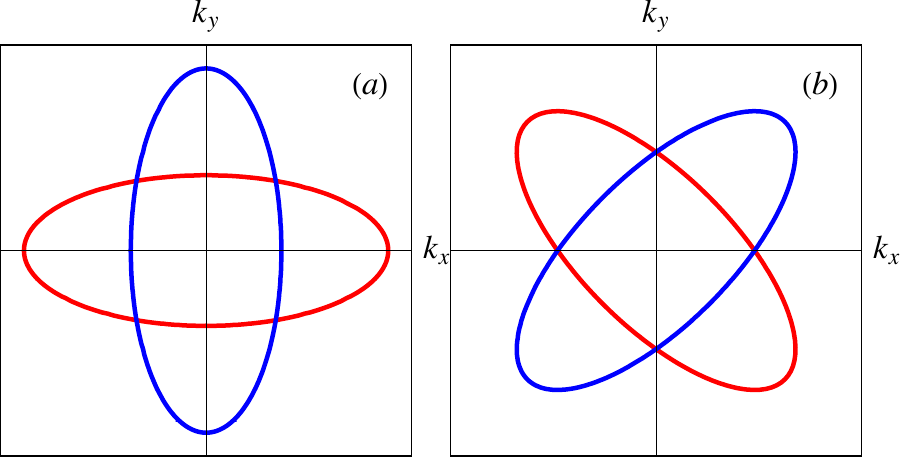}
\caption{The schematic band structure of altermagnet given in Eq.~\eqref{eq:spectra} for (a) $t_1\neq0$, $t_2=0$ and (b) $t_1=0$, $t_2\neq0$.
}
\label{fig:band}
\end{figure}

The dispersion relation of electron quasiparticles reads
\begin{equation}
\label{eq:spectra}
\varepsilon_{\lambda}=t_0 \left( k_x^2+ k_y^2 \right) +\lambda \left[ t_{1} \left( k_y^2 - k_x^2 \right) +2 t_{2} k_x k_y \right]-\mu_{\lambda}.
\vspace{1mm}
\end{equation}
The corresponding band structure is schematically shown by red ($\lambda=+$) and blue ($\lambda=-$) ellipses in Fig.~\ref{fig:band}.

\subsection{Viscoelasticity tensor}
\label{sec:Kubo}

In calculating the viscoelasticity tensor, we follow Ref.~\cite{Herasymchuk-Sukhachov-ViscoelasticResponseAnisotropic-2024}. The viscoelasticity tensor as a function of frequency $\Omega$ is defined via the stress tensor $T_{\mu\nu}$ and the strain transformation generator $J_{\mu\nu}$ as follows~\cite{Bradlyn-Read:2012}, see also Appendix~\ref{sec:App-Kubo},
\begin{equation}
\label{eq:viscoelasticity-def}
\begin{aligned}
\eta_{\mu \nu \alpha \beta}(\Omega)&=\frac{1}{\Omega+i0} \left[  \left< [T_{\mu \nu}^{\,\,\,}(0), J_{\alpha \beta}^{\,\,\,}(0)] \right>+i\delta_{\alpha \beta}^{\,\,\,} \left< T_{\mu \nu}^{\,\,\,} \right> \right.\\
&\left.-i  \delta_{\mu \nu} \delta_{\alpha \beta}\kappa^{-1} - i C_{\mu \nu \alpha \beta}(\Omega)\right],
\end{aligned}
\end{equation}
where $C_{\mu \nu \alpha \beta}(\Omega)$ is the Fourier transform of the stress-stress correlation function
\begin{equation}
\label{eq:C-corr-def}
\begin{aligned}
C_{\mu \nu \alpha \beta}(t-t') &=i \lim_{\varepsilon \to +0} \Theta(t-t') \left< [T_{\mu \nu}^{\,\,\,}(t), T_{\alpha \beta}^{\,\,\,}(t')] \right> \\
&\times e^{-\varepsilon (t-t')}.
\end{aligned}
\end{equation}
Here, $\kappa^{-1}=-V\left(\partial P/\partial V \right)_{N}$ is the inverse isentropic compressibility defined as the derivative of pressure with respect to the volume of fluid $V$ taken at fixed particle number $N$.

By using the spectral function $A(\omega;\mathbf{k})$, see Appendix~\ref{sec:App-Spectral-function} for its definition, in the stress-stress correlator, the viscoelasticity tensor can be rewritten as~\cite{Bradlyn-Read:2012}
\begin{widetext}
\begin{equation}
\label{Kubo_viscoelastic_tensor}
\begin{aligned}
\eta_{\mu \nu \alpha \beta}(\Omega)&=\frac{1}{\Omega+i0} \bigg\{ \left[  \left< [T_{\mu \nu}^{\,\,\,}(0), J_{\alpha \beta}^{\,\,\,}(0)] \right>+i\delta_{\alpha \beta}^{\,\,\,} \left< T_{\mu \nu}^{\,\,\,} \right>  -i  \delta_{\mu \nu} \delta_{\alpha \beta}\kappa^{-1} \right] \\
&- i \int_{-\infty}^{+\infty} d \omega \int_{-\infty}^{+\infty} d \omega' \frac{f(\omega)-f(\omega')}{\omega'-\omega-\Omega -i 0}  \int \frac{d^2 k}{(2\pi)^2}  \text{tr} \left[ T_{\mu \nu}^{\,\,\,} (\mathbf{k}) A(\omega;\mathbf{k}) T_{\alpha \beta}^{\,\,\,} (\mathbf{k}) A(\omega';\mathbf{k}) \right] \bigg\}.
\end{aligned}
\end{equation}
\end{widetext}
Here, $f(\omega)=1/\left(e^{\omega/T}+1\right)$ is the Fermi-Dirac distribution function and $T$ is temperature. 
Here, $\left< [T_{\mu \nu}^{\,\,\,}(0), J_{\alpha \beta}^{\,\,\,}(0)] \right>$ is an equal-time correlator; see Eq.~(\ref{eq:X-def-fourier}).

For the Hamiltonian of the model under consideration, the stress tensor is defined as
\begin{equation}
\label{eq:stress-tensor}
T_{\alpha \beta} = i\left[ J_{\alpha \beta} , H \right] =k_{\alpha} \frac{\partial H}{\partial k_{\beta}},
\end{equation}
where $J_{\alpha \beta}= -\frac{1}{2} \left \{ \hat{x}_{\beta},\hat{k}_{\alpha} \right \}$. By calculating the corresponding traces in Eq.~(\ref{Kubo_viscoelastic_tensor}), the viscoelasticity tensor simplifies to
\begin{equation}
\label{eq:viscosity-tensor-0}
\begin{aligned}
\eta_{\mu \nu \alpha \beta}(\Omega)&=\frac{1}{\Omega+i0}  \left[  \left< [T_{\mu \nu}^{\,\,\,}(0), J_{\alpha \beta}^{\,\,\,}(0)] \right>+i\delta_{\alpha \beta}^{\,\,\,} \left< T_{\mu \nu}^{\,\,\,} \right>\right.\\
&\left.  -i  \delta_{\mu \nu} \delta_{\alpha \beta}\kappa^{-1} \right],
\end{aligned}
\end{equation}
see Appendix~\ref{sec:App-Kubo} for details.

By separating real and imaginary parts in Eq.~\eqref{eq:viscosity-tensor-0}, we derive the following real part of the viscoelasticity tensor:
\begin{equation}
\label{eq:anisotropy-eta-terms}
\begin{aligned}
\text{Re}\,\eta_{\mu \nu \alpha \beta}(\Omega) &= \text{Re}\,\eta_{\mu \nu \alpha \beta}^{\text{rot}}(\Omega)+ \text{Re}\,\eta_{\mu \nu \alpha \beta}^{\text{anis}}(\Omega)\\
&+\text{Re}\,\tilde{\eta}_{\mu \nu \alpha \beta}(\Omega),
 \end{aligned}
\end{equation}
where
\begin{equation}
\label{eq:anisotropy-eta-terms-rot}
\text{Re}\,\eta_{\mu \nu \alpha \beta}^{\text{rot}}(\Omega) = \pi \delta(\Omega) \epsilon \left( \delta_{\mu \beta} \delta_{\nu \alpha} +\delta_{\mu \alpha} \delta_{\nu \beta} -\delta_{\mu \nu} \delta_{\alpha \beta} \right)
\end{equation}
is the viscoelasticity tensor of rotationally invariant 2D systems,
\begin{equation}
\label{eq:anisotropy-eta-terms-anis}
\begin{aligned}
&\text{Re}\,\eta_{\mu \nu \alpha \beta}^{\text{anis}}(\Omega) = \pi \delta(\Omega) \frac{\epsilon }{\tilde{t}_0^2} \Big\{ t_2^2 \left( \delta_{\mu \alpha} + \delta_{\nu \beta} -1\right) \\
+& t_1 t_2 \left[ \eta_{\mu \alpha} \left( 1 - \delta_{\nu \beta} \right) +\eta_{\nu \beta} \left( 1 -\delta_{\mu \alpha}  \right)  \right]\\
+& 2 t_1^2  \delta_{\mu \alpha} \delta_{\nu \beta}  \left( \delta_{\mu 2} \delta_{\nu 1}+\delta_{\mu 1} \delta_{\nu 2} \right) \Big \},
\end{aligned}
\end{equation}
which combines the terms that originate from altermagnetic anisotropy
(indices 1 and 2 correspond to $x$ and $y$ directions, respectively),
and
\begin{equation}
\label{eq:anisotropy-eta-terms-tilde}
\begin{aligned}
&\text{Re}\,\tilde{\eta}_{\mu \nu \alpha \beta}(\Omega) = \pi \delta(\Omega) \frac{t_0 \tilde{\epsilon} }{\tilde{t}_0^2} \\
\times& \left[ t_2  \left( \delta_{\mu \alpha} - \delta_{\nu \beta} \right)  + t_1 \left( \eta_{\mu \alpha} \delta_{\nu \beta} - \delta_{\mu \alpha} \eta_{\nu \beta} \right) \right]
 \end{aligned}
\end{equation}
represents terms related to different chemical potentials $\mu_{\lambda}$, i.e., proportional to $\tilde{\epsilon}=\sum_{\lambda} \lambda \epsilon_{\lambda}$. In the expressions above, we used $\eta_{\mu \nu}=\delta_{\mu \nu} \left( \delta_{\mu 1}-\delta_{\mu 2} \right)$, $\epsilon=\sum_{\lambda} \epsilon_{\lambda}$, and
\begin{equation}
\label{epsilon-lambda-def}
\epsilon_\lambda =-\frac{T^2}{4 \pi \tilde{t}_0 } \text{Li}_{2} \left( -e^{\mu_{\lambda }/T}\right)
\end{equation}
with $\mbox{Li}_n(x)$ being the polylogarithm function~\cite{Bateman-Erdelyi:book-t1} and  $\tilde{t}_0=\left(t_0^2 -t_1^2 -t_2^2\right)^{1/2}$.

Thus, as presented in Eq.~\eqref{eq:anisotropy-eta-terms}, the altermagnetic spin-splitting leads to the anisotropy of the viscoelasticity tensor. The result in Eq.~\eqref{eq:anisotropy-eta-terms} is general and applicable to other symmetries of altermagnets, see Appendix~\ref{sec:App-Kubo}. The form of separate components given in Eqs.~\eqref{eq:anisotropy-eta-terms-anis} and \eqref{eq:anisotropy-eta-terms-tilde}, however, differs.

\section{Hydrodynamic equations}
\label{sec:Hydrodynamics}

In this section, we derive hydrodynamic transport equations using the kinetic framework. In the absence of the SOC, the bands are fully spin-polarized,
the Berry curvature vanishes, and the Boltzmann kinetic equation has the standard form
\begin{equation}
\label{eq:Boltzmann-def}
\frac{\partial f_{\lambda}}{\partial t} +\left( \mathbf{v}_{\mathbf{k},\lambda} \cdot \frac{\partial f_{\lambda}}{\partial \mathbf{r}}\right)+ \left( e \mathbf{E}\cdot \frac{\partial f_{\lambda}}{\partial \mathbf{k}} \right) = I_{\text{col}}\left\{ f_{\lambda} \right\},
\end{equation}
where $f_{\lambda}=f_{\lambda}(t,\mathbf{r},\mathbf{k})$ is the distribution function, $\mathbf{v}_{\mathbf{k},\lambda} = \partial_{\mathbf{k}} \varepsilon_{\mathbf{\lambda}}$ is the quasiparticle velocity, $e<0$ is the electron charge, $\mathbf{E}$ is an electric field, and $I_{\text{col}}\left\{ f_{\lambda} \right\}=I_{\text{ee}}\left\{ f_{\lambda} \right\} +I_{\text{imp}}\left\{ f_{\lambda} \right\}$ is the collision integral which includes both momentum-conserving electron-electron $I_{\text{ee}}\left\{ f_{\lambda} \right\}$ collisions and momentum-relaxing scattering off disorder $I_{\text{imp}}\left\{ f_{\lambda} \right\}$.

Electron-electron collisions are described via the Calaway ansatz~\cite{Callaway-ModelLatticeThermal-1959}
\begin{equation}
I_{\text{ee}}\left\{ f_{\lambda} \right\} =-\frac{f_{\lambda}-f^{(\mathbf{u})}_{\lambda} }{\tau_{\text{ee}}},
\end{equation}
where $\tau_{\text{ee}}$ is the electron-electron scattering time and we introduced the hydrodynamic ansatz~\cite{Gantmakher-Levinson-CarrierScatteringMetals-1987} for the distribution function
\begin{equation}
\label{eq:hydrodynamic-ansatz}
\begin{aligned}
f^{(\mathbf{u})}_{\lambda} (t,\mathbf{r},\mathbf{k})=\frac{1}{e^{\left[  \varepsilon_{\lambda } -\left( \mathbf{u} (t,\mathbf{r} )\cdot \mathbf{k}\right) \right]/T}+1}.
\end{aligned}
\end{equation}
The velocity $\mathbf{u} (t,\mathbf{r})$ is the same for both spin projections because electron-electron collisions are, in the leading-order approximation, agnostic to spin and lead to a common drift velocity~\cite{Gurzhi-Yanovsky:2006}. In the case of weak electron-electron scattering, i.e., away from the hydrodynamic regime, drift velocities can be different with the Coulomb drag providing coupling between spin-up and spin-down species~\cite{DAmico-Vignale-TheorySpinCoulomb-2000}.

The part of the collision integral describing scattering off impurities has the following form in the relaxation time approximation
\begin{equation}
I_{\text{imp}}\left\{ f_{\lambda} \right\} =-\frac{f_{\lambda}-f^{(\mathbf{0})}_{\lambda} }{\tau},
\end{equation}
where $\tau$ is the relaxation time. Since we assumed nonmagnetic disorder, the spin-flip processes are determined by the spin-orbital coupling (see, e.g., Ref.~\cite{Matsuo:2017}), which are expected to be weak in altermagnets. In what follows, we neglect spin-flip processes.

Assuming that deviations from the ideal hydrodynamic distribution function are weak, the distribution function $f_{\lambda}$ can be approximated as
\begin{equation}
\label{eq:Boltzmann-f-app}
f_{\lambda} (t,\mathbf{r},\mathbf{k}) \approx f^{(\mathbf{u})}_{\lambda} (t,\mathbf{r},\mathbf{k})
+ \delta f_{\lambda} (t,\mathbf{r},\mathbf{k}).
\end{equation}
In addition, for slow flows, $|\mathbf{u}| \ll v_F$, we linearize the distribution function
\begin{equation}
\label{eq:Boltzmann-reduced}
f^{(\mathbf{u})}_{\lambda} (t,\mathbf{r},\mathbf{k}) \approx f^{(\mathbf{0})}_{\lambda} -  \left( \mathbf{u} (t,\mathbf{r}) \cdot \mathbf{k}\right) \frac{\partial f^{(\mathbf{0})}_{\lambda } }{\partial \varepsilon_{\lambda }}.
\end{equation}

Then, hydrodynamic equations are calculated as moments of the Boltzmann equation \eqref{eq:Boltzmann-def} multiplied by $e$ for the continuity equation and $\mathbf{k}$ for the Euler or Navier-Stokes equation. We derive the following equations for the inviscid electron fluid:
\begin{eqnarray}
\label{eq:hydrodynamics-general-spin-flip-continuity}
\frac{\partial \rho_\lambda}{\partial t}+\left(\bm{\nabla} \cdot \rho_\lambda \mathbf{u} \left(t, \mathbf{r}\right) \right)&=&0,\\
\label{eq:hydrodynamics-general-spin-flip-Euler}
\frac{\partial}{\partial t}\left[ T_{ij} u_i\left( t, \mathbf{r} \right) \mathbf{e}_j \right]+ \bm{\nabla} \epsilon - \rho \mathbf{E}&=&- \frac{1}{\tau} T_{ij} u_i\left( t, \mathbf{r} \right) \mathbf{e}_j.\,\,
\end{eqnarray}
Here the tensor $\hat{T}$ is defined as
\begin{equation}
\label{eq:altermagnetic-tensor-T-v1}
\hat{T} = t_0 \omega + t_1 \tilde{\omega} \tau_z -t_2 \tilde{\omega} \tau_x
\end{equation}
with $\tau_{x,z}$ being Pauli matrices,
and we use the following shorthand notation for the spin-resolved electron charge density:
\begin{equation}
\label{rho-lambda-def}
\rho_\lambda =-\frac{e T}{4 \pi \tilde{t}_0 } \text{Li}_{1} \left( -e^{\mu_{\lambda }/T}\right).
\end{equation}
Other notations are $\omega_{\lambda} =  \rho_{\lambda} /(2 e\tilde{t}_0^2)$, $\rho = \sum_{\lambda } \rho_{\lambda }$, $\tilde{\rho} = \sum_{\lambda } \lambda \rho_{\lambda }$, $\omega = \sum_{\lambda } \omega_{\lambda }$, $\tilde{\omega} = \sum_{\lambda } \lambda \omega_{\lambda }$, and $\mathbf{e}_i$ is the unit vector the direction $i$.

The electron fluid viscosity modifies the right-hand side of Eq.~(\ref{eq:hydrodynamics-general-spin-flip-Euler}) via the term $\eta_{ijkl} \nabla_{j} \nabla_{l} u_{k}\left( t, \mathbf{r} \right) \mathbf{e}_i$. To determine the electron viscosity, one should consider the first-order correction to the distribution function $f_{\lambda}\left(t,\mathbf{r},\mathbf{k}\right)$ due to electron-electron scattering, see the second term in Eq.~\eqref{eq:Boltzmann-f-app}. From the Boltzmann equation, we obtain the following correction assuming $\tau_{ee} \ll \tau $:
\begin{equation}
\delta f_{\lambda}=-\tau_{ee}\left[ \frac{\partial}{\partial t} +\left( \mathbf{v}_{\mathbf{k},\lambda} \cdot \frac{\partial }{\partial \mathbf{r}}\right)+ \left( e \mathbf{E}\cdot \frac{\partial  }{\partial \mathbf{k}} \right)  \right] f^{(\mathbf{u})}_{\lambda}.
\end{equation}
To obtain the electron viscosity, we pick only the terms with two spatial derivatives of $\mathbf{u}\left(t,\mathbf{r}\right)$ from the correction $\delta f_{\lambda}$ after substituting it in the Boltzmann equation. As we showed in Ref.~\cite{Herasymchuk-Sukhachov-ViscoelasticResponseAnisotropic-2024}, there is an exact correspondence between the static parts of the dissipative viscoelastic response in the Kubo and kinetic approaches. Therefore, by using the results of the Kubo approach in Eq.~(\ref{eq:anisotropy-eta-terms}) and replacing $\pi \delta(\Omega) \rightarrow \tau_{ee}$, we obtain the viscosity terms in the hydrodynamic equations. The explicit form of the corresponding viscous hydrodynamic equations is presented in Eq.~(\ref{eq:viscosity-general}).

\section{Flow of electron fluid in channel geometry}
\label{sec:Flow}

To illustrate the hydrodynamic framework, let us consider flow in the channel geometry defined as $x \in \left( -\infty, +\infty \right)$ and $y \in \left[ 0, L \right]$ with no-slip boundary conditions $u_{x}(x,y=0, L)=0$ at edges of the channel.

Using the continuity equation \eqref{eq:hydrodynamics-general-spin-flip-continuity}, we obtain
\begin{equation}
\label{eq:A-def}
u_{y} \left(y\right) =\frac{C_1}{\rho (y)}=\frac{C_2}{ \tilde{\rho} (y)},
\end{equation}
where $C_1$ and $C_2$ are constants.

In the homogeneous case with constant $\rho$, $\tilde{\rho}$, $\epsilon$, and $\tilde{\epsilon}$, which is equivalent to constant $\mu_{\lambda}$ and $T$, Eq.~\eqref{eq:viscosity-general} implies the following velocity profile:
\begin{equation}
\label{eq:ux-def}
\begin{aligned}
u_x \left( y \right)  &=2 e \tau \tilde{t}_0^2 \frac{ \rho \left(t_0 \rho- t_1 \tilde{\rho} \right)  E_x+t_2 \rho \tilde{\rho} E_y }{ t_0^2 \rho^2- t_1^2 \tilde{\rho}^2  - t_2^2 \tilde{\rho}^2 } \\
&\times \left[ 1-\frac{\cosh \left(  \frac{L- 2y }{2 \lambda^{\text{G}}} \right)}{\cosh \left( \frac{L}{2 \lambda^{\text{G}}} \right)}\right],
\end{aligned}
\end{equation}
where the modified Gurzhi length in altermagnets is
\begin{equation}
\begin{aligned}
\lambda^{\text{G}}= \lambda_0^{\text{G}} \sqrt{ \frac{ t_0^2 + t_1^2 + 2 t_0 t_1   \tilde{\epsilon} /\epsilon   }{t_0^2  + t_0 t_1 \tilde{\rho} / \rho} }
\end{aligned}
\end{equation}
and $\lambda_0^{\text{G}}= \sqrt{2 e \tau \tau_{ee} t_0 \epsilon/\rho}$ is the Gurzhi length in the absence of altermagnetism. Using Eq.~(\ref{eq:viscosity-general}), we obtain the following constraint for values of $\rho$, $\tilde{\rho}$, $\epsilon$, and $\tilde{\epsilon}$:
\begin{equation}
\label{eq:A-constraint}
t_2 \left[t_0 \left( \epsilon \tilde{\rho} - \tilde{\epsilon} \rho \right) -t_1 \left( \epsilon \rho - \tilde{\epsilon} \tilde{\rho}\right)\right]=0 .
\end{equation}

Similarly, by using Eq.~(\ref{eq:viscosity-general}), we find the $y$-component of the fluid velocity
\begin{equation}
\label{eq:uy-def}
u_y = 2 e\tau \tilde{t}_0^2  \frac{t_2 \rho \tilde{\rho} E_x+  \rho \left(t_0 \rho +  t_1  \tilde{\rho} \right) E_y}{ t_0^2 \rho^2- t_1^2 \tilde{\rho}^2  - t_2^2 \tilde{\rho}^2 }.
\end{equation}
As follows from Eq.~\eqref{eq:A-constraint}, there are no constraints for $\rho$, $\tilde{\rho}$, $\epsilon$, and $\tilde{\epsilon}$ at $t_2=0$. This originates from the fact that the projections of Eq.~(\ref{eq:viscosity-general}) on the $x$ and $y$ directions decouple. However, since achieving pure altermagnetic splitting with $t_2 = 0$ can be challenging, we treat $t_2=0$ as the limit $t_{2} \rightarrow 0$.

For $t_1=0$, Eq.~(\ref{eq:A-constraint}) has only trivial solution $\mu_{+}=\mu_{-}$, where $\mu_{\lambda}=\mu + \lambda \tilde{\mu}$. Hence, there is no spin density generation in the channel. A nonzero parameter $t_1$, i.e., splitting perpendicular to edges of the channel, allows for $\tilde{\mu}\neq0$. For instance, assuming $T \rightarrow 0$ and at $\mu_{\lambda}>0$, we have
\begin{equation}
T^s \text{Li}_{s}\left( -e^{\mu_{\lambda }/ T}\right) \underset{T \rightarrow 0 }{\rightarrow}-\frac{\mu_{\lambda}^s }{\Gamma \left( s+1 \right)}
\end{equation}
in Eqs.~\eqref{epsilon-lambda-def} and \eqref{rho-lambda-def}. Therefore, Eq.~(\ref{eq:A-constraint}) has the following solution:
\begin{equation}
\label{tilde-mu-sol}
\tilde{\mu} = -\frac{t_1}{t_0} \mu
\end{equation}
signifying nonzero spin density in the system. This spin density originates from the incompatibility of the channel geometry and the symmetry of altermagnet; the generation of the edge magnetization considered in Ref.~\cite{Hodt-Linder-InterfaceinducedMagnetizationAltermagnets-2024} has a similar symmetry requirement, albeit a different microscopic origin. Note also that, according to Eqs.~\eqref{rho-lambda-def} and (\ref{tilde-mu-sol}), $\rho / e >0$ and $\tilde{\rho}/ e <0$. Therefore, the electric $\mathbf{j}=\rho \mathbf{u}$ and spin $\tilde{\mathbf{j}}=\tilde{\rho} \mathbf{u}$ currents flow in different directions.

The electric current density is
\begin{eqnarray}
\label{eq:el-current-x}
j_x(y)&=& \left[ 1-\frac{\cosh \left(  \frac{L-2y}{2 \lambda^{\text{G}}} \right)}{\cosh \left( \frac{L}{2\lambda^{\text{G}}} \right)}\right]  \frac{ 2 e \tau  \tilde{t}_0^2   \rho^2  \left(t_0 \rho- t_1 \tilde{\rho} \right)  }{ t_0^2 \rho^2- t_1^2 \tilde{\rho}^2  - t_2^2 \tilde{\rho}^2 } E_x \nonumber\\
&+& \left[ 1-\frac{\cosh \left(  \frac{L-2y}{2 \lambda^{\text{G}}} \right)}{\cosh \left( \frac{L}{2\lambda^{\text{G}}} \right)}\right]   \frac{  2 e \tau \tilde{t}_0^2 t_2 \rho^2  \tilde{\rho} }{ t_0^2 \rho^2- t_1^2 \tilde{\rho}^2  - t_2^2 \tilde{\rho}^2 }  E_y,  \nonumber\\
\\
\label{eq:el-current-y}
j_y(y)&=&  \frac{2 e \tau \tilde{t}_0^2 \rho^2  }{ t_0^2 \rho^2- t_1^2 \tilde{\rho}^2  - t_2^2 \tilde{\rho}^2 } \left[t_2 \tilde{\rho}  E_x+  \left(t_0 \rho +  t_1 \tilde{\rho} \right) E_y \right].\nonumber\\
\end{eqnarray}
The spin current density is given by the same expressions, albeit multiplied by $\tilde{\rho}/\rho$.

In the channel geometry, there should be no current through the boundary unless we apply contacts at the edges of the channel. By setting $j_y=0$, we find that the Hall field $E_y$ equals
\begin{equation}
E_y=-\frac{ t_2 \tilde{\rho}  }{t_0 \rho +  t_1 \tilde{\rho} } E_x.
\end{equation}
In this case, Eq.~\eqref{eq:el-current-x} simplifies as follows:
\begin{equation}
j_x(y)= \left[ 1-\frac{\cosh \left(  \frac{L-2y}{2 \lambda^{\text{G}}} \right)}{\cosh \left( \frac{L}{2\lambda^{\text{G}}} \right)}\right]  \frac{ 2 e \tau  \tilde{t}_0^2   \rho^2   }{t_0 \rho+ t_1 \tilde{\rho}} E_x.
\end{equation}
For the spin current, one has to multiply the above expression by $\tilde{\rho}/\rho$.

There are a few effects described by the above expressions. As expected for hydrodynamic flows, currents along the channel have Poiseuille profiles, i.e., they attain maximum in the middle of the channel and vanish at boundaries. The presence of altermagnetic splitting allows for Hall-like electric current $\mathbf{j}\perp\mathbf{E}$ if electric contacts are applied to $y=0,L$ edges or Hall-like voltage if there are no contacts, see the last term in Eq.~\eqref{eq:el-current-x} and the first term in Eq.~\eqref{eq:el-current-y}.

Furthermore, the generation of the spin density $\tilde{\rho}$ allows for the spin current. In addition to the longitudinal spin current, altermagnets whose lobes are misaligned with respect to the channel, i.e., at $t_2\neq0$, support the hydrodynamic version of the spin-splitter effect, which is the spin current perpendicular to the applied field; this current is allowed when contacts are applied to all edges of the channel. Unlike the spin-splitter effect discussed in Refs.~\cite{Gonzalez-Hernandez-Zelezny:2021, Karube-Nitta:2022, Bai-Song:2022, Bose-Ralph:2022}, the spin current requires altermagnets with $t_1\neq0$ and $t_2\neq0$. Moreover, the spin current component along the channel acquires the characteristic Poiseuille profile.

\section{Nonlocal responses}
\label{sec:nonlocal-responses}

As we show in this section, in addition to the electric and spin currents in the channel, the anisotropy of the Fermi surfaces of altermagnets is also manifested in the nonlocal transport. We consider the same channel geometry as in Sec.~\ref{sec:Flow} and focus on the stationary transport regime. We assume that deviations caused by external sources are weak, hence, the transport equations can be linearized in $E$, $u$, and $\delta \mu_{\lambda}$. This allows us to represent the charge and energy densities as
$\rho_{\lambda} \approx \rho_{\lambda}^{(0)}+\delta \rho_{\lambda}$ and $\epsilon_{\lambda} \approx \epsilon_{\lambda}^{(0)}+\delta \epsilon_{\lambda}$, respectively. Here, $\delta \rho_{\lambda} \propto \delta \mu_{\lambda}$ and the following relation is useful:
\begin{equation}
\label{eq:nonlocal-deps}
\delta \epsilon_{\lambda}=\frac{\partial \epsilon_{\lambda}}{\partial \mu_{\lambda } } \delta \mu_{\lambda}=-\frac{\rho_{\lambda}^{(0)}}{e} \delta \mu_{\lambda}.
\end{equation}

Performing linearization and using the relaxation-time approximation, the hydrodynamic equations given in Eq.~(\ref{eq:viscosity-general}) simplify as follows:
\begin{equation}
\left(\bm{\nabla} \cdot \mathbf{u} \left(\mathbf{r}\right) \right)=0,\\
\end{equation}
\begin{widetext}
\begin{equation}
\label{eq:euler-nonlocal}
\begin{aligned}
\rho^{(0)} \bm{\nabla} \bar{\phi}^{\text{el}}
&=-\frac{t_0 \omega^{(0)}}{\tau}   \mathbf{u}  -\frac{t_1 \tilde{\omega}^{(0)}}{\tau}  \left( u_x \mathbf{e}_x -  u_y    \mathbf{e}_y \right)  +\frac{t_2 \tilde{\omega}^{(0)}}{\tau}   \left( u_y   \mathbf{e}_x+ u_x   \mathbf{e}_y \right)  +\tau_{ee} \epsilon^{(0)} \Delta \mathbf{u }   \\
&+ 2\frac{\tau_{ee} t_1^2}{\tilde{t}_0^2} \epsilon^{(0)} \left( \nabla_{x}^2 u_{y} \mathbf{e}_y +\nabla_{y}^2 u_x   \mathbf{e}_x \right)+ \frac{\tau_{ee} t_2^2}{\tilde{t}_0^2} \epsilon^{(0)}  \left[  \Delta \mathbf{u } - 2\nabla_x \nabla_{y} u_y \mathbf{e}_x-2\nabla_x \nabla_{y} u_x  \mathbf{e}_y \right] \\
    &+ \frac{\tau_{ee} t_1 t_2}{\tilde{t}_0^2} \epsilon^{(0)} \left[ 2 \nabla_{x}\nabla_y \left( u_x  \mathbf{e}_x -u_y  \mathbf{e}_y \right) +\left( \nabla_{x}^2-\nabla_{y}^2 \right) \left( u_y   \mathbf{e}_x+u_x   \mathbf{e}_y  \right)  \right] \\
    &+ 2 \frac{\tau_{ee} t_0 t_1}{\tilde{t}_0^2}\tilde{\epsilon}^{(0)} \left[ \nabla_{y}^2 u_x \mathbf{e}_x - \nabla_{x}^2 u_y  \mathbf{e}_y \right] +  \frac{\tau_{ee} t_0 t_2}{\tilde{t}_0^2}\tilde{\epsilon}^{(0)}   \left[ 2\nabla_x \nabla_{y}  \mathbf{u}-  \Delta \left(u_y  \mathbf{e}_x + u_x  \mathbf{e}_y \right) \right],  \\
\end{aligned}
\end{equation}
where $\bar{\phi}^{\text{el}}=-\delta \mu / e + \phi$ is the electrochemical potential. In this section, we focus on a ribbon geometry with $x\in \left(-\infty, \infty\right)$. The results for a rectangular sample are presented in Sec.~\ref{sec:visco-fem}.

In 2D incompressible fluids, it is convenient to introduce the scalar stream function $\psi\left( \mathbf{r} \right)$ as~\cite{Landau:t6-2013}
\begin{equation}
\mathbf{u} \left(\mathbf{r}\right) =\left\{ -\nabla_{y} \psi\left( \mathbf{r}\right), \nabla_{x} \psi\left( \mathbf{r}\right) \right\}.
\end{equation}

By taking the curl of Eq.~(\ref{eq:euler-nonlocal}) we obtain the equation for $\psi(\mathbf{r})$
\begin{equation}
\label{eq:euler-nonlocal-psi-1}
\begin{aligned}
0&=-\frac{1}{\tau}\left[  t_0 \omega^{(0)} \nabla^2  + t_1 \tilde{\omega}^{(0)} \left( \nabla_y^2 -\nabla_x^2 \right)  +2 t_2 \tilde{\omega}^{(0)} \nabla_x \nabla_y  \right] \psi +\tau_{ee} \epsilon^{(0)} \nabla^4 \psi+ 2\frac{\tau_{ee} t_1^2}{\tilde{t}_0^2} \epsilon^{(0)} \left( \nabla_{x}^4  +\nabla_{y}^4 \right) \psi \\
& + 4\frac{\tau_{ee} t_1 t_2}{\tilde{t}_0^2} \epsilon^{(0)}  \nabla_{x}\nabla_y \left( \nabla_y^2- \nabla_x^2 \right) \psi + \frac{\tau_{ee} t_2^2}{\tilde{t}_0^2} \epsilon^{(0)}  \left( \nabla^4 +4\nabla_x^2 \nabla_y^2 \right)\psi + 2 \frac{\tau_{ee} t_0 t_1}{\tilde{t}_0^2}\tilde{\epsilon}^{(0)} \left( \nabla_{y}^4  - \nabla_{x}^4 \right) \psi+ 4 \frac{\tau_{ee} t_0 t_2}{\tilde{t}_0^2} \tilde{\epsilon}^{(0)}\nabla^2 \nabla_x \nabla_y \psi,  \\
\end{aligned}
\end{equation}
where $\nabla^4=(\nabla^2)^2$ is the biharmonic operator.

Since our system is infinite in the $x$-direction, we perform the Fourier transform with respect to $x$,
\begin{equation}
\label{eq:euler-psi-k-def}
\psi \left(\mathbf{r} \right) =\frac{1}{2\pi} \int_{-\infty}^{+\infty} dk \, e^{ikx} \psi_{k}(y).
\end{equation}
Then $\psi_{k}(y)$ satisfies the following equation:
\begin{equation}
\label{eq:euler-nonlocal-psi-2}
\begin{aligned}
0&= -\frac{t_0 \omega^{(0)}}{\tau}  \left( \nabla_y^2-k^2 \right) \psi_{k}(y) -\frac{ t_1 \tilde{\omega}^{(0)} }{\tau}\left( \nabla_y^2 +k^2 \right) \psi_{k}(y) - 2i\frac{t_2 \tilde{\omega}^{(0)} }{\tau} k \nabla_y \psi_{k}(y) +\tau_{ee} \epsilon^{(0)} \left( \nabla_y^2-k^2\right)^2 \psi_{k}(y) \\
&+ 2\frac{\tau_{ee} t_1^2}{\tilde{t}_0^2} \epsilon^{(0)} \left( k^4  +\nabla_{y}^4 \right) \psi_{k}(y) + 4i \frac{\tau_{ee} t_1 t_2}{\tilde{t}_0^2} \epsilon^{(0)} k \nabla_y \left( \nabla_y^2+ k^2 \right) \psi_{k}(y) + \frac{\tau_{ee} t_2^2}{\tilde{t}_0^2} \epsilon^{(0)}  \left[ \left( \nabla_y^2-k^2\right)^2 -4k^2 \nabla_y^2 \right]\psi_{k}(y)  \\
    &+ 2 \frac{\tau_{ee} t_0 }{\tilde{t}_0^2}\tilde{\epsilon}^{(0)} \left[ t_1 \left( \nabla_{y}^4  - k^4 \right) + 2 i t_2 k \nabla_y \left( \nabla_y^2-k^2 \right) \right] \psi_k(y).
\end{aligned}
\end{equation}

We consider the no-slip boundary conditions $u_{x}(x,y=0,L)=0$ at the edges of the channel and define the source and drain of the electric current as $\rho^{(0)} u_{y}(x,y=0) = j_{1}(x)$ and $\rho^{(0)} u_{y}(x,y=L) = j_{2}(x)$. Therefore, the boundary conditions for Eq.~(\ref{eq:euler-nonlocal-psi-2}) read as follows
\begin{equation}
\label{eq:bc-nonlocal}
\nabla_y \psi_k(0)=0,  \quad \nabla_y \psi_k(L)=0, \quad
i k \rho^{(0)} \psi_k(0)=j_{1} (k),  \quad i k \rho^{(0)} \psi_k(L)=j_{2} (k).
\end{equation}
The electrochemical potential $\bar{\phi}^{\text{el}}\left( \mathbf{r} \right)$ can be obtained by substituting $\psi\left( \mathbf{r} \right)$ into Eq.~(\ref{eq:euler-nonlocal}). We have
\begin{eqnarray}
\label{eq:euler-phi-def}
\bar{\phi}^{\text{el}} \left(\mathbf{r} \right) &=&\frac{1}{2\pi \rho^{(0)}} \int_{-\infty}^{+\infty} dk \, \frac{e^{ikx}}{ik} \Big\{   \frac{t_0 \omega^{(0)} +t_1 \tilde{\omega}^{(0)}}{\tau} \nabla_y   + \frac{ik t_2 \tilde{\omega}^{(0)} }{\tau} -\tau_{ee} \tilde{\epsilon}^{(0)} \left[\frac{t_0 t_2}{\tilde{t}^2_0} ik \left( k^2-3 \nabla_y^2\right) -2\frac{t_0 t_1}{\tilde{t}^2_0} \nabla_y^3 \right] \nonumber\\
&-& \tau_{ee} \epsilon^{(0)} \left[ \nabla_{y} \left(\nabla_{y}^2 -k^2 \right) +2 \frac{t_1^2}{\tilde{t}_0^2} \nabla_{y}^3 +\frac{t_2^2}{\tilde{t}^2_0} \nabla_y\left(\nabla_{y}^2-3k^2 \right)+\frac{t_1 t_2}{\tilde{t}^2_0} i k \left(\nabla_{y}^2+3k^2 \right)\right]  \Big\} \psi_{k}(y).
\end{eqnarray}

In the purely viscous case, we set $\tau\to \infty$ while keeping $\tau_{ee}$ constant. Then Eq.~\eqref{eq:euler-nonlocal-psi-2} reads

\begin{eqnarray}
\label{eq:euler-nonlocal-psi-viscous}
0&=&\tau_{ee} \epsilon^{(0)} \left( \nabla_y^2-k^2\right)^2 \psi_{k}(y) + 2\frac{\tau_{ee} t_1^2}{\tilde{t}_0^2} \epsilon^{(0)} \left( \nabla_{y}^4  +  k^4  \right) \psi_{k}(y) + \frac{\tau_{ee} t_1 t_2}{\tilde{t}_0^2} \epsilon^{(0)} \left[ 4 i k\nabla_y \left( \nabla_y^2+ k^2 \right) \right] \psi_{k}(y) \nonumber \\
&+& \frac{\tau_{ee} t_2^2}{\tilde{t}_0^2} \epsilon^{(0)}  \left[ \left( \nabla_y^2-k^2\right)^2 -4k^2 \nabla_y^2 \right]\psi_{k}(y)
+ 2 \frac{\tau_{ee} t_0 t_1}{\tilde{t}_0^2}\tilde{\epsilon}^{(0)}  \left( \nabla_{y}^4  - k^4 \right) \psi_k(y)
+ \frac{\tau_{ee} t_0 t_2}{\tilde{t}_0^2}\tilde{\epsilon}^{(0)} \left[ 4 i k  \nabla_y \left( \nabla_y^2-k^2 \right) \right] \psi_k(y).\nonumber\\
\end{eqnarray}
\end{widetext}

We seek the solution in the form $\psi_{k}(y) = \sum_{i=1,4} A_i e^{\varkappa_{i} y}$ and determine eigenvalues $\varkappa_{i}$. For $\tilde{\epsilon}^{(0)}=0$, these eigenvalues are given by
\begin{equation}
\label{eq:euler-nonlocal-lambda}
\begin{aligned}
\varkappa_{1,2} &=-k\frac{t_2 \pm \sqrt{t_0^2+t_1^2+t_2^2}}{t_0- i t_1}, \\
\varkappa_{3,4} &=k\frac{t_2 \mp\sqrt{t_0^2+t_1^2+t_2^2}}{t_0+ i t_1}.
\end{aligned}
\end{equation}

By substituting the general solution into the boundary conditions \eqref{eq:bc-nonlocal}, we find
\begin{equation}
\begin{aligned}
A_i&=\frac{(-1)^{i}}{\text{det} \, K}\underset{\left(j,k,l\right)}{\mathfrak{S}} \Bigg[ \varkappa_{j} \left(\varkappa_{k}-\varkappa_{l}\right)   \\
& \times \left( \frac{j_{2}(k)}{i k \rho} e^{\varkappa_{j} L}  +\frac{j_{1}(k)}{i k \rho} e^{\left(\varkappa_{k} + \varkappa_{l}\right) L} \right)\Bigg], \\
\end{aligned}
\end{equation}
where $i\neq j,k,l$, $j<k<l$,
\begin{equation}
\begin{aligned}
\text{det}\,K &=\left( \varkappa_{1} -\varkappa_{2}\right)\left( \varkappa_{3} -\varkappa_{4}\right) \left( e^{\left(\varkappa_{1}+\varkappa_{2}\right)L}+e^{\left(\varkappa_{3}+\varkappa_{4}\right)L}\right)\\
&-\left( \varkappa_{1} -\varkappa_{3}\right)\left( \varkappa_{2} -\varkappa_{4}\right) \left( e^{\left(\varkappa_{1}+\varkappa_{3}\right)L}+e^{\left(\varkappa_{2}+\varkappa_{4}\right)L}\right)\\
&+\left( \varkappa_{1} -\varkappa_{4}\right)\left( \varkappa_{2} -\varkappa_{3}\right) \left( e^{\left(\varkappa_{1}+\varkappa_{4}\right)L}+e^{\left(\varkappa_{2}+\varkappa_{3}\right)L}\right),
\end{aligned}
\end{equation}
and $\underset{\left(j,k,l\right)}{\mathfrak{S}}$ denotes the sum over cyclic permutations. Finally, $\psi \left(\mathbf{r} \right)$ follows from Eq.~\eqref{eq:euler-psi-k-def} and the electrochemical potential is determined by Eq.~\eqref{eq:euler-phi-def}.

We consider two configurations of point-like contacts: two contacts on opposite sides and two contacts on the same side of the sample. In both cases, for simplicity, we focus on the case $\tilde{\epsilon}^{(0)}=0$.

\subsubsection{Contacts on the opposite sides}
\label{sec:nonlocal-responses-viscous-1}

In the setup where the source and drain are located symmetrically on opposite sides of the sample, we have
\begin{equation}
j_{1}(x)= j_{2}(x) =I\delta\left(x\right).
\end{equation}

\begin{figure}[t]
\subfigure[]{\includegraphics[width=.45\textwidth]{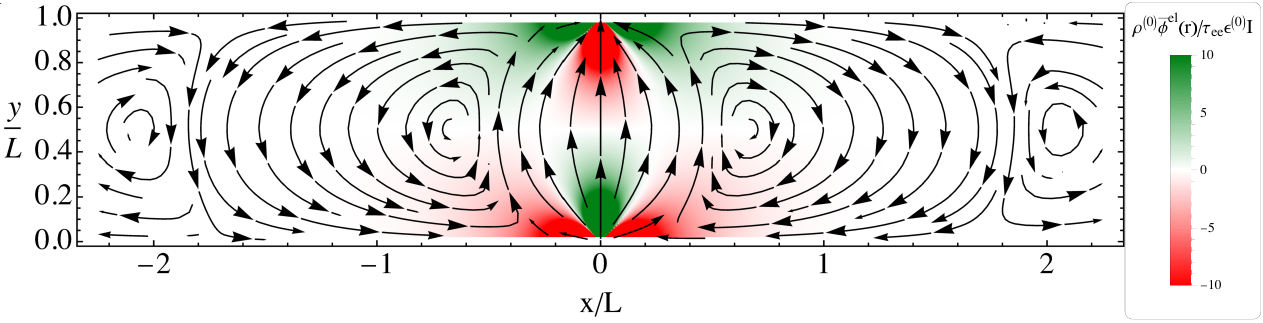}}
\subfigure[]{\includegraphics[width=.45\textwidth]{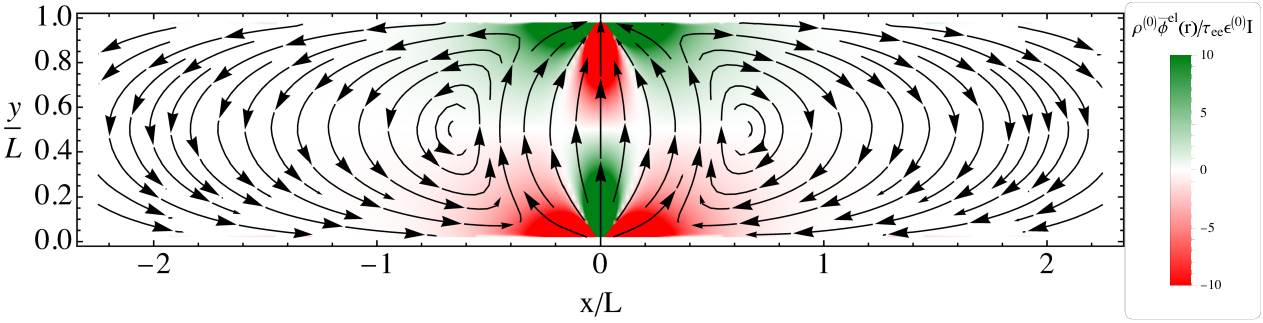}}
\subfigure[]{\includegraphics[width=.45\textwidth]{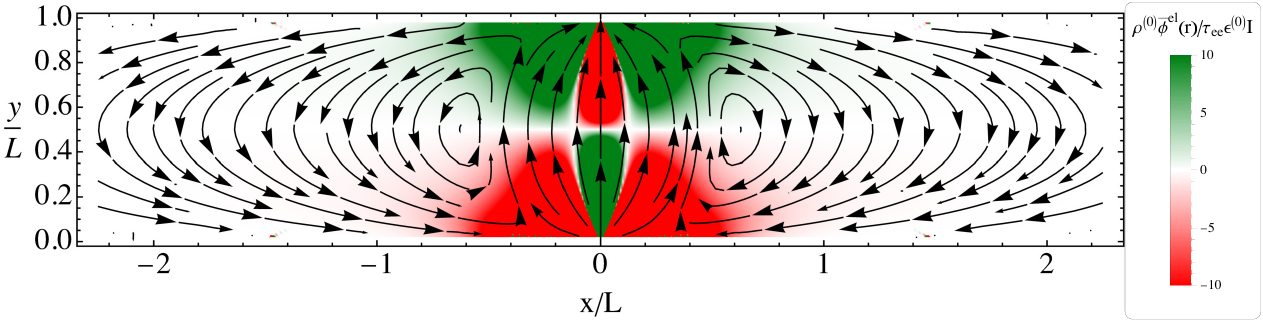}}
\caption{
The current streamlines ($\psi(\mathbf{r}) = \text{const}$) for (a) $t_2=0$, (b) $t_2=0.5\,t_0$, and (c) $t_2=0.9\,t_0$. The electrochemical potential $\rho^{(0)} \bar{\phi}^{\text{el}} \left(\mathbf{r}\right) / \tau_{ee} \epsilon^{(0)} I$ is shown in color. The green and red colors correspond to $\bar{\phi}^{\text{el}}(\mathbf{r})>0$ and $\bar{\phi}^{\text{el}}(\mathbf{r})<0$, respectively. In all panels, we set $t_1=0$ and $\tilde{\epsilon}^{(0)}=0$.}
\label{fig:streamlines1}
\end{figure}

At $t_1=0$ and $t_2 \neq 0$, the eigenvalues $\varkappa_{i}$ given in Eq.~\eqref{eq:euler-nonlocal-lambda} are simplified as follows:
\begin{equation}
\label{eq:eigenvalues_t2}
\varkappa_{i}=\pm k \left(\sqrt{1+\frac{t_2^2}{t_0^2} } \pm \frac{t_2}{t_0} \right).
\end{equation}
The current streamlines ($\psi(\mathbf{r}) = \text{const}$) and the electrochemical potential distributions are shown in Fig.~\ref{fig:streamlines1}.

The results for $t_{1,2}=0$ agree with the results presented in Refs.~\cite{Levitov-Falkovich:2016, Falkovich-Levitov:2017} for graphene. As one can see, the viscous flow generates vortices, which give rise to backflows near contacts. In agreement with Ref.~\cite{Danz-Narozhny:2019}, there are two well-formed pairs of vortices. Increasing the ratio $t_2/t_0$, the size of vortices increases. 
In general, the position of the center of the vortices is determined by the conditions
\begin{equation}
\begin{cases}
u_x(x, y)=0\\
u_y(x, y)=0
\end{cases} \Leftrightarrow\,\,\,\, \begin{cases}
\nabla_{y}\psi(x, y)=0\\
\nabla_{x}\psi(x, y)=0
\end{cases}.
\end{equation}
In the case with $\tilde{\epsilon}=0$, we have that the solutions for the center of the vortices are located on the line $y=L/2$. Hence, their positions are determined by $u_y(x, L/2)=0$ or $\nabla_{x}\psi(x, L/2)=0$.
The centers of vortices move further away from the axis $x=0$. 
While the electrochemical potential distribution is affected by the altermagnetic splitting, the effect is quantitative rather than qualitative.

\begin{figure}[t]
\subfigure[]{\includegraphics[width=.45\textwidth]{flow_viscous_t1=000_t2=000}}
\subfigure[]{\includegraphics[width=.45\textwidth]{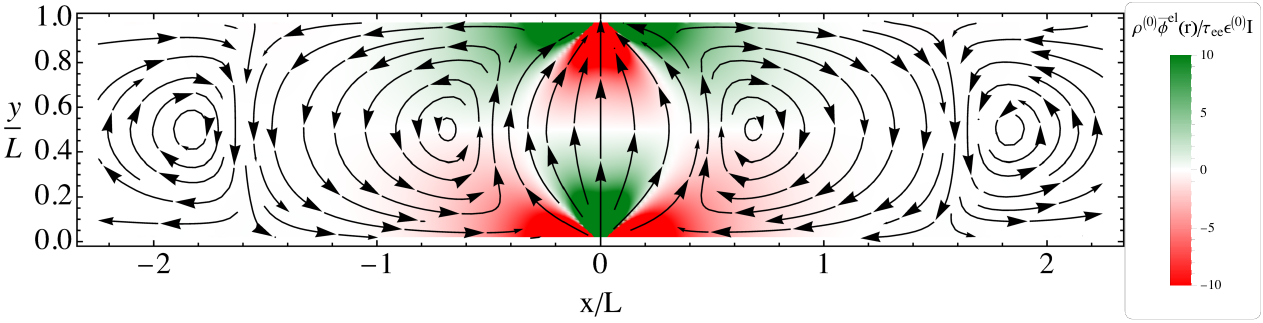}}
\subfigure[]{\includegraphics[width=.45\textwidth]{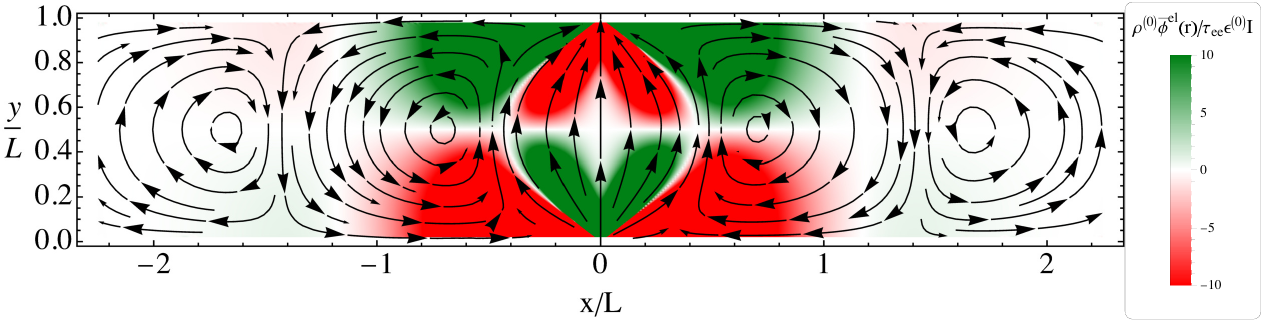}}
\caption{The current streamlines ($\psi(\mathbf{r}) = \text{const}$) for (a) $t_1=0$, (b) $t_1=0.5\,t_0$, and (c) $t_1=0.9\,t_0$. The electrochemical potential $\rho^{(0)} \bar{\phi}^{\text{el}} \left(\mathbf{r}\right) / \tau_{ee} \epsilon^{(0)} I$ is shown in color. The green and red colors correspond to $\bar{\phi}^{\text{el}}(\mathbf{r})>0$ and $\bar{\phi}^{\text{el}}(\mathbf{r})<0$, respectively. In all panels, we set $t_2=0$ and $\tilde{\epsilon}^{(0)}=0$.}
\label{fig:streamlines3}
\end{figure}

In the case $t_1\neq0$ and $t_2 = 0$, the eigenvalues $\varkappa_{i}$ given in Eq.~\eqref{eq:euler-nonlocal-lambda} simplify,
\begin{equation}
\label{eq:eigenvalues_t1}
\varkappa_{i}= \pm k \frac{ t_0 \pm i t_1}{\sqrt{t_0^2+t_1^2 }}.
\end{equation}
The corresponding current streamlines and electrochemical potential distributions are shown in Fig.~\ref{fig:streamlines3}. Unlike the case $t_2=0$, the size of vortices decreases as the ratio $t_1/t_0$ increases. The centers of vortices move closer to the axis $x=0$. As for the electrochemical potential $\bar{\phi}^{\text{el}} \left(\mathbf{r}\right)$, the altermagnetic splitting affects its shape, leading to the splitting of the central maximum.

\begin{figure}[t]
\subfigure[]{\includegraphics[width=.45\textwidth]{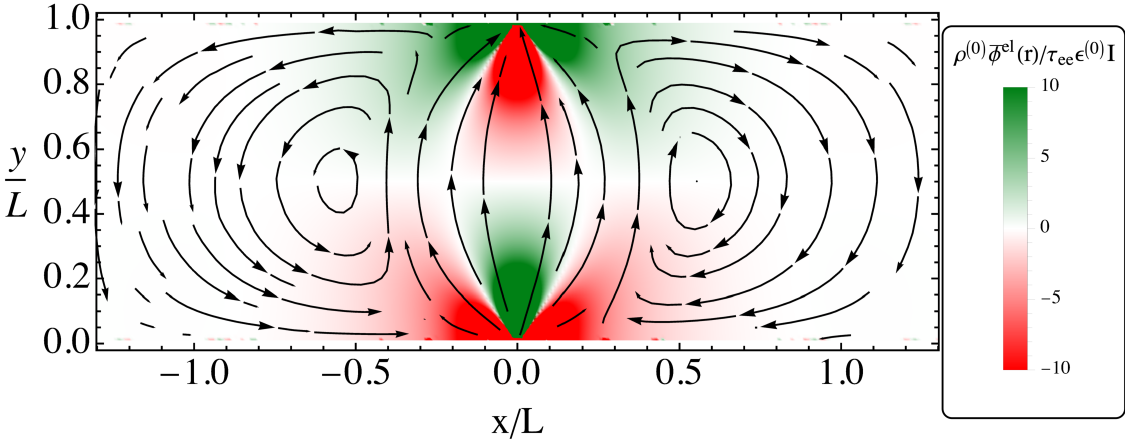}}
\subfigure[]{\includegraphics[width=.45\textwidth]{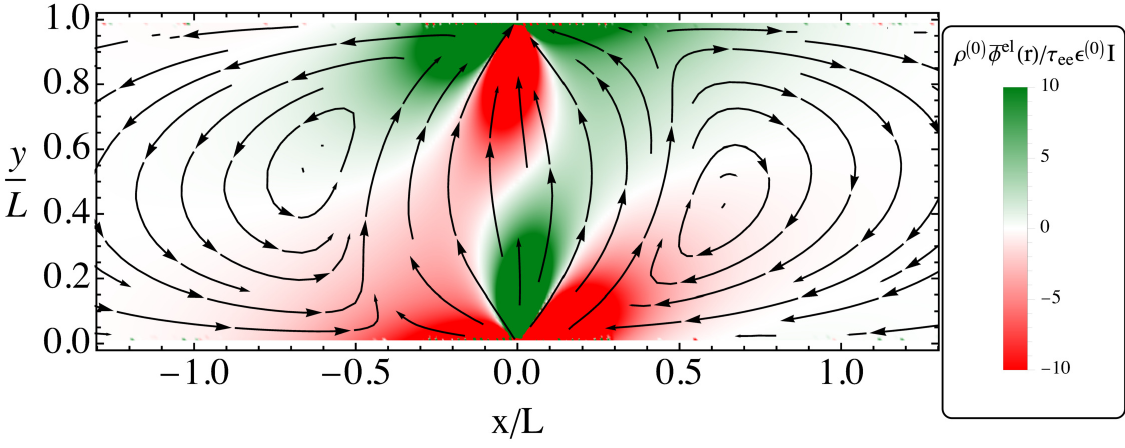}}
\caption{
The current streamlines ($\psi(\mathbf{r}) = \text{const}$) for $\tilde{\epsilon}^{(0)}=0.5 \epsilon^{(0)}$ (a) $t_1=0.5\,t_0$, $t_2=0$ and (b) $t_1=0$, $t_2=0.5\,t_0$. The electrochemical potential $\rho^{(0)} \bar{\phi}^{\text{el}} \left(\mathbf{r}\right) / \tau_{ee} \epsilon^{(0)} I$ is shown in color. The green and red colors correspond to $\bar{\phi}^{\text{el}}(\mathbf{r})>0$ and $\bar{\phi}^{\text{el}}(\mathbf{r})<0$, respectively.
}
\label{fig:streamlines1tilde}
\end{figure}

In the case $\tilde{\epsilon} \neq 0$, the effects of the altermagnetic spin splitting on the current streamlines and electric potential differ for the cases $t_1\neq 0$ and $t_2\neq 0$. 
The eigenvalues $\varkappa_i$ can be determined numerically from Eq.~(\ref{eq:euler-nonlocal-psi-viscous}) for each $\tilde{\epsilon}$. The current streamlines ($\psi(\mathbf{r}) = \text{const}$), including only the first vortices, and the electrochemical potential distributions are shown in Fig.~\ref{fig:streamlines1tilde}. The effects of the altermagnetic splitting $t_1\neq0$ are quantitative, with the splitting of the central maximum less pronounced. In the case $t_2\neq 0$, $\tilde{\epsilon}$ leads to the asymmetry of the electrochemical potential distribution. Additionally, the centers of the vortices are no longer located on the line $y=L/2$ for $t_2\neq 0$.

\subsubsection{Contacts on the same side}
\label{sec:nonlocal-responses-viscous-2}

When the source and drain are located on the same side of the channel (vicinity geometry), the boundary conditions are
\begin{equation}
j_{1}(x)=0, \quad
j_{2}(x)= I\delta\left(x-x_0\right)-I\delta\left(x+x_0\right).
\end{equation}
Their Fourier transform is given by
\begin{equation}
j_{1}(k)=0, \quad
j_{2}(k) =2iI\sin \left( k x_0 \right).
\end{equation}

\begin{figure}[t]
\subfigure[]{\includegraphics[width=.45\textwidth]{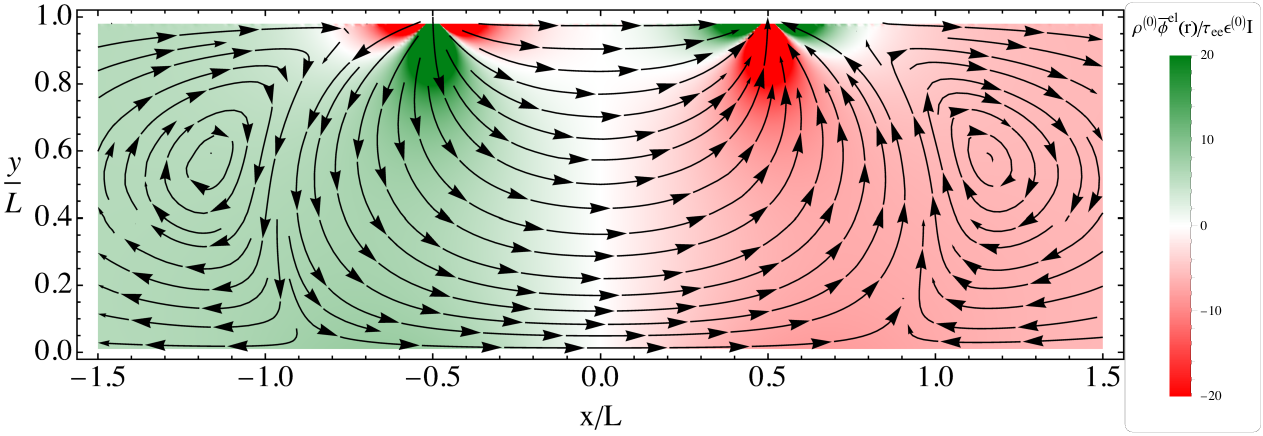}}
\subfigure[]{\includegraphics[width=.45\textwidth]{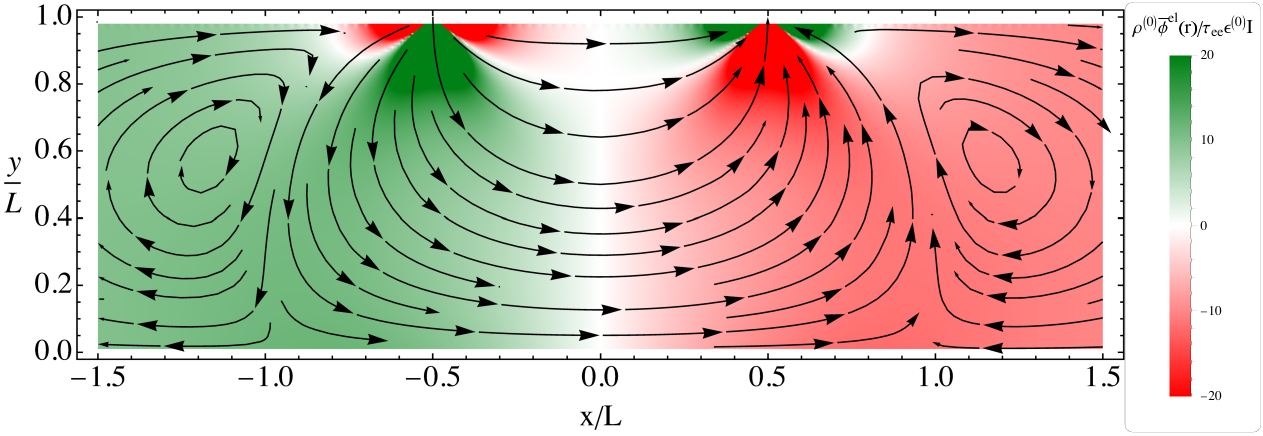}}
\subfigure[]{\includegraphics[width=.45\textwidth]{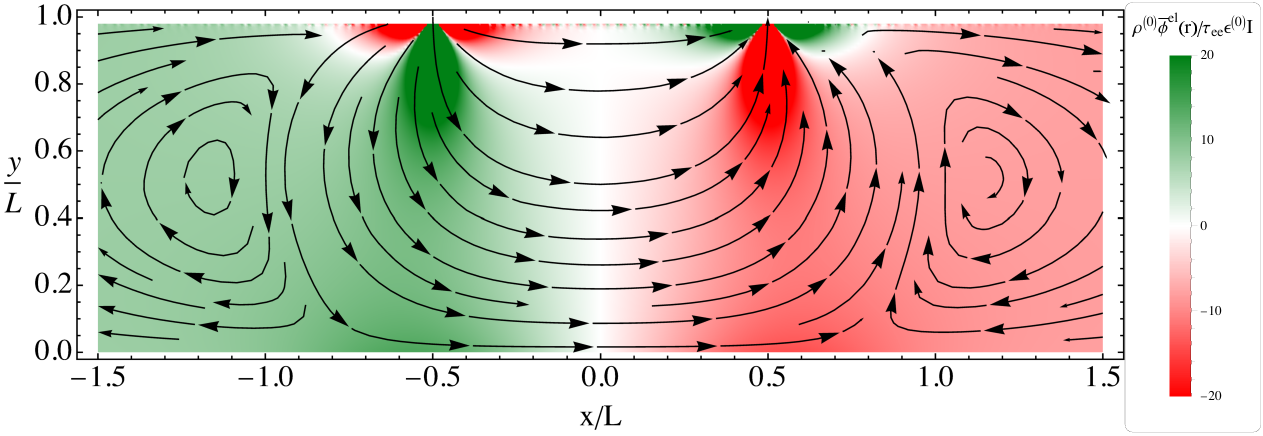}}
\caption{The current streamlines ($\psi(\mathbf{r}) = \text{const}$) for (a) $t_1=0,\,\,t_2=0$, (b) $t_1=0.5\, t_0,\,\,t_2=0$, and (c) $t_1=0,\,\,t_2=0.5\,t_0$ for $x_0=0.5\,L$. The electrochemical potential $\rho^{(0)} \bar{\phi}^{\text{el}} \left(\mathbf{r}\right) / \tau_{ee} \epsilon^{(0)} I$ is shown in color. The green and red colors correspond to $\bar{\phi}^{\text{el}} \left(\mathbf{r}\right)>0$ and $\bar{\phi}^{\text{el}} \left(\mathbf{r}\right)<0$, respectively. In all panels, we set $\tilde{\epsilon}^{(0)}=0$.
}
\label{fig:streamlines2}
\end{figure}

Using the eigenvalues $\varkappa_{i}$ given in Eqs.~(\ref{eq:eigenvalues_t2}) and (\ref{eq:eigenvalues_t1}), we plot the corresponding current streamlines in Fig.~\ref{fig:streamlines2} for a few combinations of altermagnetic parameters.

The streamlines and electrochemical potential at $t_{1,2}=0$ reproduce the results presented in Ref.~\cite{Torre-Polini:2015}. The equipotential lines are deformed at $t_2\neq0$. The changes are, however, quantitative rather than qualitative. At $t_1\neq0$, the electrochemical potential near contacts changes similarly to the case in Sec.~\ref{sec:nonlocal-responses-viscous-1}.

\section{Role of spin-orbital coupling}
\label{sec:SOC}

In the previous sections, we ignored the effects of SOC in altermagnets, which is usually weak compared with the nonrelativistic spin-splitting~\cite{Smejkal-Jungwirth-ConventionalFerromagnetismAntiferromagnetism-2022}. For the sake of completeness, we include the Rashba SOC term~\cite{Bychkov-Rashba-OscillatoryEffectsMagnetic-1984}
\begin{equation}
\label{eq:hamiltonian-soc-0}
H_{\text{soc}}=\alpha \left( \sigma_{x} k_y -\sigma_{y} k_x\right)
\end{equation}
in the Hamiltonian \eqref{eq:hamiltonian} and address its effects on the viscoelastic response and hydrodynamic flows of altermagnets. The details are provided in Appendixes~\ref{sec:visco-soc} and \ref{sec:App-Hydro-soc}.

Let us start with the viscoelastic tensor. The SOC leads to the emergence of dynamical and Hall viscosities in addition to modifying the static part in Eq.~\eqref{eq:anisotropy-eta-terms}. Indeed, the static part of the viscoelasticity tensor has a similar structure as in the case without the SOC, see Eq.~(\ref{eq:stat-viscoelasticity-soc}). Additional components of the viscoelasticity tensor appear due to asymmetry between bands. For example, the Hall component is realized at $\tilde{\mu}\neq0$,
\begin{equation}
\label{kubo-eta-1-soc-Hall-approx-napp}
\begin{aligned}
&\text{Re}\,\eta_{\mu \nu \alpha \beta}^{\text{Hall}}(0)  \approx -\frac{\alpha^2}{32\pi t_0^2} \left( \delta_{\nu 2} \delta_{\beta 1}-\delta_{\nu 1} \delta_{\beta 2} \right) \\
&\times  \sum_{\eta=\pm} \frac{ \eta \mu_{\eta}^2}{\tilde{\mu}^2}  \bigg[ \delta_{\mu \alpha} + \eta_{\mu \alpha}  \frac{\eta  \left(3 \tilde{\mu} - 4 \mu_{\eta} \right)}{3 \tilde{\mu}}  \frac{t_1}{t_0} \\
&-\left(1-\delta_{\mu \alpha}\right) \frac{\eta  \left(3 \tilde{\mu} - 4 \mu_{\eta} \right)}{3 \tilde{\mu}}  \frac{t_2}{t_0}+o\left( \frac{t_{1,2}}{t_0} \right) \bigg].
\end{aligned}
\end{equation}
Here $\eta$ is the band index, $\mu_{\eta}=\mu+\eta \tilde{\mu}$ quantifies the asymmetry between the bands, and we assumed weak SOC and altermagnetic spin splitting. Note that the band index $\eta$ is not equivalent to the spin projection index $\lambda$ in the presence of SOC.

While the hydrodynamic equations are modified by the SOC via the viscosity tensor, the form of the solutions given in Eqs.~(\ref{eq:ux-def}) and (\ref{eq:uy-def}) is not affected, and the effects of the SOC can be absorbed by redefining the parameters in these equations. Ignoring the SOC-induced corrections in the viscosity tensor, there is no effect on the constraint (\ref{eq:A-constraint}) and its solution (\ref{tilde-mu-sol}). On the other hand, the SOC leads to additional terms in the electric current. As demonstrated in Appendix~\ref{sec:App-Hydro-soc}, the SOC-induced Berry curvature results in an AHE current, see Eq.~\eqref{eq:ahe-soc-def}. 

The AHE, which is nontrivial due to emergent $\tilde{\mu}$ in the channel geometry,  modifies the electric current component $j_{y}$ as
\begin{equation}
j_y^{\text{el}}(y) =\frac{2 e \tau \tilde{t}_0^2 \rho^2 \left[t_2 \tilde{\rho}  E_x+  \left(t_0 \rho +  t_1 \tilde{\rho} \right) E_y \right] }{ t_0^2 \rho^2- (t_1^2 + t_2^2) \tilde{\rho}^2} -E_x \sigma^{\text{AHE}},
\end{equation}
where $\sigma^{\text{AHE}}$ is the AHE conductivity, see Eq.~(\ref{eq:ahe-soc-def}) and the text below. Note that the quantities with tilde have a different meaning in the presence of the SOC: instead of spin-imbalance, they quantify the imbalance between the bands. The Hall field $E_y$ reads as:
\begin{equation}
E_y=-\frac{t_2 \tilde{\rho}}{t_0 \rho +t_1 \tilde{\rho}} E_x+\frac{t_0^2 \rho^2 -(t_1^2+t_2^2) \tilde{\rho}^2 } {2e\tau \tilde{t}_0^2 \rho^2 \left( t_0 \rho +t_1 \tilde{\rho} \right)} \sigma^{\text{AHE}}.
\end{equation}
The effects of the SOC in the viscosity tensor can be included, but result in more complicated equations, especially in the presence of the asymmetry between the bands.

\section{Discussion and summary}
\label{sec:summary}

In this work, we analyzed the manifestations of altermagnetic spin splitting in viscoelasticity tensor and electron flows in the hydrodynamic transport regime of altermagnets. Using the effective low-energy model and the Kubo formalism, we calculated the viscoelasticity tensor. As with other anisotropic systems, the viscoelasticity tensor inherits the anisotropy of the dispersion relation of altermagnets, see Eq.~\eqref{eq:anisotropy-eta-terms}.

Employing the kinetic theory, we formulated the hydrodynamic framework in altermagnets, applied it to study the flow of the electron fluid in the channel geometry, and calculated the nonlocal responses. Since the channel geometry defines a preferred direction, electron and spin currents are affected by the direction of the altermagnetic lobes. In the case where the lobes are elongated asymmetrically with respect to the edges of the channel, the altermagnetic spin splitting allows for a nonzero spin density. Hence, both electric and spin currents emerge in this setup. Such currents contain longitudinal and Hall-like parts, see Eqs.~\eqref{eq:el-current-x} and \eqref{eq:el-current-y}, allowing for the hydrodynamic spin-splitter effect where the spin current flows perpendicularly to the applied electric field. The hydrodynamic regime is manifested in the characteristic Poiseuille profile of currents along the channel. The SOC affects the Hall part of the response via the AHE, but has a minor effect on the steady hydrodynamic equations.

In the nonlocal response regime, current streamlines are also affected by both the magnitude of the spin splitting and the direction of the altermagnetic lobes. If the channel geometry respects the symmetry of the altermagnetic lobes, then the electrochemical potential distribution is affected only quantitatively by the spin splitting. On the other hand, the current streamlines show a stronger effect because the spin splitting affects the position of vortices, see Fig.~\ref{fig:streamlines1}. In the case where the channel geometry breaks the symmetry of the lobes, the effect is the opposite: streamlines are affected quantitatively, and the electrochemical potential distribution shows noticeable deviations in shape with the characteristic split maximum near the source, see Figs.~\ref{fig:streamlines3} and \ref{fig:streamlines2}(b).

The proposed effects, namely, the Poiseuille profile of the electric and spin currents as well as the modification of electric potential in the nonlocal transport response, provide an efficient means to pinpoint the hydrodynamic regime in altermagnets. Furthermore, since altermagnets have vanishing net magnetization, they allow one to study the interplay of spin currents and hydrodynamics without stray magnetic fields. This is beneficial for methods of imagining electron flows that are based on quantum spin magnetometry~\cite{Maze-Lukin-NanoscaleMagneticSensing-2008, Levine-Walsworth-PrinciplesTechniquesQuantum-2019, Barry-Walsworth:2019-QSM, Marchiori-Poggio-NanoscaleMagneticField-2021, Rovny-Leon-NewOpportunitiesCondensed-2024}. In particular, such magnetometry methods were already successfully used to imagine electron hydrodynamic flows in graphene~\cite{Ku-Walsworth:2019, Kumar-Ilani:2021, Jenkins-BleszynskiJayich-ImagingBreakdownOhmic-2022, Aharon-Steinberg2022}. The distribution of the electrochemical potential can be probed via scanning tunneling potentiometry~\cite{Baddorf-ScanningTunnelingPotentiometry-2007}. Local spin currents could be probed via techniques similar to the nonlocal spin valve measurement~\cite{Johnson-Silsbee-InterfacialChargespinCoupling-1985, Jedema-VanWees-ElectricalSpinInjection-2001}.

Potential candidates to observe electron hydrodynamic regime are metallic altermagnets such as ultrathin films of RuO$_2$, Mn$_5$Si$_3$, CrSb, and intercalated transition metal dichalcogenides CoNb$_4$Se$_8$. A general strategy in finding a suitable material is to search for clean altermagnets with weak electro-phonon interactions. Intercalated transition metal dichalcogenides are promising candidates in view of their quasi-2D structure, albeit their high residual resistivity~\cite{Regmi-Ghimire-AltermagnetismLayeredIntercalated-2024, Sakhya-Neupane-ElectronicStructureLayered-2025} implies that these altermagnets are disordered. While bulk RuO$_2$ is a good conductor, the thin-film geometry in which signatures of altermagnetism were observed makes it susceptible to surface disorder and grain boundaries. A prospective candidate for observing electron hydrodynamics is a 3D altermagnet CrSb, which has a relatively low resistivity~\cite{Rai-Kumar-DirectionDependentConductionPolarity-2025}.

Momentum-dependent spin splitting of altermagnets allows for a nontrivial nonlocal response even beyond the hydrodynamic regime. As we demonstrate in Ref.~\cite{Herasymchuk-Sukhachov-ElectricSpinCurrent-2025}, altermagnets may support swirling electric and spin currents in the Ohmic transport regime for certain configurations of the contacts.

While, in the present study, we focused on 2D altermagnets, the obtained results can be straightforwardly generalized to 3D case. The generalization to other symmetries of altermagnets, namely, $g$- and $i$-wave is also possible, albeit, as we show for the viscoelastic tensor in Appendix~\ref{sec:App-Kubo}, leads to more cumbersome expressions without affecting the structure of the hydrodynamic equations and, hence, the qualitative signatures of the flows. Furthermore, while we used the simple relaxation-time approximation and assumed that the electron hydrodynamic regime is reachable in altermagnets, more realistic calculations and estimates of the corresponding hydrodynamic window in candidate materials should be performed. These studies are, however, beyond the scope of our work and will be reported elsewhere.

\begin{acknowledgments}
A.A.H. acknowledges support from the Project No. 0122U000887 of the Department of Physics and Astronomy of the NAS of Ukraine.
The work of E.V.G. was supported by the Program “Dynamics of particles and collective excitations in high-energy physics, astrophysics and quantum macrosystems” (No. 0121U109612) of the Department of Physics and Astronomy of the NAS of Ukraine. P.O.S. acknowledges useful communications with J.~Linder and A.~Qaiumzadeh. P.O.S. was supported by the Research Council of Norway through Grant No.~323766 and its Centres of Excellence funding scheme Grant No.~262633 “QuSpin” during his stay at NTNU.
\end{acknowledgments}

\appendix

\vspace{5mm}
\begin{widetext}

\section{Calculation of viscoelasticity tensor in Kubo approach}
\label{sec:App-Kubo}

Let us derive the viscoelasticity tensor in the Kubo approach. In this derivation, we follow Ref.~\cite{Bradlyn-Read:2012}. Under deformations quantified by the displacement vector $\mathbf{u}(\mathbf{x}, t)$, coordinates transform as $\mathbf{x} \rightarrow \mathbf{x}'= \mathbf{x} + \mathbf{u}(\mathbf{x}, t) =\mathbf{x}+\mathbf{e}_{\mu}  x_{\nu} \partial u_{\mu}/(\partial x_{\nu}) +o(|\mathbf{x}|)$, which can be rewritten in terms of homogeneous but time-dependent invertible matrix $\Lambda(t)$ with positive determinant $\mathbf{x}'= \Lambda(t) \mathbf{x}$. The matrix $\Lambda(t)$ is expressed in terms of the strain tensor $\lambda(t)$ as $\Lambda(t)=\exp{(\lambda(t))}=1+ \lambda(t)+o(\lambda)$. The viscoelasticity tensor is defined by the linear response relation
\begin{equation}
\label{eq:tau-visco-response}
\left< \tau_{\mu \nu} \right>_{\mathbf{x}'}-\left< \tau_{\mu \nu} \right>_{\mathbf{x}} = \int_{-\infty}^{+\infty} d t'  \sum_{\alpha \beta} \left[ \eta_{\mu \nu \alpha \beta}^{\,\,\,\,\,\,\,}(t-t') -\delta_{\mu \nu} \delta_{\alpha \beta} \kappa^{-1}\Theta(t-t^{\prime}) \right] \frac{\partial \lambda^{\alpha \beta}_{}}{\partial t},
\end{equation}
where $\tau_{\mu \nu}(\mathbf{x})$ is the stress tensor operator, $\left< \ldots \right>_{\mathbf{x}'}$ and $\left< \ldots \right>_{\mathbf{x}}$ denotes the expectation value in the presence and the absence of strain, respectively.

Let us define the integral stress tensor $T_{\mu \nu}=\int d\mathbf{x} \, \tau_{\mu \nu}(\mathbf{x})/V$, where $V$ is the total volume of the fluid in the absence of strain. 
For the integral stress tensor, we have the following expression:
\begin{equation}
\label{eq:itau-X-response}
\left<T_{\mu \nu}^{\,\,\,}\right>_{\mathbf{x}'}-\left<T_{\mu \nu}^{\,\,\,}\right>_{\mathbf{x}} =\int_{-\infty}^{+\infty} d t  \sum_{\alpha \beta}X_{\mu \nu \alpha \beta}(t-t') \frac{\partial \lambda_{\,\,\, }^{\alpha \beta}}{\partial t},
\end{equation}
where
\begin{equation}
\label{eq:itau-X-def}
X_{\mu \nu \alpha \beta}(t-t') =-i \lim_{\varepsilon \to +0} \Theta(t-t') \left< [T_{\mu \nu}^{\,\,\,}(t), J_{\alpha \beta}^{\,\,\,}(t')] \right> e^{-\varepsilon (t-t')}
\end{equation}
is the retarded response function. Performing the Fourier transform in Eq.~\eqref{eq:itau-X-def}, we obtain
\begin{equation}
\label{eq:X-def-fourier}
\begin{aligned}
X_{\mu \nu \alpha \beta}(\Omega)&=-i \lim_{\varepsilon \to +0} \int_{-\infty}^{+\infty} dt e^{i(\Omega+i\varepsilon)(t-t')} \Theta(t-t') \left< [T_{\mu \nu}^{\,\,\,}(t), J_{\alpha \beta}^{\,\,\,}(t')] \right>\\
&=\frac{1}{\Omega+i0} \int_{-\infty}^{+\infty} dt \left( \delta(t-t') \left< [T_{\mu \nu}^{\,\,\,}(t), J_{\alpha \beta}^{\,\,\,}(t')] \right>+ \Theta(t-t') \left< [T_{\mu \nu}^{\,\,\,}(t), T_{\alpha \beta}^{\,\,\,}(t')] \right> \right)e^{i(\Omega+i0)(t-t')}\\
&=\frac{1}{\Omega+i0} \left(  \left< [T_{\mu \nu}^{\,\,\,}(0), J_{\alpha \beta}^{\,\,\,}(0)] \right>+ \int_{0}^{+\infty} \left< [T_{\mu \nu}^{\,\,\,}(t), T_{\alpha \beta}^{\,\,\,}(0)] \right> e^{i(\Omega+i0)t}\right)\\
&=\frac{1}{\Omega+i0} \left(  \left< [T_{\mu \nu}^{\,\,\,}(0), J_{\alpha \beta}^{\,\,\,}(0)] \right> -i C_{\mu \nu \alpha \beta}(\Omega)\right).\\
\end{aligned}
\end{equation}
Here $C_{\mu \nu \alpha \beta}(\Omega)$ is the Fourier transform of the stress-stress correlation function given in Eq.~(\ref{eq:C-corr-def}).

Integrating Eq.~(\ref{eq:tau-visco-response}) over $\mathbf{x}$, and using Eq.~(\ref{eq:itau-X-response}), we obtain the relation between the viscoelasticity $\eta_{\mu \nu \alpha \beta}$ and the correlation function $X_{\mu \nu \alpha \beta}$ 
\begin{equation}
\eta_{\mu \nu \alpha \beta}(t) = X_{\mu \nu \alpha \beta}(t)+ \delta_{\mu \nu} \delta_{\alpha \beta} \kappa^{-1}\Theta(t) -\delta_{\alpha \beta}^{\,\,\,} \left< T_{\mu \nu}^{\,\,\,} \right>_{\mathbf{x}} \Theta(t).
\end{equation}
Performing the Fourier transform, we obtain the viscoelasticity tensor given in Eq.~(\ref{eq:viscoelasticity-def}). To simplify the notations in the main text, we drop the subscript $\mathbf{x}$ in the expectation value $\left< T_{\mu \nu}^{\,\,\,} \right>_{\mathbf{x}}$. The viscoelasticity tensor can be rewritten as in Eq.~(\ref{Kubo_viscoelastic_tensor}), whose explicit form is
\begin{equation}
\label{app-kubo-eta}
\begin{aligned}
\eta_{\mu \nu \alpha \beta}(\Omega)
&=-\frac{i}{\Omega+i0}  \int_{-\infty}^{+\infty} d \omega \int_{-\infty}^{+\infty} d \omega' \frac{f(\omega)-f(\omega')}{\omega'-\omega-\Omega-i0}  \int \frac{d^2k}{(2\pi)^2}  \text{tr} \left[ T_{\mu \nu}^{\,\,\,} (\mathbf{k}) A(\omega;\mathbf{k}) T_{\alpha \beta}^{\,\,\, } (\mathbf{k}) A(\omega';\mathbf{k}) \right]\\
&+\frac{1}{\Omega+i0} \left\{ i \delta_{\alpha \beta}  \int_{-\infty}^{+\infty} d\omega f(\omega) \int \frac{d^2 k}{(2\pi)^2} \text{tr}\left[  T_{\mu \nu}(\mathbf{k}) A(\omega; \mathbf{k}) \right] -i \kappa^{-1} \delta_{\mu \nu} \delta_{\alpha \beta} \right.\\
&\left. + \,
\int_{-\infty}^{+\infty} d\omega f(\omega) \int \frac{d^2 k}{(2\pi)^2}  \text{tr}\left\{ [T_{\mu \nu}(\mathbf{k}), J_{\alpha \beta}^{\,\,\,} ] A(\omega; \mathbf{k}) \right\} \right\}
\\
&=\eta_{\mu \nu \alpha \beta}^{(1)}(\Omega)+\eta_{\mu \nu \alpha \beta}^{(2)}(\Omega)-\frac{i}{\Omega+i0}  \kappa^{-1} \delta_{\mu \nu} \delta_{\alpha \beta} +\eta_{\mu \nu \alpha \beta}^{(3)}(\Omega).
\end{aligned}
\end{equation}

We have the following traces for the low-energy model of altermagnets:
\begin{eqnarray}
\text{tr} \left[ T_{\mu \nu}^{\,\,\,} (\mathbf{k}) A(\omega;\mathbf{k}) \right] &=& \sum_{\lambda} k_{\mu } \frac{\partial \varepsilon_{\lambda}}{\partial k_{\nu}} \delta \left( \omega -\varepsilon_{\lambda} \right),\\
\text{tr} \left\{ \left[T_{\mu \nu}(\mathbf{k}) ,J_{\alpha \beta} \right]  A(\omega;\mathbf{k})  \right\} &=& i \sum_{\lambda} \left(  k_{\mu } k_{\alpha}\frac{\partial^2 \varepsilon_{\lambda}}{\partial k_{\nu} \partial k_{\beta}}+ \delta_{\mu \beta}  k_{\alpha} \frac{\partial \varepsilon_{\lambda}}{\partial k_{\nu}} \right)\delta \left( \omega -\varepsilon_{\lambda} \right), \\
\text{tr} \left[ T_{\mu \nu}^{\,\,\,} (\mathbf{k}) A(\omega;\mathbf{k}) T_{\alpha \beta}^{\,\,\,} (\mathbf{k}) A(\omega';\mathbf{k}) \right] &=& \sum_{\lambda} k_{\mu} k_{\alpha} \frac{\partial \varepsilon_{\lambda}}{\partial k_{\nu}} \frac{\partial \varepsilon_{\lambda}}{\partial k_{\beta}} \delta \left( \omega -\varepsilon_{\lambda} \right)\delta \left( \omega' -\varepsilon_{\lambda} \right).
\end{eqnarray}

\subsection{$d$-wave altermagnets}
\label{sec:App-Kubo-dwave}

In what follows, we use the low-energy model of $d$-wave altermagnets given in Eq.~\eqref{eq:hamiltonian}. Hence, for the first term in Eq.~(\ref{app-kubo-eta}), we obtain
\begin{equation}
\label{app-kubo-eta-1}
\begin{aligned}
\eta_{\mu \nu \alpha \beta}^{(1)}(\Omega)&=-\frac{i}{\Omega+i0} \int_{-\infty}^{+\infty} d \omega \int_{-\infty}^{+\infty} d \omega' \frac{f(\omega)-f(\omega')}{\omega'-\omega-\Omega -i 0}  \int \frac{d^2 k}{(2\pi)^2} \sum_{\lambda} k_{\mu} k_{\alpha} \frac{\partial \varepsilon_{\lambda}}{\partial k_{\nu}} \frac{\partial \varepsilon_{\lambda}}{\partial k_{\beta}} \delta \left( \omega -\varepsilon_{\lambda} \right)\delta \left( \omega' -\omega \right) = 0.
\end{aligned}
\end{equation}
For $\eta^{(2)}_{\mu \nu \alpha \beta}$, we find
\begin{equation}
\label{app-kubo-eta-2}
\begin{aligned}
\eta_{\mu \nu \alpha \beta}^{(2)}(\Omega)&=\frac{ i \delta_{\alpha \beta}}{\Omega+i0} \int_{-\infty}^{+\infty}d\omega f(\omega) \int \frac{d^2k}{(2\pi)^2}\sum_{\lambda} k_{\mu } \frac{\partial \varepsilon_{\lambda}}{\partial k_{\nu}} \delta \left( \omega -\varepsilon_{\lambda} \right)= \frac{ i \delta_{\alpha \beta} }{\Omega+i0} \int \frac{d^2k}{(2\pi)^2}\sum_{\lambda} f\left( \varepsilon_{\lambda} \right) k_{\mu } \frac{\partial \varepsilon_{\lambda}}{\partial k_{\nu}} \\
&=-\frac{ i \delta_{\mu \nu }\delta_{\alpha \beta} }{\Omega+i0} \sum_{\lambda} \frac{T^2}{4 \pi \tilde{t}_0 } \text{Li}_2 \left( -e^{\mu_{\lambda}/T}\right) =\frac{ i \delta_{\mu \nu }\delta_{\alpha \beta} }{\Omega+i0}  \sum_{\lambda} \epsilon_{\lambda} = \frac{ i \delta_{\mu \nu }\delta_{\alpha \beta} }{\Omega+i0}  \epsilon,
\end{aligned}
\end{equation}
where
\begin{equation}
\epsilon=\int_{-\infty}^{+\infty}d\omega f(\omega) \int \frac{d^2k}{(2\pi)^2}\sum_{\lambda} \varepsilon_{\lambda} \delta \left( \omega -\varepsilon_{\lambda} \right)=- \sum_{\lambda} \frac{T^2}{4 \pi \tilde{t}_0 } \text{Li}_2 \left( -e^{\mu_{\lambda}/T}\right).
\end{equation}
For $\kappa^{-1}$, we have
\begin{equation}
\label{app-kubo-eta-23}
\kappa^{-1}=-V\frac{\partial P}{\partial V}=\left(\epsilon+P\right)\frac{\partial P}{\partial \epsilon }=-\sum_{\lambda} \frac{T^2}{2 \pi \tilde{t}_0 } \text{Li}_2 \left( -e^{\mu_{\lambda}/T}\right).
\end{equation}
Finally, for $\eta_{\mu \nu \alpha \beta}^{(3)}(\Omega)$, we obtain
\begin{equation}
\label{app-kubo-eta-3}
\begin{aligned}
\eta_{\mu \nu \alpha \beta}^{(3)}(\Omega)&=\frac{i}{\Omega+i0}  \int_{-\infty}^{+\infty}d\omega f(\omega) \int \frac{d^2k}{(2\pi)^2}\sum_{\lambda=\pm} \left(  k_{\mu } k_{\alpha}\frac{\partial^2 \varepsilon_{\lambda}}{\partial k_{\nu} \partial k_{\beta}}+ \delta_{\mu \beta}  k_{\alpha} \frac{\partial \varepsilon_{\lambda}}{\partial k_{\nu}} \right)\delta \left( \omega -\varepsilon_{\lambda} \right) \\
&=\frac{i}{\Omega+i0} \int \frac{d^2k}{(2\pi)^2}\sum_{\lambda=\pm} f\left( \varepsilon_{\lambda} \right) \left(  k_{\mu } k_{\alpha}\frac{\partial^2 \varepsilon_{\lambda}}{\partial k_{\nu} \partial k_{\beta}}+ \delta_{\mu \beta}  k_{\alpha} \frac{\partial \varepsilon_{\lambda}}{\partial k_{\nu}} \right) =- \frac{i \delta_{\mu \beta} \delta_{\nu \alpha} T^2}{4 \tilde{t}_0 \left( \Omega+i0 \right)} \sum_{\lambda} \text{Li}_2 \left( -e^{\mu_{\lambda}/T}\right) \\
&- \frac{i T^2}{4\tilde{t}_0 \left( \Omega+i0 \right) }\sum_{\lambda} \text{Li}_2 \left( -e^{\mu_{\lambda}/T}\right) A_{\nu \beta} (\lambda) B_{\mu \alpha}(\lambda) = \frac{i}{\Omega+i0} \left[ \delta_{\mu \beta} \delta_{\nu \alpha} \epsilon + \sum_{\lambda=\pm} \epsilon_{\lambda}  A_{\nu \beta} (\lambda) B_{\mu \alpha}(\lambda) \right], \\
\end{aligned}
\end{equation}
where the matrices $A_{\mu \nu} (\lambda)$ and $B_{\mu \nu} (\lambda)$ are defined as follows:
\begin{eqnarray}
\label{eq:anisotropy-matrix-A}
 \hat{A}(\lambda) &=& t_0 -\lambda t_1 \tau_z +\lambda t_2 \tau_x,\\
\label{eq:anisotropy-matrix-B}
 \hat{B}(\lambda) &=& \frac{1}{\tilde{t}_0^2} \left( t_0 + \lambda t_1 \tau_z -\lambda t_2 \tau_x \right).
\end{eqnarray}
Here $\tilde{t}_0=\left(t_0^2 -t_1^2 -t_2^2\right)^{1/2}$ and $\tau_{x,z}$ are Pauli matrices. By combining the results in Eqs.~(\ref{app-kubo-eta-1}), (\ref{app-kubo-eta-2}), (\ref{app-kubo-eta-23}), and (\ref{app-kubo-eta-3}), we obtain the viscoelasticity tensor given in Eq.~(\ref{eq:anisotropy-eta-terms}).

\subsection{$g$- and $i$-wave altermagnets}
\label{sec:App-Kubo-giwave}

Similarly to $d$- wave altermagnets, one may consider altermagnets with $g$-, and $i$- symmetry. In this case, the dispersion relation reads as
\begin{equation}
\varepsilon_{\lambda}= t_0 k^2 + \lambda J(k_x , k_y),
\end{equation}
where, for $g$-wave altermagnets, the function $J(\mathbf{k})$ reads as follows~\cite{Smejkal-Jungwirth-ConventionalFerromagnetismAntiferromagnetism-2022}
\begin{equation}
\begin{aligned}
J(\mathbf{k})=t_1 k_x k_y \left( k_x^2-k_y^2\right)+ \frac{1}{4} t_2\left[ \left( k_x^2-k_y^2\right)^2 -4k_x^2k_y^2\right],
\end{aligned}
\end{equation}
and for $i$-wave altermagnets, we have~\cite{Smejkal-Jungwirth-ConventionalFerromagnetismAntiferromagnetism-2022}
\begin{equation}
\begin{aligned}
J(\mathbf{k})=t_1 k_xk_y\left(3k_x^2-k_y^2\right)\left(3k_y^2-k_x^2\right) +\frac{1}{2} t_2 \left( k_x^2-k_y^2\right) \left[ \left( k_x^2+k_y^2\right)^2 -16 k_x^2k_y^2\right].
\end{aligned}
\end{equation}
The viscoelasticity tensor in this case is
\begin{equation}
\label{app-eta-gi}
\begin{aligned}
\eta_{\mu \nu \alpha \beta} \left( \Omega \right)=\frac{i}{\Omega+i0} \left[\delta_{\alpha \beta} I^{(1)}_{\mu \nu} + \delta_{\mu \beta} I^{(1)}_{\alpha \nu}  + I^{(2)}_{\mu \nu \alpha \beta} -  \delta_{\mu \nu}\delta_{\alpha \beta} \left(1+a\right) P \right],
\end{aligned}
\end{equation}
where
\begin{equation}
\begin{aligned}
I^{(1)}_{\mu \nu}&=\delta_{\mu \nu} P=\int \frac{d^2k}{(2\pi)^2}\sum_{\lambda=\pm} f\left( \varepsilon_{\lambda} \right)   k_{\mu} \frac{\partial \varepsilon_{\lambda}}{\partial k_{\nu}},\quad
I^{(2)}_{\mu \nu \alpha \beta}=\int \frac{d^2k}{(2\pi)^2}\sum_{\lambda=\pm} f\left( \varepsilon_{\lambda} \right)  k_{\mu } k_{\alpha}\frac{\partial^2 \varepsilon_{\lambda}}{\partial k_{\nu} \partial k_{\beta}},\quad
a=\frac{\partial P}{\partial \epsilon}=\sum_{\lambda=\pm} \frac{\partial P}{\partial \mu_{\lambda}} \left(\frac{\partial \epsilon}{\partial \mu_{\lambda}} \right)^{-1}.
\end{aligned}
\end{equation}

To simplify the calculations, we set the limit $T \rightarrow 0$. For example, in the case of $g$-symmetry, for $I^{(1)}_{\mu \nu}$, $I^{(2)}_{\mu \nu \alpha \beta}$, and $\alpha$, we obtain:
\begin{equation}
\begin{aligned}
I^{(1)}_{\mu \nu}&=\delta_{\mu \nu} \epsilon+\delta_{\mu \nu} \sum_{\lambda=\pm} \int \frac{d^2k}{(2\pi)^2} \frac{\lambda t k^4}{4} \sin(4\phi +4\phi_0) f\left( \varepsilon_{\lambda} \right), \\
\end{aligned}
\end{equation}
\begin{equation}
\begin{aligned}
I^{(2)}_{\mu \nu \alpha \beta}&=\delta_{\mu \alpha} \delta_{\nu \beta} \epsilon
+ \left[ \eta_{\mu \alpha} \left(1- \delta_{\nu \beta}\right)+ \eta_{\nu \beta} \left( 1-\delta_{\mu \alpha}\right) \right] \sum_{\lambda=\pm} \int \frac{d^2k}{(2\pi)^2}  3\lambda t k^4  \cos^2 (\phi ) \cos (2 \phi +4 \phi_0)  f\left( \varepsilon_{\lambda} \right) \\
&+ \left( 1-\delta_{\mu \alpha}\right) \left( 1-\delta_{\nu \beta}\right) \sum_{\lambda=\pm} \int \frac{d^2k}{(2\pi)^2}   3\lambda t k^4 \sin ( \phi ) \cos (\phi)   \cos (2 \phi +4 \phi_0)  f\left( \varepsilon_{\lambda} \right) \\
&-  \sum_{\lambda=\pm} \int \frac{d^2k}{(2\pi)^2}  \frac{\lambda t k^4}{4}  \left[ \delta_{\mu \alpha} \delta_{\nu \beta} \sin (4 \phi +4 \phi_0) - \eta_{\mu \alpha} \eta_{\nu \beta} 12 \cos^2(\phi) \sin (2 \phi +4 \phi_0) \right]  f\left( \varepsilon_{\lambda} \right),\\
\end{aligned}
\end{equation}
\begin{equation}
\begin{aligned}
a&=1+\sum_{\lambda=\pm} \left[\int_{0}^{+\infty} dk \, \frac{ \left( \mu_{\lambda} -t_0 k^2 \right)k }{\sqrt{\left(t k^4/ 4\right)^2-\left(\mu_{\lambda}-t_0k^2\right)^2}} \Theta\left( \left|t k^4\right|-4\left|\mu_{\lambda}-t_0 k^2 \right|\right)\right]\\
&\times \left[ \int_{0}^{+\infty} dk \, \frac{ \mu_{\lambda} k  }{\sqrt{\left(t k^4/ 4\right)^2-\left(\mu_{\lambda}-t_0k^2\right)^2}} \Theta\left( \left|t k^4\right|-4\left|\mu_{\lambda}-t_0 k^2 \right|\right) \right]^{-1},
\end{aligned}
\end{equation}
where we find it convenient to use the following parametrization: $t_1 = t \cos{(4\phi_0)}$ and $t_2 = t \sin{(4\phi_0)}$. The viscoelasticity tensor $\text{Re}\,\eta_{\mu \nu \alpha \beta}(\Omega)$ can be presented in the form given in Eq.~(\ref{eq:anisotropy-eta-terms}) with
\begin{equation}
\begin{aligned}
\text{Re}\,\eta_{\mu \nu \alpha \beta}^{\text{anis}}(\Omega) &= \pi \delta(\Omega) \Bigg\{  \left(1-a\right) \epsilon \delta_{\mu \nu} \delta_{\alpha \beta} +\left( \delta_{\mu \beta} \delta_{ \nu \alpha} -a \delta_{\mu \nu} \delta_{\alpha \beta} -\delta_{\mu \alpha} \delta_{ \nu \beta}\right)\sum_{\lambda=\pm} \int \frac{d^2k}{(2\pi)^2} \frac{\lambda t k^4}{4} \sin(4\phi +4\phi_0) f\left( \varepsilon_{\lambda} \right)  \\
&+ \left( 1-\delta_{\mu \alpha}\right) \left( 1-\delta_{\nu \beta}\right) \sum_{\lambda=\pm} \int \frac{d^2k}{(2\pi)^2}  3\lambda t k^4 \sin ( \phi ) \cos (\phi) \cos (2 \phi +4 \phi_0)  \frac{ f\left( \varepsilon_{\lambda} \right) + f\left( \varepsilon_{\lambda} + \mu_{\lambda}-\mu_{-\lambda} \right)  }{ 2 } \\
&+   \eta_{\mu \alpha} \eta_{\nu \beta}  \sum_{\lambda=\pm} \int \frac{d^2k}{(2\pi)^2}  3 \lambda t k^4   \cos^2(\phi) \sin (2 \phi +4 \phi_0) \frac{ f\left( \varepsilon_{\lambda} \right) + f\left( \varepsilon_{\lambda} + \mu_{\lambda}-\mu_{-\lambda} \right)  }{ 2 }  \Bigg \},
 \end{aligned}
\end{equation}
\begin{equation}
\begin{aligned}
\text{Re}\,\tilde{\eta}_{\mu \nu \alpha \beta}(\Omega) &= \pi \delta(\Omega) \Bigg\{ \left[ \eta_{\mu \alpha} \left(1- \delta_{\nu \beta}\right)+ \eta_{\nu \beta} \left( 1-\delta_{\mu \alpha}\right) \right] \sum_{\lambda=\pm} \int \frac{d^2k}{(2\pi)^2}  3\lambda t k^4  \cos^2 (\phi ) \cos (2 \phi +4 \phi_0)  f\left( \varepsilon_{\lambda} \right) \\
&+ \left( 1-\delta_{\mu \alpha}\right) \left( 1-\delta_{\nu \beta}\right) \sum_{\lambda=\pm} \int \frac{d^2k}{(2\pi)^2}  3\lambda t k^4 \sin ( \phi ) \cos (\phi) \cos (2 \phi +4 \phi_0)  \frac{ f\left( \varepsilon_{\lambda} \right) -f\left( \varepsilon_{\lambda} + \mu_{\lambda}-\mu_{-\lambda} \right)  }{ 2 } \\
&+   \eta_{\mu \alpha} \eta_{\nu \beta}  \sum_{\lambda=\pm} \int \frac{d^2k}{(2\pi)^2}  3 \lambda t k^4   \cos^2(\phi) \sin (2 \phi +4 \phi_0) \frac{ f\left( \varepsilon_{\lambda} \right) -f\left( \varepsilon_{\lambda} + \mu_{\lambda}-\mu_{-\lambda} \right)  }{ 2 } \Bigg\},
 \end{aligned}
\end{equation}
and the isotropic component defined in Eq.~\eqref{eq:anisotropy-eta-terms-rot}.

The viscoelastic tensor given in Eq.~\eqref{app-eta-gi} leads to the Navier-Stokes equation of a similar structure as that for $d$-wave altermagnets at $t_1\neq0$ and $t_2\neq0$. Therefore, we do not expect drastic changes in the current streamlines and the electrochemical potential.

\subsection{Role of SOC}
\label{sec:visco-soc}

In this Section, we present the details of the calculations related to the spin-orbital coupling (SOC). The Rashba SOC is described by an additional term \eqref{eq:hamiltonian-soc-0} in the altermagnetic Hamiltonian given in Eq.~(\ref{eq:hamiltonian}).

The dispersion relation of electron quasiparticles reads
\begin{equation}
\label{eq:eigenvalue-soc}
\varepsilon_{\eta}= t_0 k^2-\mu +\eta \sqrt{\left[ J(\mathbf{k}) - \tilde{\mu} \right]^2+\alpha^2 k^2},
\end{equation}
where $\mu=\left(\mu_{+}+\mu_{-}\right)/2$, $\tilde{\mu}=\left(\mu_{+}-\mu_{-}\right)/2$, and  we focus on a $d$-wave altermagnet with $J\left(\mathbf{k}\right)=t_1 \left(k_y^2-k_x^2\right) + 2 t_2 k_x k_y$. Note that, in the presence of SOC, $\tilde{\mu}$ loses its direct interpretation as a spin-imbalance chemical potential and contributes to the asymmetry between the two bands. For the total Hamiltonian $H_{\text{tot}}=H+H_{\text{soc}}$, the spectal function is given by:
\begin{equation}
\label{eq:spectral-fucntion-soc}
A\left(\omega;\mathbf{k}\right)= \frac{1}{2} \left[ \delta\left(\omega-\varepsilon_{+} \right)+ \delta\left(\omega-\varepsilon_{-} \right)\right] + \frac{H_{\text{tot}}-\left( \varepsilon_{+} + \varepsilon_{-}\right)/2}{\varepsilon_{+} -\varepsilon_{-} } \left[ \delta\left(\omega-\varepsilon_{+} \right)- \delta\left(\omega-\varepsilon_{-} \right)\right]
\end{equation}
The normalized wavefunctions $\psi_{\eta}\left( \mathbf{k}\right)$ read as follows
\begin{equation}
\label{eq:wave-fucntion-soc}
\begin{aligned}
\psi_{+}\left( \mathbf{k}\right)&= \frac{1}{\sqrt{ \left(  \left[ J(\mathbf{k}) - \tilde{\mu} \right] +   \sqrt{\left[ J(\mathbf{k}) - \tilde{\mu} \right]^2+\alpha^2 k^2} \right)^2+\alpha^2 k^2}}
\left(\begin{array}{c}
i \left[ J(\mathbf{k}) - \tilde{\mu} \right] + i  \sqrt{\left[ J(\mathbf{k}) - \tilde{\mu} \right]^2+\alpha^2 k^2}  \\
\alpha \left( k_x+ik_y\right)
\end{array}\right), \\
\psi_{-}\left( \mathbf{k}\right)&= \frac{1}{\sqrt{ \left(  \left[ J(\mathbf{k}) - \tilde{\mu} \right] +   \sqrt{\left[ J(\mathbf{k}) - \tilde{\mu} \right]^2+\alpha^2 k^2} \right)^2+\alpha^2 k^2}}
\left(\begin{array}{c}
\alpha \left( k_x-ik_y\right)\\
i \left[ J(\mathbf{k}) - \tilde{\mu} \right] + i  \sqrt{\left[ J(\mathbf{k}) - \tilde{\mu} \right]^2+\alpha^2 k^2}  \\
\end{array}\right).
\end{aligned}
\end{equation}

The stress tensor $T_{\alpha\beta}$ acquires an additional term
\begin{equation}
\label{eq:hamiltonian-soc}
T_{\alpha\beta}^{\text{soc}}=k_{\alpha} \frac{\partial H_{\text{soc}}}{\partial k_{\beta}}.
\end{equation}

Diagonalization the total Hamiltonian of an altermagnet with SOC, we obtain the following traces in Eq.~\eqref{app-kubo-eta}:
\begin{eqnarray}
\text{tr} \left[ T_{\mu \nu}^{\,\,\,} (\mathbf{k}) A(\omega;\mathbf{k}) \right] &=& \lim_{ \mathbf{k}^{\prime} \to \mathbf{k}  }  k_{\mu} \frac{\partial }{\partial k_{\nu}}  \text{tr} \left[H_{\text{tot}}(\mathbf{k})  A\left(\omega;\mathbf{k}^{\prime}\right) \right] = \sum_{\eta} k_{\mu } \frac{\partial \varepsilon_{\eta}}{\partial k_{\nu}} \delta \left( \omega -\varepsilon_{\eta} \right),\\
\text{tr} \left\{ \left[T_{\mu \nu}(\mathbf{k}) ,J_{\alpha \beta} \right]  A(\omega;\mathbf{k})  \right\} &=& \lim_{  \mathbf{k}^{\prime} \to \mathbf{k} } i \left(  k_{\mu } k_{\alpha}\frac{\partial^2 }{\partial k_{\nu} \partial k_{\beta}}+ \delta_{\mu \beta}  k_{\alpha} \frac{\partial }{\partial k_{\nu}} \right) \text{tr} \left[H_{\text{tot}}(\mathbf{k})  A\left(\omega;\mathbf{k}^{\prime}\right) \right]  \nonumber\\
&=& i \sum_{\eta} \left(  k_{\mu } k_{\alpha}\frac{\partial^2 \varepsilon_{\eta}}{\partial k_{\nu} \partial k_{\beta}}+ \delta_{\mu \beta}  k_{\alpha} \frac{\partial \varepsilon_{\eta}}{\partial k_{\nu}} \right)\delta \left( \omega -\varepsilon_{\eta} \right),
\end{eqnarray}
\begin{equation}
\begin{aligned}
\text{tr} \left[ T_{\mu \nu}^{\,\,\,} (\mathbf{k}) A(\omega;\mathbf{k}) T_{\alpha \beta}^{\,\,\,} (\mathbf{k}) A(\omega';\mathbf{k}) \right] &=\sum_{\eta} \delta \left( \omega -\varepsilon_{\eta} \right)\delta \left( \omega' -\varepsilon_{\eta} \right) k_{\mu} k_{\alpha} \frac{\partial \varepsilon_{\eta}}{\partial k_{\nu}} \frac{\partial \varepsilon_{\eta}}{\partial k_{\beta}} +\sum_{\eta,\eta^{\prime }} \delta \left( \omega -\varepsilon_{\eta} \right)\delta \left( \omega' -\varepsilon_{\eta^{\prime }} \right)   \frac{\eta \eta^{\prime } k_{\mu} k_{\alpha}}{2}  \\
&\times \Bigg[ \left\{  \frac{2 t_0 k_{\nu} \left[J(\mathbf{k})-\tilde{\mu}\right]}{\sqrt{ [ J(\mathbf{k})- \tilde{\mu } ]^2 +\alpha ^2 k^2}} +\eta \frac{\partial J(\mathbf{k})}{\partial k_{\nu}}  \right\} \left\{  \frac{2 t_0 k_{\beta} \left[J(\mathbf{k})-\tilde{\mu}\right]}{\sqrt{ [ J(\mathbf{k})- \tilde{\mu } ]^2 +\alpha ^2 k^2}} +\eta^{\prime } \frac{\partial J(\mathbf{k})}{\partial k_{\beta}}  \right\}  \\
&- \frac{\partial \varepsilon_{\eta}}{\partial k_{\nu}} \frac{\partial \varepsilon_{\eta^{\prime } }}{\partial k_{\beta}} \Bigg] +\alpha ^2 \sum_{\eta,\eta^{\prime}} \delta \left( \omega -\varepsilon_{\eta} \right)\delta \left( \omega' -\varepsilon_{\eta^{\prime }} \right) \frac{\eta\eta^{\prime} k_{\mu} k_{\alpha}}{2}\frac{k_{\nu} k_{\beta} - \delta_{\nu \beta} k^2 }{[ J(\mathbf{k})- \tilde{\mu } ]^2 +\alpha ^2 k^2}   \\
&+\alpha ^2\sum_{\eta,\eta^{\prime}} \delta \left( \omega -\varepsilon_{\eta} \right)\delta \left( \omega' -\varepsilon_{\eta^{\prime }} \right) \frac{k_{\mu} k_{\alpha}}{2} \Bigg\{ \delta_{\nu \beta }+ \left(\eta+\eta^{\prime}\right) \frac{ 2 t_0 k_{\nu} k_{\beta}}{\sqrt{ [ J(\mathbf{k})- \tilde{\mu } ]^2 +\alpha ^2 k^2} }  \\
&+\frac{ 4 \eta \eta^{\prime} t_0^2 k^2 k_{\nu} k_{\beta}}{[ J(\mathbf{k})- \tilde{\mu } ]^2 +\alpha ^2 k^2} \Bigg\}  + \sum_{\eta}  \frac{ i \eta \alpha ^2 k_{\mu} k_{\alpha} \delta\left( \omega -\varepsilon_{\eta} \right) \delta\left( \omega' -\varepsilon_{-\eta} \right) }{\sqrt{ [ J(\mathbf{k})- \tilde{\mu } ]^2 +\alpha ^2 k^2} }    \\
&\times \left[  \left( \delta_{\nu 2} \delta_{\beta 1}-\delta_{\nu 1} \delta_{\beta 2} \right)+ \left( k_x \delta_{\beta 2}-k_y \delta_{\beta 1}\right)\frac{\partial }{\partial k_{\nu}} +\left( k_y \delta_{\nu 1}-k_x \delta_{\nu 2}\right) \frac{\partial}{\partial k_{\beta}}  \right] \left[ J(\mathbf{k}) -\tilde{\mu} \right].
\end{aligned}    
\end{equation}

By using the above equations, we obtain the following first term in Eq.~(\ref{app-kubo-eta}):
\begin{equation}
\label{app-kubo-eta-1-soc}
\begin{aligned}
\eta_{\mu \nu \alpha \beta}^{(1)}(\Omega)&=-\frac{i}{\Omega+i0} \int_{-\infty}^{+\infty} d \omega \int_{-\infty}^{+\infty} d \omega' \frac{f(\omega)-f(\omega')}{\omega'-\omega-\Omega -i 0}  \int \frac{d^2 k}{(2\pi)^2} \sum_{\eta=\pm} \frac{k_{\mu} k_{\alpha}}{2} \delta \left( \omega -\varepsilon_{\eta} \right)\delta \left( \omega' -\varepsilon_{-\eta} \right) \\
&\times \Bigg[\frac{\partial \varepsilon_{\eta}}{\partial k_{\nu}} \frac{\partial \varepsilon_{-\eta }}{\partial k_{\beta}} -  \left\{  \frac{2 t_0 k_{\nu} \left[J(\mathbf{k})-\tilde{\mu}\right]}{\sqrt{ [ J(\mathbf{k})- \tilde{\mu } ]^2 +\alpha ^2 k^2}} +\eta \frac{\partial J(\mathbf{k})}{\partial k_{\nu}}  \right\} \left\{  \frac{2 t_0 k_{\beta} \left[J(\mathbf{k})-\tilde{\mu}\right]}{\sqrt{ [ J(\mathbf{k})- \tilde{\mu } ]^2 +\alpha ^2 k^2}} -\eta \frac{\partial J(\mathbf{k})}{\partial k_{\beta}}  \right\}   \\
&-\alpha ^2 \frac{ k_{\nu} k_{\beta} - \delta_{\nu \beta} k^2  }{[ J(\mathbf{k})- \tilde{\mu } ]^2 +\alpha ^2 k^2} + \alpha ^2\Bigg\{ \delta_{\nu \beta }-\frac{ 4 t_0^2 k^2 k_{\nu} k_{\beta}}{[ J(\mathbf{k})- \tilde{\mu } ]^2 +\alpha ^2 k^2} \Bigg\} + \sum_{\eta}  \frac{ i \eta \alpha ^2 k_{\mu} k_{\alpha} \delta\left( \omega -\varepsilon_{\eta} \right) \delta\left( \omega' -\varepsilon_{-\eta} \right) }{\sqrt{ [ J(\mathbf{k})- \tilde{\mu } ]^2 +\alpha ^2 k^2} }    \\
&\times \left[  \left( \delta_{\nu 2} \delta_{\beta 1}-\delta_{\nu 1} \delta_{\beta 2} \right)+ \left( k_x \delta_{\beta 2}-k_y \delta_{\beta 1}\right)\frac{\partial }{\partial k_{\nu}} +\left( k_y \delta_{\nu 1}-k_x \delta_{\nu 2}\right) \frac{\partial}{\partial k_{\beta}}  \right] \left[ J(\mathbf{k}) -\tilde{\mu} \right] \Bigg].
\end{aligned}
\end{equation}
The SOC allows for a dynamic part of the viscoelasticity tensor and the Hall viscosity. Indeed, for the real part of the $\eta^{(1)}_{\mu \nu \alpha \beta}(\Omega)$, we obtain
\begin{equation}
\label{app-kubo-eta-1-soc-stat}
\begin{aligned}
\text{Re}\,\eta_{\mu \nu \alpha \beta}^{(1),\,\text{stat}}(\Omega)&=\pi \delta (\Omega)  \int \frac{d^2 k}{(2\pi)^2} \sum_{\eta=\pm} \frac{k_{\mu} k_{\alpha}}{2} \frac{ f\left(\varepsilon_{\eta} \right)-f\left( \varepsilon_{-\eta}\right) }{\varepsilon_{\eta}  - \varepsilon_{-\eta} }  \Bigg[\frac{\partial \varepsilon_{\eta}}{\partial k_{\nu}} \frac{\partial \varepsilon_{-\eta }}{\partial k_{\beta}} -  \left\{  \frac{2 t_0 k_{\nu} \left[J(\mathbf{k})-\tilde{\mu}\right]}{\sqrt{ [ J(\mathbf{k})- \tilde{\mu } ]^2 +\alpha ^2 k^2}} +\eta \frac{\partial J(\mathbf{k})}{\partial k_{\nu}}  \right\} \\
&\times \left\{  \frac{2 t_0 k_{\beta} \left[J(\mathbf{k})-\tilde{\mu}\right]}{\sqrt{ [ J(\mathbf{k})- \tilde{\mu } ]^2 +\alpha ^2 k^2}} - \eta \frac{\partial J(\mathbf{k})}{\partial k_{\beta}}  \right\}  -\alpha ^2 \frac{\left( 1+4t_0^2 k^2 \right)k_{\nu} k_{\beta} - \delta_{\nu \beta} k^2 }{[ J(\mathbf{k})- \tilde{\mu } ]^2 +\alpha ^2 k^2}+ \alpha^2  \delta_{\nu \beta }  \Bigg],
\end{aligned}
\end{equation}
\begin{equation}
\label{app-kubo-eta-1-soc-dyn}
\begin{aligned}
\text{Re}\,\eta_{\mu \nu \alpha \beta}^{\text{dyn}}(\Omega)&=\frac{\pi}{\Omega}  \int \frac{d^2 k}{(2\pi)^2} \sum_{\eta=\pm} \frac{k_{\mu} k_{\alpha}}{2} \left[ f\left(\varepsilon_{\eta} \right)-f\left( \varepsilon_{\eta}+\Omega \right) \right]  \Bigg[\frac{\partial \varepsilon_{\eta}}{\partial k_{\nu}} \frac{\partial \varepsilon_{-\eta }}{\partial k_{\beta}} -  \left\{  \frac{4 t_0 k_{\nu} \left[J(\mathbf{k})-\tilde{\mu}\right]}{\Omega} - \frac{\partial J(\mathbf{k})}{\partial k_{\nu}}  \right\} \\
&\times \left\{  \frac{4 t_0 k_{\beta} \left[J(\mathbf{k})-\tilde{\mu}\right]}{\Omega} + \frac{\partial J(\mathbf{k})}{\partial k_{\beta}}  \right\}  -\alpha ^2 \frac{4\left( 1+4t_0^2 k^2 \right)k_{\nu} k_{\beta} - 4 \delta_{\nu \beta} k^2 }{\Omega^2}+ \alpha^2  \delta_{\nu \beta }   \Bigg] \delta \left( \Omega +\varepsilon_{\eta}-\varepsilon_{-\eta}\right),
\end{aligned}
\end{equation}
\begin{equation}
\label{app-kubo-eta-1-soc-Hall}
\text{Re}\,\eta_{\mu \nu \alpha \beta}^{\text{Hall}}(\Omega) = \left( \delta_{\nu 1} \delta_{\beta 2}-\delta_{\nu 2} \delta_{\beta 1} \right) \int \frac{d^2 k}{(2\pi)^2}  \frac{f\left(\varepsilon_{+} \right)-f\left( \varepsilon_{-} \right)}{4 \left\{ [ J(\mathbf{k})- \tilde{\mu } ]^2 +\alpha ^2 k^2 \right\}-\Omega^2}  \frac{ 2 \alpha ^2 k_{\mu} k_{\alpha} [ J(\mathbf{k}) + \tilde{\mu } ] }{\sqrt{ [ J(\mathbf{k})- \tilde{\mu } ]^2 +\alpha ^2 k^2} }, 
\end{equation}
where we used that $\left( \mathbf{k} \cdot \partial_{\mathbf{k}} J(\mathbf{k}) \right)=2 J(\mathbf{k})$ for $d$-wave altermagnets in the last expression. For $\eta^{(2)}_{\mu \nu \alpha \beta}(\Omega)$, we find
\begin{equation}
\label{app-kubo-eta-2-soc}
\eta_{\mu \nu \alpha \beta}^{(2)}(\Omega) =\frac{ i \delta_{\alpha \beta}}{\Omega+i0} \int_{-\infty}^{+\infty}d\omega f(\omega) \int \frac{d^2k}{(2\pi)^2}\sum_{\eta=\pm} k_{\mu } \frac{\partial \varepsilon_{\eta}}{\partial k_{\nu}} \delta \left( \omega -\varepsilon_{\eta} \right)= \frac{ i \delta_{\alpha \beta} }{\Omega+i0} \int \frac{d^2k}{(2\pi)^2}\sum_{\eta=\pm} f\left( \varepsilon_{\eta} \right) k_{\mu } \frac{\partial \varepsilon_{\eta}}{\partial k_{\nu}}.
\end{equation}
For $\kappa^{-1}$, we have
\begin{equation}
\label{app-kubo-eta-23-soc}
\kappa^{-1}=-V\frac{\partial P}{\partial V}=\left(1+a\right)P,
\end{equation}
where
\begin{equation}
\label{app-kubo-eta-alpha-soc}
a=\frac{\partial P}{\partial \epsilon }=\frac{\partial P}{\partial \mu } \left(\frac{\partial \epsilon}{\partial \mu } \right)^{-1}+\frac{\partial P}{\partial \tilde{\mu } } \left(\frac{\partial \epsilon}{\partial \tilde{\mu} } \right)^{-1}.\\
\end{equation}
Finally, for $\eta_{\mu \nu \alpha \beta}^{(3)}(\Omega)$, we obtain
\begin{equation}
\label{app-kubo-eta-3-soc}
\begin{aligned}
\eta_{\mu \nu \alpha \beta}^{(3)}(\Omega)&=\frac{i}{\Omega+i0}  \int_{-\infty}^{+\infty}d\omega f(\omega) \int \frac{d^2k}{(2\pi)^2}\sum_{\eta=\pm} \left(  k_{\mu } k_{\alpha}\frac{\partial^2 \varepsilon_{\eta}}{\partial k_{\nu} \partial k_{\beta}}+ \delta_{\mu \beta}  k_{\alpha} \frac{\partial \varepsilon_{\eta}}{\partial k_{\nu}} \right)\delta \left( \omega -\varepsilon_{\eta} \right) \\
&=\frac{i}{\Omega+i0} \int \frac{d^2k}{(2\pi)^2}\sum_{\eta=\pm} f\left( \varepsilon_{\eta} \right) \left(  k_{\mu } k_{\alpha}\frac{\partial^2 \varepsilon_{\eta}}{\partial k_{\nu} \partial k_{\beta}}+ \delta_{\mu \beta}  k_{\alpha} \frac{\partial \varepsilon_{\eta}}{\partial k_{\nu}} \right). 
\end{aligned}
\end{equation}

As one can see from the above equations, the SOC preserves the structure of the static part of the viscoelasticity tensor, cf. Eq.~(\ref{eq:anisotropy-eta-terms}). Indeed, by combining Eqs.~(\ref{app-kubo-eta-1-soc-stat}), (\ref{app-kubo-eta-2-soc}), (\ref{app-kubo-eta-23-soc}), and (\ref{app-kubo-eta-3-soc}), we obtain the following general structure of the static part of the viscoelasticity tensor:
\begin{equation}
\label{eq:stat-viscoelasticity-soc}
\text{Re}\,\eta_{\mu \nu \alpha \beta}^{\text{stat}}(\Omega) = \pi \delta(\Omega)  \left[ \delta_{\mu \beta} \delta_{\nu \alpha} P  + \sum_{\eta} C_{\mu \alpha}^{(1)}(\eta)C_{\nu \beta}^{(2)}(\eta)- a \delta_{\mu \nu} \delta_{\alpha \beta} P\right],
\end{equation}
where
\begin{equation}
C_{\mu \nu}^{(i)}(\eta) = c^{(i)}_0 \delta_{\mu \nu}+c^{(i)}_1 \eta_{\mu \nu }+c^{(i)}_2 \left(1-\delta_{\mu \nu}\right).
\end{equation}

The Hall part of the viscoelasticity tensor is simplified at $\tilde{\mu}=0$, 
\begin{equation}
\label{app-kubo-eta-1-soc-Hall-0}
\text{Re}\,\eta_{\mu \nu \alpha \beta}^{\text{Hall}}(\Omega) = \left( \delta_{\nu 1} \delta_{\beta 2}-\delta_{\nu 2} \delta_{\beta 1} \right) \int \frac{d^2 k}{(2\pi)^2}  \frac{f\left(\varepsilon_{+} \right)-f\left( \varepsilon_{-} \right)}{4 \left[ J(\mathbf{k}) ^2 +\alpha ^2 k^2 \right]-\Omega^2}  \frac{ 2 \alpha ^2 k_{\mu} k_{\alpha}  J(\mathbf{k}) }{\sqrt{ J(\mathbf{k})^2 +\alpha ^2 k^2} }.
\end{equation}
Due to symmetry reasons, $\text{Re}\,\eta_{\mu \nu \mu \beta}^{\text{Hall}}(\Omega)=0$. Another simplification can be achieved in the case of weak SOC and small altermagnetic splitting,
\begin{equation}
\label{app-kubo-eta-1-soc-Hall-approx-2}
\begin{aligned}
\text{Re}\,\eta_{\mu \nu \alpha \beta}^{\text{Hall}}(0)& \approx \alpha^2 \left( \delta_{\nu 2} \delta_{\beta 1}-\delta_{\nu 1} \delta_{\beta 2} \right) \int \frac{d^2 k}{(2\pi)^2}  \frac{f\left(\varepsilon_{+}^{(0)} \right)-f\left( \varepsilon_{-}^{(0)} \right)}{2 \left[ J(\mathbf{k}) -\tilde{\mu} \right]^3}  k_{\mu} k_{\alpha} \left[ J(\mathbf{k})+\tilde{\mu} \right]= \left| \left| t_{1,2} \ll t_0 \right| \right| \\
&=-\frac{\alpha^2}{32\pi t_0^2} \left( \delta_{\nu 2} \delta_{\beta 1}-\delta_{\nu 1} \delta_{\beta 2} \right) \sum_{\eta=\pm} \frac{ \eta \mu_{\eta}^2}{\tilde{\mu}^2} \Bigg\{ \delta_{\mu \alpha} +\frac{\eta  \left(3 \tilde{\mu} - 4 \mu_{\eta} \right)}{3 \tilde{\mu}} \left[ \eta_{\mu \alpha}  \frac{t_1}{t_0}-\left(1-\delta_{\mu \alpha}\right)\frac{t_2}{t_0} \right]  \\
&+\delta_{\mu \alpha}\frac{6 \tilde{\mu}^2-16 \tilde{\mu}  \mu_{\eta} +9 \mu_{\eta}^2 }{4 \tilde{\mu} ^2 } \frac{t_1^2+t_2^2}{t_0^2} +o\left( \frac{t_{1,2}^2}{t_0^2} \right)\Bigg\},
\end{aligned}
\end{equation}
where we also set $T=0$ at $\Omega=0$.

To show that the SOC does not strongly affect the viscoelasticity tensor, we show the dependence of the $xyxy$-component of the viscoelasticity tensor on the chemical potential for a few values of $\alpha$ in Fig.~\ref{fig:stat-viscoelasticity-soc}. As one can see, the SOC does not lead to any new features in the dependence on $\mu$.

\begin{figure}[h]
\includegraphics[width=.45\textwidth]{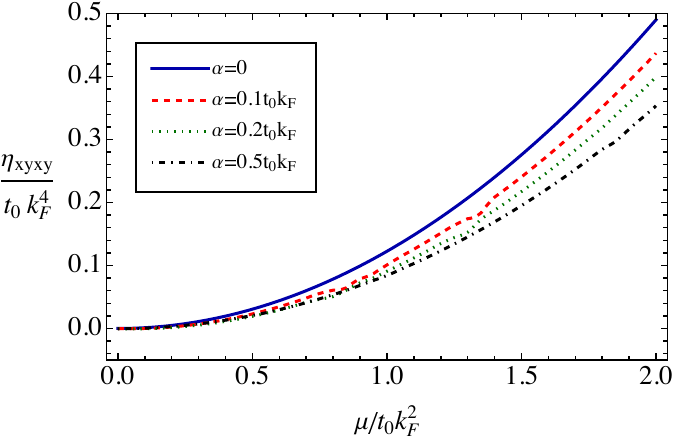}
\caption{The dependence of the $xyxy$-component of the viscoelasticity tensor on $\mu/t_0 k_F^2$ at a few values of $\alpha$. We set $\tilde{\mu}=0$.
}
\label{fig:stat-viscoelasticity-soc}
\end{figure}

\section{Spectral function}
\label{sec:App-Spectral-function}

The retarded $(+)$ and advanced $(-)$ Green functions read\\
\begin{equation}
G\left( \omega \pm i0,\mathbf{k}\right)=\frac{i}{\omega\pm i0 -H(\mathbf{k})},
\end{equation}
where $\omega$ is frequency, $\mathbf{k}$ is momentum, and $H(\mathbf{k})$ is the Hamiltonian.

The spectral function is defined as the difference between the advanced and retarded Green functions at vanishing chemical potential
\begin{equation}
A\left( \omega ;\mathbf{k}\right)=\frac{1}{2\pi}\left[G\left( \omega + i0,\mathbf{k}\right)-G\left( \omega - i0,\mathbf{k}\right)\right]_{\mu_{\lambda}=0}.
\end{equation}

For the low-energy model of altermagnets, see Eq.~(\ref{eq:hamiltonian}), the spectral function is given by
\begin{equation}
A\left(\omega;\mathbf{k}\right) = \text{diag}\left[ \delta\left( \omega-\varepsilon_{+}(\mathbf{k})\right),\delta\left(\omega-\varepsilon_{-} (\mathbf{k}) \right) \right].
\end{equation}
Here, $\varepsilon_{\lambda}(\mathbf{k})$ is the dispersion relation of quasiparticles with spin $\lambda$.

\section{Viscous hydrodynamic equations}
\label{sec:App-Hydro}

By substituting the viscoelasticity tensor into Eqs.~(\ref{eq:hydrodynamics-general-spin-flip-continuity}) and (\ref{eq:hydrodynamics-general-spin-flip-Euler}), we find the following set of hydrodynamic equations:
\begin{eqnarray}
\label{continuity_eq_1}
\frac{\partial \rho}{\partial t}+\left(\bm{\nabla} \cdot \rho  \mathbf{u} \left( t, \mathbf{r} \right)  \right)=0,\\
\label{continuity_eq_2}
\frac{\partial \tilde{\rho}}{\partial t}+\left(\bm{\nabla} \cdot \tilde{\rho}  \mathbf{u} \left( t, \mathbf{r} \right)  \right)=0,
\end{eqnarray}
and
\begin{equation}
\label{eq:viscosity-general}
\begin{aligned}
&\frac{\partial}{\partial t}\left[  t_0 \omega \mathbf{u}\left( t, \mathbf{r} \right) + t_1 \tilde{\omega} \left(  u_x \left( t, \mathbf{r} \right) \mathbf{e}_x - u_y \left( t, \mathbf{r} \right)  \mathbf{e}_y \right) - t_2 \tilde{\omega} \left( u_y \left( t, \mathbf{r} \right)  \mathbf{e}_x+ u_x \left( t, \mathbf{r} \right)  \mathbf{e}_y \right) \right]+ \bm{\nabla} \epsilon - \rho \mathbf{E}\\
=&-\frac{1}{\tau}\left[  t_0 \omega \mathbf{u}\left( t, \mathbf{r} \right) + t_1 \tilde{\omega}  u_x \left( t, \mathbf{r} \right) \mathbf{e}_x -  t_1 \tilde{\omega} u_y \left( t, \mathbf{r} \right)  \mathbf{e}_y  -  t_2 \tilde{\omega} u_y \left( t, \mathbf{r} \right) \mathbf{e}_x-  t_2 \tilde{\omega} u_x \left( t, \mathbf{r} \right) \mathbf{e}_y  \right]  +\tau_{ee} \epsilon \Delta \mathbf{u }\left( t, \mathbf{r} \right)\\
+& 2\frac{\tau_{ee} t_1^2}{\tilde{t}_0^2} \epsilon \left[ \nabla_{x}^2 u_{y}\left( t, \mathbf{r} \right)\mathbf{e}_y +\nabla_{y}^2 u_x \left( t, \mathbf{r} \right) \mathbf{e}_x \right] - \frac{\tau_{ee} t_2^2}{\tilde{t}_0^2} \epsilon  \left[ 2\nabla_x \nabla_{y} u_y \left( t,\mathbf{r}\right) \mathbf{e}_x + 2\nabla_x \nabla_{y} u_x \left( t,\mathbf{r}\right) \mathbf{e}_y -\Delta \mathbf{u} \left( t,\mathbf{r}\right) \right] \\
+& \frac{\tau_{ee} t_1 t_2}{\tilde{t}_0^2} \epsilon \left\{ 2 \nabla_{x}\nabla_y \left[ u_x \left( t, \mathbf{r} \right) \mathbf{e}_x -u_y \left( t, \mathbf{r} \right) \mathbf{e}_y \right] +\left( \nabla_{x}^2-\nabla_{y}^2 \right) \left[] u_y \left( t, \mathbf{r} \right) \mathbf{e}_x+u_x \left( t, \mathbf{r} \right) \mathbf{e}_y  \right]  \right\} \\
+& 2 \frac{\tau_{ee} t_0 t_1}{\tilde{t}_0^2}\tilde{\epsilon} \left[ \nabla_{y}^2 u_x\left( t, \mathbf{r} \right) \mathbf{e}_x - \nabla_{x}^2 u_y \left( t, \mathbf{r} \right) \mathbf{e}_y \right] +  \frac{\tau_{ee} t_0 t_2}{\tilde{t}_0^2}\tilde{\epsilon}   \left[ 2\nabla_x \nabla_{y}  \mathbf{u}\left( t,\mathbf{r}\right) -  \Delta u_y \left( t,\mathbf{r}\right) \mathbf{e}_x - \Delta u_x \left( t,\mathbf{r}\right) \mathbf{e}_y \right].   \\
\end{aligned}
\end{equation}

These are the continuity equations for the electric and spin currents as well as the Navier-Stokes equation.

For the electron fluid, the above equations should be supplemented with Maxwell's equations. In the quasistatic approximation,
\begin{equation}
\left(\bm{\nabla} \cdot \mathbf{E} \right)=4\pi \delta \rho, \quad \quad \left[ \bm{\nabla} \times \mathbf{E} \right] =0,
\end{equation}
where $\delta \rho = \rho -\rho^{(0)}$ is the electric charge density deviation from the equilibrium value.

\subsection{Finite geometry}
\label{sec:visco-fem}

While in the main text, we focused on the case of a ribbon geometry, similar distributions of the electric currents can be obtained in finite samples. In the stationary regime, we solve the following set of equations:
\begin{eqnarray}
\label{eq:hydro-FEM}
\bm{\nabla}\cdot \mathbf{u}\left( \mathbf{r}\right) &=& 0,\\
\rho \nabla_{x} \bar{\phi}_{\text{el}} \left( \mathbf{r} \right) &=&  \eta_{1 jkl}\nabla_{j}\nabla_{l} u_{k}\left( \mathbf{r} \right),\\
\rho \nabla_{y} \bar{\phi}_{\text{el}} \left( \mathbf{r} \right) &=& \eta_{2 jkl}\nabla_{j}\nabla_{l} u_{k}\left( \mathbf{r} \right) 
\end{eqnarray}
with the no-slip boundary conditions $u_{x}(x,y=0,L_y)=0$, $u_{y}(x=\pm L_x,y)=0$ at the edges of the channel and fixed electrochemical potential $\bar{\phi}_{\text{el}}$ at the contacts
\begin{eqnarray}
\label{eq:phi-lambda-gen-bc-FEM}
\bar{\phi}_{\text{el}} \left(x,y=0\right)&=& \phi_0 ,\,\,|x|<w,\\
\label{eq:phi-lambda-gen-bc-FEM-2}
\bar{\phi}_{\text{el}} \left(x,y=L_y\right)&=& -\phi_0,\,\,|x|<w,
\end{eqnarray}
where $2w$ is the size of the contacts required for numerics. The normal components of the electric current through all the surfaces except the contacts should also vanish. In our numerical calculations, we use the finite element method implemented in Wolfram Mathematica~\cite{Wolfram}.

The numerical results in the sample with $x \in \left[ -2L ,2L \right]$ and $y \in \left[ 0 , L \right]$ are presented in the Fig.~\ref{fig:streamlines-finite}. As one can see by comparing with the results in Figs.~\ref{fig:streamlines1}(a), \ref{fig:streamlines3}(c), and \ref{fig:streamlines1}(c), the results in finite samples share the same key features, such as the sign-changing electrochemical potential near the contacts and the vortices near the middle of the sample.

\begin{figure*}[h]
\subfigure[]{\includegraphics[width=.45\textwidth]{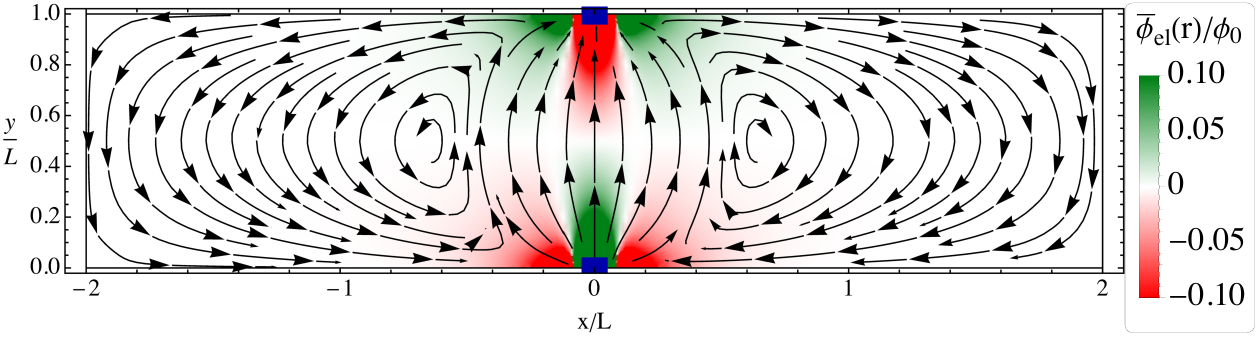}}
\subfigure[]{\includegraphics[width=.45\textwidth]{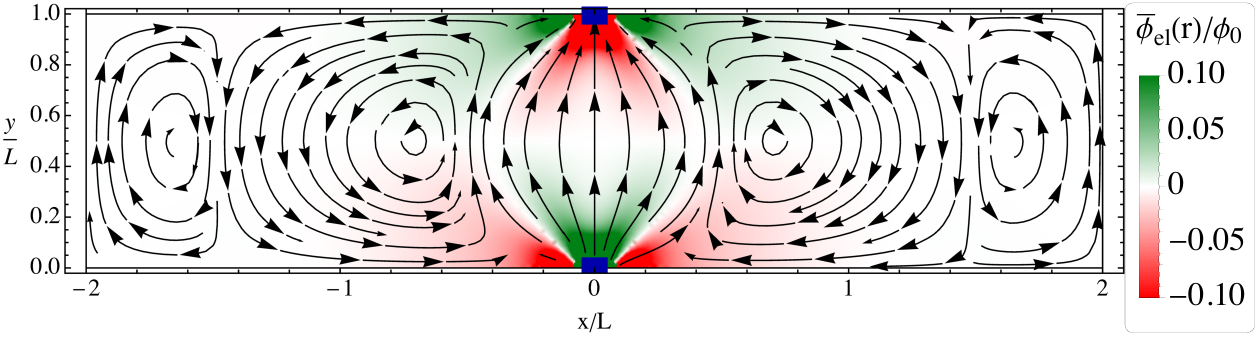}}
\subfigure[]{\includegraphics[width=.45\textwidth]{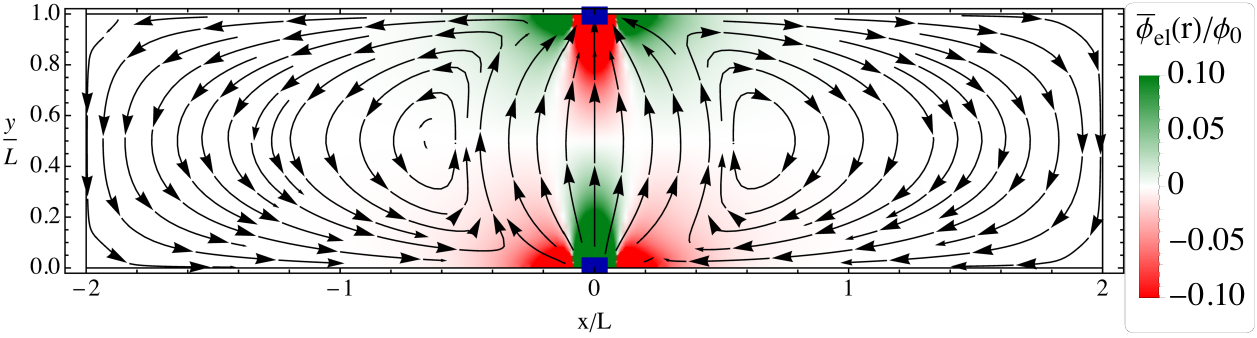}}
\caption{
The current streamlines ($\psi(\mathbf{r}) = \text{const}$) for (a) $t_{1,2} = 0$, (b) $t_{1}=0.9 \, t_0$, $t_2=0$, (c) $t_{1}=0$, $t_2=0.9 \, t_0$. In all panels, we set $\mu = t_0 k_F^2 $, $w=0.1L$, and $\tilde{\mu} =0$.
The electrochemical potential $\bar{\phi}_{\text{el}} /\phi_0$ is shown in color. The green and red colors correspond to $\bar{\phi}_{\text{el}}(\mathbf{r})>0$ and $\bar{\phi}_{\text{el}}(\mathbf{r})<0$, respectively. Blue rectangles denote the position of the source and drain.
}
\label{fig:streamlines-finite}
\end{figure*}

Similarly to the ribbon geometry, the splitting of the central maxima in Fig.~\ref{fig:streamlines-finite}(b) becomes more prominent as the ratio $t_1/t_0$ increases.

\subsection{Role of SOC}
\label{sec:App-Hydro-soc}

Let us address the effects of the SOC on the transport equations. We start with the Boltzmann kinetic equation that contains the anomalous velocity term,
\begin{equation}
\label{eq:Boltzmann-def-soc}
\frac{\partial f_{\eta}}{\partial t} +\left( \mathbf{v}_{\mathbf{k},\eta} + e\left[ \mathbf{E} \times \bm{\Omega}_{\eta} \right]\right)\cdot \frac{\partial f_{\eta}}{\partial \mathbf{r}}+ \left( e \mathbf{E}\cdot \frac{\partial f_{\eta}}{\partial \mathbf{k}} \right) = I_{\text{col}}\left\{ f_{\eta} \right\},
\end{equation}
see the term $\propto \left[ \mathbf{E} \times \bm{\Omega}_{\eta} \right]$. The Berry curvature $ \bm{\Omega}_{\eta}$ reads 
\begin{equation}
\bm{\Omega}_{\eta} = -\eta \alpha ^2  \frac{ J(\mathbf{k}) -\tilde{\mu }-k_x \partial_{k_x} J(\mathbf{k}) - k_y \partial_{k_y} J(\mathbf{k})  }{2 \left\{ \left[J(\mathbf{k})-\tilde{\mu }\right]^2+\alpha ^2 k^2\right\}^{3/2}} \mathbf{e}_{z}.
\end{equation}

The anomalous velocity leads to the anomalous Hall effect (AHE) current
\begin{equation}
\label{eq:ahe-soc-def}
\begin{aligned}
\mathbf{j}_{\eta}^{\text{AHE}}&=e^2\int \frac{d^2 k}{\left(2\pi\right)^2} \left[ \mathbf{E} \times \bm{\Omega}_{\eta}  \right] f_{\eta}^{(\mathbf{0})}=\eta \alpha ^2 e^2 \left( \mathbf{e}_x E_y - \mathbf{e}_y E_x\right) \int \frac{d^2 k}{\left( 2 \pi \right)^2}  \frac{ J(\mathbf{k}) +\tilde{\mu } }{2 \left\{ \left[J(\mathbf{k})-\tilde{\mu }\right]^2+\alpha ^2 k^2\right\}^{3/2}} f\left( \varepsilon_{\eta} \right)\\
&\equiv \left( \mathbf{e}_x E_y - \mathbf{e}_y E_x\right) \sigma_{\eta}^{\text{AHE}}.
\end{aligned}
\end{equation}
Here we used the relation $\left( \mathbf{k} \cdot \partial_{\mathbf{k}} J(\mathbf{k}) \right)=2 J(\mathbf{k})$ valid for $d$-wave altermagnets, which we assume below. The AHE conductivity is obtained by summing over all bands, $\sigma^{\text{AHE}}=\sum_{\eta} \sigma_{\eta}^{\text{AHE}}$.

At $\tilde{\mu}=0$, the AHE current vanishes,
\begin{equation}
\label{eq:ahe-soc-def-0}
\mathbf{j}_{\eta}^{\text{AHE}} =\eta \alpha ^2 e^2 \left( \mathbf{e}_x E_y - \mathbf{e}_y E_x\right) \int \frac{d^2 k}{\left( 2 \pi \right)^2}  \frac{ t_2 \sin (2\phi)-t_1 \cos(2\phi) }{2 k \left\{ k^2 \left[t_2 \sin (2\phi)-t_1 \cos(2\phi) \right]^2+\alpha ^2 \right\}^{3/2}} f\left( \varepsilon_{\eta} \right)=0.
\end{equation}
Heuristically, this follows from the fact that altermagnetic spin-splitting that determines the Berry curvature averages to zero. For small spin-orbital coupling $\alpha k_F/ \mu \ll 1$ and small parameters of altermagnetic spin-splitting, the anomalous Hall current can be calculated analytically, 
\begin{equation}
\mathbf{j}_{\eta}^{\text{AHE}} \approx - \frac{ \alpha ^2 e^2 }{16 \pi t_0} \left( \mathbf{e}_x E_y - \mathbf{e}_y E_x\right) \left[ \frac{2 \eta \mu_{\eta}}{\tilde{\mu}^2}-  \frac{\eta \mu_{\eta} \left( \tilde{\mu} - \mu_{\eta} \right) \left( \tilde{\mu} - 3 \mu_{\eta} \right) }{ \tilde{\mu}^4} \frac{t_1^2+t_2^2}{t_0^2} +o\left( \frac{t_{1,2}^2}{t_0^2} \right) \right].
\end{equation}

In what follows, we consider the effects of the SOC in the channel and ribon geometries. The channel geometry is defined as $x\in\left( -\infty, +\infty \right)$ and $y\in\left[ 0, L \right]$ with no-slip boundary conditions $u_{x}(x,y=0, L)=0$ at edges of the channel as discussed in Sec.~\ref{sec:Flow}. The electric current continuity equation and the components of the Navier-Stokes equation read
\begin{equation}
\label{app:eq-cont}
 \nabla_{y} \left[\rho(y) u_{y}(y) \right] - \sum_{\eta} \nabla_{y} \left[ E_x \sigma_{\eta}^{\text{AHE}}(y)\right] = 0
\end{equation}
and
\begin{equation}
\label{app:eq-navier-stokes-x-flow}
-\frac{1}{2 e \tau \tilde{t}_0^2 } \left[ \left( t_0 \rho+t_1 \tilde{\rho} \right) u_x - t_2 \tilde{\rho} u_y \right]+\tau_{\text{ee}} \epsilon \frac{t_0^2+t_1^2}{\tilde{t}_0^2} \nabla_{y}^2 u_x - \tau_{\text{ee}} \epsilon \frac{t_1 t_2}{\tilde{t}_0^2} \nabla_{y}^2 u_y + 2 \tau_{\text{ee}} \tilde{\epsilon} \frac{t_0 t_1 }{\tilde{t}_0^2} \nabla_{y}^2 u_x -\tau_{\text{ee}} \tilde{\epsilon} \frac{t_0 t_2 }{\tilde{t}_0^2} \nabla_{y}^2 u_y=-\rho E_x,
\end{equation}
\begin{equation}
\label{app:eq-navier-stokes-y-flow}
-\frac{1}{2 e \tau \tilde{t}_0^2 } \left[ \left( t_0 \rho-t_1 \tilde{\rho} \right) u_y - t_2 \tilde{\rho} u_x \right]+\tau_{\text{ee}} \epsilon \frac{t_0^2-t_1^2}{\tilde{t}_0^2} \nabla_{y}^2 u_y - \tau_{\text{ee}} \epsilon \frac{t_1 t_2}{\tilde{t}_0^2} \nabla_{y}^2 u_x  -\tau_{\text{ee}} \tilde{\epsilon} \frac{t_0 t_2 }{\tilde{t}_0^2} \nabla_{y}^2 u_x=-\rho E_y,
\end{equation}
respectively.

Using the continuity equation \eqref{app:eq-cont}, we obtain:
\begin{equation}
\rho(y)  u_{y} = C_1+ E_x \sigma^{\text{AHE}},
\end{equation}
where $C_1$ is a constant and $\sigma^{\text{AHE}}=\sum_{\eta} \sigma_{\eta}^{\text{AHE}}$. By using the same approximations as in Sec.~\ref{sec:Flow}, i.e., uniform $\mu_{\eta}$ and $T$, Eqs.~(\ref{app:eq-navier-stokes-x-flow}) and (\ref{app:eq-navier-stokes-y-flow}) can be rewritten
\begin{equation}
\tau_{\text{ee}} \frac{\left( t_0^2+t_1^2 \right) \epsilon  +2 t_0 t_1 \tilde{\epsilon}}{\tilde{t}_0^2} \nabla_{y}^2 u_x  -\frac{t_0 \rho+t_1 \tilde{\rho}}{2 e \tau \tilde{t}_0^2 }  u_x  =-\rho E_x- \frac{t_2 \tilde{\rho} }{2 e \tau \tilde{t}_0^2 } u_y
\end{equation}
\begin{equation}
\tau_{\text{ee}} \frac{t_2 \left(t_0 \tilde{\epsilon} +t_1 \epsilon\right) }{\tilde{t}_0^2} \nabla_{y}^2 u_x-\frac{t_2 \tilde{\rho}}{2 e \tau \tilde{t}_0^2 } u_{x}  = \rho E_y - \frac{t_0 \rho-t_1 \tilde{\rho} }{2 e \tau \tilde{t}_0^2 } u_y.
\end{equation}
These equations have the same solutions as those given in Eqs.~(\ref{eq:ux-def}) and (\ref{eq:uy-def}) as well as the constraint (\ref{eq:A-constraint}) for the parameters $\rho$, $\tilde{\rho}$, $\epsilon$, and $\tilde{\epsilon}$. The latter follows from the consistency of the solution (\ref{eq:ux-def}) for each projection of the Navier-Stokes equation given in Eqs.~(\ref{app:eq-navier-stokes-x-flow}) and (\ref{app:eq-navier-stokes-y-flow}). The Hall viscosity given in Eq.~\eqref{app-kubo-eta-1-soc-Hall-0} does not modify the Navier-Stokes equation; this follows from the symmetry of the viscosity tensor.

In the case of a ribbon geometry with point contacts, the flows are not qualitatively affected by the SOC at $\tilde{\mu}=0$. This follows from the fact that the structure of the Navier-Stokes equations is preserved.
To demonstrate this, we solve Eq.~(\ref{eq:hydro-FEM}) in the finite sample with the boundary conditions given in Eqs.~(\ref{eq:phi-lambda-gen-bc-FEM}) and \eqref{eq:phi-lambda-gen-bc-FEM-2} for the modified by SOC viscoelasticity tensor presented in Eq.~(\ref{eq:stat-viscoelasticity-soc}). In this case, the numerical results in the sample with $x \in \left[ -2L ,2L \right]$ and $y \in \left[ 0 , L \right]$ for $\mu = t_0 k_F^2 $ and $w=0.1L$ are presented in Fig.~\ref{fig:streamlines-finite-soc}.

\begin{figure*}[h]
\subfigure[]{\includegraphics[width=.45\textwidth]{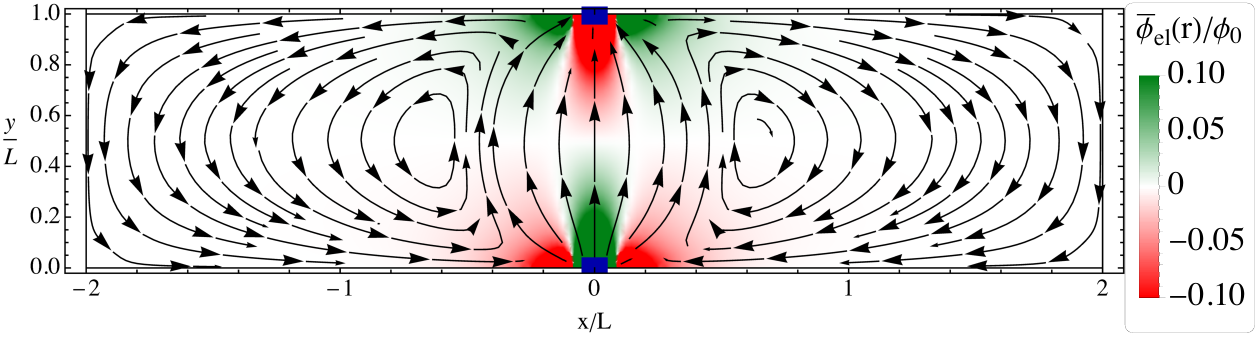}}
\subfigure[]{\includegraphics[width=.45\textwidth]{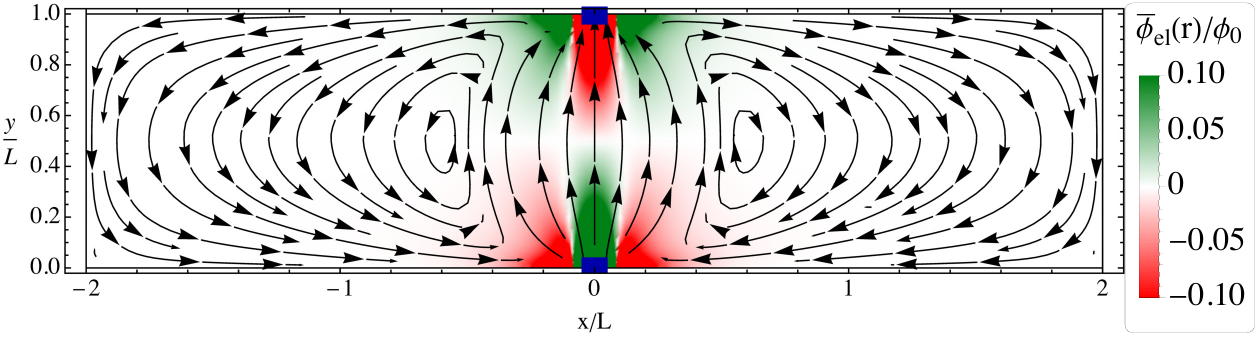}}\\
\caption{
The current streamlines ($\psi(\mathbf{r}) = \text{const}$) for (a) $\alpha = 0$, (b) $\alpha=0.5 t_0 k_F$.
In all panels, we set $t_2=0.5\, t_0$, $\tilde{\mu}=0$, and $\mu = t_0 k_F^2 $. The electrochemical potential $\bar{\phi}_{\text{el}} /\phi_0$ is shown in color. The green and red colors correspond to $\bar{\phi}_{\text{el}}(\mathbf{r})>0$ and $\bar{\phi}_{\text{el}}(\mathbf{r})<0$, respectively. Blue rectangles denote the position of the source and drain.
}
\label{fig:streamlines-finite-soc}
\end{figure*}

In Figs.~\ref{fig:streamlines-finite-soc}(a) and \ref{fig:streamlines-finite-soc}(b), we can see that the centers of vortices move slightly closer to the axis $x=0$ as the SOC parameter $\alpha$ increases. These are, however, quantitative rather than qualitative features.

\end{widetext}

\bibliography{library-short}

\begin{thebibliography}{107}%
\makeatletter
\providecommand \@ifxundefined [1]{%
 \@ifx{#1\undefined}
}%
\providecommand \@ifnum [1]{%
 \ifnum #1\expandafter \@firstoftwo
 \else \expandafter \@secondoftwo
 \fi
}%
\providecommand \@ifx [1]{%
 \ifx #1\expandafter \@firstoftwo
 \else \expandafter \@secondoftwo
 \fi
}%
\providecommand \natexlab [1]{#1}%
\providecommand \enquote  [1]{``#1''}%
\providecommand \bibnamefont  [1]{#1}%
\providecommand \bibfnamefont [1]{#1}%
\providecommand \citenamefont [1]{#1}%
\providecommand \href@noop [0]{\@secondoftwo}%
\providecommand \href [0]{\begingroup \@sanitize@url \@href}%
\providecommand \@href[1]{\@@startlink{#1}\@@href}%
\providecommand \@@href[1]{\endgroup#1\@@endlink}%
\providecommand \@sanitize@url [0]{\catcode `\\12\catcode `\$12\catcode `\&12\catcode `\#12\catcode `\^12\catcode `\_12\catcode `\%12\relax}%
\providecommand \@@startlink[1]{}%
\providecommand \@@endlink[0]{}%
\providecommand \url  [0]{\begingroup\@sanitize@url \@url }%
\providecommand \@url [1]{\endgroup\@href {#1}{\urlprefix }}%
\providecommand \urlprefix  [0]{URL }%
\providecommand \Eprint [0]{\href }%
\providecommand \doibase [0]{https://doi.org/}%
\providecommand \selectlanguage [0]{\@gobble}%
\providecommand \bibinfo  [0]{\@secondoftwo}%
\providecommand \bibfield  [0]{\@secondoftwo}%
\providecommand \translation [1]{[#1]}%
\providecommand \BibitemOpen [0]{}%
\providecommand \bibitemStop [0]{}%
\providecommand \bibitemNoStop [0]{.\EOS\space}%
\providecommand \EOS [0]{\spacefactor3000\relax}%
\providecommand \BibitemShut  [1]{\csname bibitem#1\endcsname}%
\let\auto@bib@innerbib\@empty
\bibitem [{\citenamefont {Gurzhi}(1963)}]{Gurzhi:1963}%
  \BibitemOpen
  \bibfield  {author} {\bibinfo {author} {\bibfnamefont {R.}~\bibnamefont {Gurzhi}},\ }\bibfield  {title} {\bibinfo {title} {Minimum of resistance in impurity-free conductors},\ }\href {http://www.jetp.ras.ru/cgi-bin/e/index/e/17/2/p521?a=list} {\bibfield  {journal} {\bibinfo  {journal} {JETP}\ }\textbf {\bibinfo {volume} {17}},\ \bibinfo {pages} {521} (\bibinfo {year} {1963})}\BibitemShut {NoStop}%
\bibitem [{\citenamefont {Gurzhi}(1968)}]{Gurzhi:1968}%
  \BibitemOpen
  \bibfield  {author} {\bibinfo {author} {\bibfnamefont {R.~N.}\ \bibnamefont {Gurzhi}},\ }\bibfield  {title} {\bibinfo {title} {Hydrodynamic effects in solids at low temperature},\ }\href {https://doi.org/10.1070/PU1968v011n02ABEH003815} {\bibfield  {journal} {\bibinfo  {journal} {Sov. Phys. Uspekhi}\ }\textbf {\bibinfo {volume} {11}},\ \bibinfo {pages} {255} (\bibinfo {year} {1968})}\BibitemShut {NoStop}%
\bibitem [{\citenamefont {Lucas}\ and\ \citenamefont {Fong}(2018)}]{Lucas-Fong:rev-2017}%
  \BibitemOpen
  \bibfield  {author} {\bibinfo {author} {\bibfnamefont {A.}~\bibnamefont {Lucas}}\ and\ \bibinfo {author} {\bibfnamefont {K.~C.}\ \bibnamefont {Fong}},\ }\bibfield  {title} {\bibinfo {title} {Hydrodynamics of electrons in graphene},\ }\href {https://doi.org/10.1088/1361-648X/aaa274} {\bibfield  {journal} {\bibinfo  {journal} {J. Phys. Condens. Matter}\ }\textbf {\bibinfo {volume} {30}},\ \bibinfo {pages} {053001} (\bibinfo {year} {2018})},\ \Eprint {https://arxiv.org/abs/1710.08425} {arXiv:1710.08425} \BibitemShut {NoStop}%
\bibitem [{\citenamefont {Narozhny}(2019)}]{Narozhny:rev-2019}%
  \BibitemOpen
  \bibfield  {author} {\bibinfo {author} {\bibfnamefont {B.~N.}\ \bibnamefont {Narozhny}},\ }\bibfield  {title} {\bibinfo {title} {Electronic hydrodynamics in graphene},\ }\href {https://doi.org/10.1016/j.aop.2019.167979} {\bibfield  {journal} {\bibinfo  {journal} {Ann. Phys.}\ }\textbf {\bibinfo {volume} {411}},\ \bibinfo {pages} {167979} (\bibinfo {year} {2019})},\ \Eprint {https://arxiv.org/abs/1905.09686} {arXiv:1905.09686} \BibitemShut {NoStop}%
\bibitem [{\citenamefont {Narozhny}(2022)}]{Narozhny:rev-2022}%
  \BibitemOpen
  \bibfield  {author} {\bibinfo {author} {\bibfnamefont {B.~N.}\ \bibnamefont {Narozhny}},\ }\bibfield  {title} {\bibinfo {title} {Hydrodynamic approach to two-dimensional electron systems},\ }\href {https://doi.org/10.1007/s40766-022-00036-z} {\bibfield  {journal} {\bibinfo  {journal} {Riv. Nuovo Cimento}\ }\textbf {\bibinfo {volume} {45}},\ \bibinfo {pages} {661} (\bibinfo {year} {2022})},\ \Eprint {https://arxiv.org/abs/2207.10004} {arXiv:2207.10004} \BibitemShut {NoStop}%
\bibitem [{\citenamefont {Fritz}\ and\ \citenamefont {Scaffidi}(2024)}]{Fritz-Scaffidi-HydrodynamicElectronicTransport-2024}%
  \BibitemOpen
  \bibfield  {author} {\bibinfo {author} {\bibfnamefont {L.}~\bibnamefont {Fritz}}\ and\ \bibinfo {author} {\bibfnamefont {T.}~\bibnamefont {Scaffidi}},\ }\bibfield  {title} {\bibinfo {title} {Hydrodynamic {{Electronic Transport}}},\ }\href {https://doi.org/10.1146/annurev-conmatphys-040521-042014} {\bibfield  {journal} {\bibinfo  {journal} {Annu. Rev. Condens. Matter Phys.}\ }\textbf {\bibinfo {volume} {15}},\ \bibinfo {pages} {17} (\bibinfo {year} {2024})}\BibitemShut {NoStop}%
\bibitem [{\citenamefont {Molenkamp}\ and\ \citenamefont {{de Jong}}(1994)}]{Molenkamp-Jong:1994}%
  \BibitemOpen
  \bibfield  {author} {\bibinfo {author} {\bibfnamefont {L.}~\bibnamefont {Molenkamp}}\ and\ \bibinfo {author} {\bibfnamefont {M.}~\bibnamefont {{de Jong}}},\ }\bibfield  {title} {\bibinfo {title} {Observation of {{Knudsen}} and {{Gurzhi}} transport regimes in a two-dimensional wire},\ }\href {https://doi.org/10.1016/0038-1101(94)90244-5} {\bibfield  {journal} {\bibinfo  {journal} {Solid-State Electron.}\ }\textbf {\bibinfo {volume} {37}},\ \bibinfo {pages} {551} (\bibinfo {year} {1994})}\BibitemShut {NoStop}%
\bibitem [{\citenamefont {De~Jong}\ and\ \citenamefont {Molenkamp}(1995)}]{Jong-Molenkamp:1995}%
  \BibitemOpen
  \bibfield  {author} {\bibinfo {author} {\bibfnamefont {M.~J.}\ \bibnamefont {De~Jong}}\ and\ \bibinfo {author} {\bibfnamefont {L.~W.}\ \bibnamefont {Molenkamp}},\ }\bibfield  {title} {\bibinfo {title} {Hydrodynamic electron flow in high-mobility wires},\ }\href {https://doi.org/10.1103/PhysRevB.51.13389} {\bibfield  {journal} {\bibinfo  {journal} {Phys. Rev. B}\ }\textbf {\bibinfo {volume} {51}},\ \bibinfo {pages} {13389} (\bibinfo {year} {1995})},\ \Eprint {https://arxiv.org/abs/cond-mat/9411067} {arXiv:cond-mat/9411067} \BibitemShut {NoStop}%
\bibitem [{\citenamefont {Crossno}\ \emph {et~al.}(2016)\citenamefont {Crossno}, \citenamefont {Shi}, \citenamefont {Wang}, \citenamefont {Liu}, \citenamefont {Harzheim}, \citenamefont {Lucas}, \citenamefont {Sachdev}, \citenamefont {Kim}, \citenamefont {Taniguchi}, \citenamefont {Watanabe}, \citenamefont {Ohki},\ and\ \citenamefont {Fong}}]{Crossno-Fong:2016}%
  \BibitemOpen
  \bibfield  {author} {\bibinfo {author} {\bibfnamefont {J.}~\bibnamefont {Crossno}}, \bibinfo {author} {\bibfnamefont {J.~K.}\ \bibnamefont {Shi}}, \bibinfo {author} {\bibfnamefont {K.}~\bibnamefont {Wang}}, \bibinfo {author} {\bibfnamefont {X.}~\bibnamefont {Liu}}, \bibinfo {author} {\bibfnamefont {A.}~\bibnamefont {Harzheim}}, \bibinfo {author} {\bibfnamefont {A.}~\bibnamefont {Lucas}}, \bibinfo {author} {\bibfnamefont {S.}~\bibnamefont {Sachdev}}, \bibinfo {author} {\bibfnamefont {P.}~\bibnamefont {Kim}}, \bibinfo {author} {\bibfnamefont {T.}~\bibnamefont {Taniguchi}}, \bibinfo {author} {\bibfnamefont {K.}~\bibnamefont {Watanabe}}, \bibinfo {author} {\bibfnamefont {T.~A.}\ \bibnamefont {Ohki}},\ and\ \bibinfo {author} {\bibfnamefont {K.~C.}\ \bibnamefont {Fong}},\ }\bibfield  {title} {\bibinfo {title} {Observation of the {{Dirac}} fluid and the breakdown of the {{Wiedemann-Franz}} law in graphene},\ }\href {https://doi.org/10.1126/science.aad0343} {\bibfield  {journal} {\bibinfo  {journal} {Science}\
  }\textbf {\bibinfo {volume} {351}},\ \bibinfo {pages} {1058} (\bibinfo {year} {2016})},\ \Eprint {https://arxiv.org/abs/1509.04713} {arXiv:1509.04713} \BibitemShut {NoStop}%
\bibitem [{\citenamefont {Ghahari}\ \emph {et~al.}(2016)\citenamefont {Ghahari}, \citenamefont {Xie}, \citenamefont {Taniguchi}, \citenamefont {Watanabe}, \citenamefont {Foster},\ and\ \citenamefont {Kim}}]{Ghahari-Kim:2016}%
  \BibitemOpen
  \bibfield  {author} {\bibinfo {author} {\bibfnamefont {F.}~\bibnamefont {Ghahari}}, \bibinfo {author} {\bibfnamefont {H.-Y.}\ \bibnamefont {Xie}}, \bibinfo {author} {\bibfnamefont {T.}~\bibnamefont {Taniguchi}}, \bibinfo {author} {\bibfnamefont {K.}~\bibnamefont {Watanabe}}, \bibinfo {author} {\bibfnamefont {M.~S.}\ \bibnamefont {Foster}},\ and\ \bibinfo {author} {\bibfnamefont {P.}~\bibnamefont {Kim}},\ }\bibfield  {title} {\bibinfo {title} {Enhanced thermoelectric power in graphene: {{Violation}} of the {{Mott}} relation by inelastic scattering},\ }\href {https://doi.org/10.1103/PhysRevLett.116.136802} {\bibfield  {journal} {\bibinfo  {journal} {Phys. Rev. Lett.}\ }\textbf {\bibinfo {volume} {116}},\ \bibinfo {pages} {136802} (\bibinfo {year} {2016})},\ \Eprint {https://arxiv.org/abs/1601.05859} {arXiv:1601.05859} \BibitemShut {NoStop}%
\bibitem [{\citenamefont {Krishna~Kumar}\ \emph {et~al.}(2017)\citenamefont {Krishna~Kumar}, \citenamefont {Bandurin}, \citenamefont {Pellegrino}, \citenamefont {Cao}, \citenamefont {Principi}, \citenamefont {Guo}, \citenamefont {Auton}, \citenamefont {Ben~Shalom}, \citenamefont {Ponomarenko}, \citenamefont {Falkovich}, \citenamefont {Watanabe}, \citenamefont {Taniguchi}, \citenamefont {Grigorieva}, \citenamefont {Levitov}, \citenamefont {Polini},\ and\ \citenamefont {Geim}}]{Krishna-Falkovich:2017}%
  \BibitemOpen
  \bibfield  {author} {\bibinfo {author} {\bibfnamefont {R.}~\bibnamefont {Krishna~Kumar}}, \bibinfo {author} {\bibfnamefont {D.~A.}\ \bibnamefont {Bandurin}}, \bibinfo {author} {\bibfnamefont {F.~M.~D.}\ \bibnamefont {Pellegrino}}, \bibinfo {author} {\bibfnamefont {Y.}~\bibnamefont {Cao}}, \bibinfo {author} {\bibfnamefont {A.}~\bibnamefont {Principi}}, \bibinfo {author} {\bibfnamefont {H.}~\bibnamefont {Guo}}, \bibinfo {author} {\bibfnamefont {G.~H.}\ \bibnamefont {Auton}}, \bibinfo {author} {\bibfnamefont {M.}~\bibnamefont {Ben~Shalom}}, \bibinfo {author} {\bibfnamefont {L.~A.}\ \bibnamefont {Ponomarenko}}, \bibinfo {author} {\bibfnamefont {G.}~\bibnamefont {Falkovich}}, \bibinfo {author} {\bibfnamefont {K.}~\bibnamefont {Watanabe}}, \bibinfo {author} {\bibfnamefont {T.}~\bibnamefont {Taniguchi}}, \bibinfo {author} {\bibfnamefont {I.~V.}\ \bibnamefont {Grigorieva}}, \bibinfo {author} {\bibfnamefont {L.~S.}\ \bibnamefont {Levitov}}, \bibinfo {author} {\bibfnamefont {M.}~\bibnamefont {Polini}},\ and\ \bibinfo
  {author} {\bibfnamefont {A.~K.}\ \bibnamefont {Geim}},\ }\bibfield  {title} {\bibinfo {title} {Superballistic flow of viscous electron fluid through graphene constrictions},\ }\href {https://doi.org/10.1038/nphys4240} {\bibfield  {journal} {\bibinfo  {journal} {Nat. Phys.}\ }\textbf {\bibinfo {volume} {13}},\ \bibinfo {pages} {1182} (\bibinfo {year} {2017})},\ \Eprint {https://arxiv.org/abs/1703.06672} {arXiv:1703.06672} \BibitemShut {NoStop}%
\bibitem [{\citenamefont {Berdyugin}\ \emph {et~al.}(2019)\citenamefont {Berdyugin}, \citenamefont {Xu}, \citenamefont {Pellegrino}, \citenamefont {Krishna~Kumar}, \citenamefont {Principi}, \citenamefont {Torre}, \citenamefont {Ben~Shalom}, \citenamefont {Taniguchi}, \citenamefont {Watanabe}, \citenamefont {Grigorieva}, \citenamefont {Polini}, \citenamefont {Geim},\ and\ \citenamefont {Bandurin}}]{Berdyugin-Bandurin:2018}%
  \BibitemOpen
  \bibfield  {author} {\bibinfo {author} {\bibfnamefont {A.~I.}\ \bibnamefont {Berdyugin}}, \bibinfo {author} {\bibfnamefont {S.~G.}\ \bibnamefont {Xu}}, \bibinfo {author} {\bibfnamefont {F.~M.~D.}\ \bibnamefont {Pellegrino}}, \bibinfo {author} {\bibfnamefont {R.}~\bibnamefont {Krishna~Kumar}}, \bibinfo {author} {\bibfnamefont {A.}~\bibnamefont {Principi}}, \bibinfo {author} {\bibfnamefont {I.}~\bibnamefont {Torre}}, \bibinfo {author} {\bibfnamefont {M.}~\bibnamefont {Ben~Shalom}}, \bibinfo {author} {\bibfnamefont {T.}~\bibnamefont {Taniguchi}}, \bibinfo {author} {\bibfnamefont {K.}~\bibnamefont {Watanabe}}, \bibinfo {author} {\bibfnamefont {I.~V.}\ \bibnamefont {Grigorieva}}, \bibinfo {author} {\bibfnamefont {M.}~\bibnamefont {Polini}}, \bibinfo {author} {\bibfnamefont {A.~K.}\ \bibnamefont {Geim}},\ and\ \bibinfo {author} {\bibfnamefont {D.~A.}\ \bibnamefont {Bandurin}},\ }\bibfield  {title} {\bibinfo {title} {Measuring {{Hall}} viscosity of graphene's electron fluid},\ }\href
  {https://doi.org/10.1126/science.aau0685} {\bibfield  {journal} {\bibinfo  {journal} {Science}\ }\textbf {\bibinfo {volume} {364}},\ \bibinfo {pages} {162} (\bibinfo {year} {2019})},\ \Eprint {https://arxiv.org/abs/1806.01606} {arXiv:1806.01606} \BibitemShut {NoStop}%
\bibitem [{\citenamefont {Bandurin}\ \emph {et~al.}(2018)\citenamefont {Bandurin}, \citenamefont {Shytov}, \citenamefont {Levitov}, \citenamefont {Kumar}, \citenamefont {Berdyugin}, \citenamefont {Ben~Shalom}, \citenamefont {Grigorieva}, \citenamefont {Geim},\ and\ \citenamefont {Falkovich}}]{Bandurin-Falkovich:2018}%
  \BibitemOpen
  \bibfield  {author} {\bibinfo {author} {\bibfnamefont {D.~A.}\ \bibnamefont {Bandurin}}, \bibinfo {author} {\bibfnamefont {A.~V.}\ \bibnamefont {Shytov}}, \bibinfo {author} {\bibfnamefont {L.~S.}\ \bibnamefont {Levitov}}, \bibinfo {author} {\bibfnamefont {R.~K.}\ \bibnamefont {Kumar}}, \bibinfo {author} {\bibfnamefont {A.~I.}\ \bibnamefont {Berdyugin}}, \bibinfo {author} {\bibfnamefont {M.}~\bibnamefont {Ben~Shalom}}, \bibinfo {author} {\bibfnamefont {I.~V.}\ \bibnamefont {Grigorieva}}, \bibinfo {author} {\bibfnamefont {A.~K.}\ \bibnamefont {Geim}},\ and\ \bibinfo {author} {\bibfnamefont {G.}~\bibnamefont {Falkovich}},\ }\bibfield  {title} {\bibinfo {title} {Fluidity onset in graphene},\ }\href {https://doi.org/10.1038/s41467-018-07004-4} {\bibfield  {journal} {\bibinfo  {journal} {Nat. Commun.}\ }\textbf {\bibinfo {volume} {9}},\ \bibinfo {pages} {4533} (\bibinfo {year} {2018})},\ \Eprint {https://arxiv.org/abs/1806.03231} {arXiv:1806.03231} \BibitemShut {NoStop}%
\bibitem [{\citenamefont {Ku}\ \emph {et~al.}(2020)\citenamefont {Ku}, \citenamefont {Zhou}, \citenamefont {Li}, \citenamefont {Shin}, \citenamefont {Shi}, \citenamefont {Burch}, \citenamefont {Anderson}, \citenamefont {Pierce}, \citenamefont {Xie}, \citenamefont {Hamo}, \citenamefont {Vool}, \citenamefont {Zhang}, \citenamefont {Casola}, \citenamefont {Taniguchi}, \citenamefont {Watanabe}, \citenamefont {Fogler}, \citenamefont {Kim}, \citenamefont {Yacoby},\ and\ \citenamefont {Walsworth}}]{Ku-Walsworth:2019}%
  \BibitemOpen
  \bibfield  {author} {\bibinfo {author} {\bibfnamefont {M.~J.~H.}\ \bibnamefont {Ku}}, \bibinfo {author} {\bibfnamefont {T.~X.}\ \bibnamefont {Zhou}}, \bibinfo {author} {\bibfnamefont {Q.}~\bibnamefont {Li}}, \bibinfo {author} {\bibfnamefont {Y.~J.}\ \bibnamefont {Shin}}, \bibinfo {author} {\bibfnamefont {J.~K.}\ \bibnamefont {Shi}}, \bibinfo {author} {\bibfnamefont {C.}~\bibnamefont {Burch}}, \bibinfo {author} {\bibfnamefont {L.~E.}\ \bibnamefont {Anderson}}, \bibinfo {author} {\bibfnamefont {A.~T.}\ \bibnamefont {Pierce}}, \bibinfo {author} {\bibfnamefont {Y.}~\bibnamefont {Xie}}, \bibinfo {author} {\bibfnamefont {A.}~\bibnamefont {Hamo}}, \bibinfo {author} {\bibfnamefont {U.}~\bibnamefont {Vool}}, \bibinfo {author} {\bibfnamefont {H.}~\bibnamefont {Zhang}}, \bibinfo {author} {\bibfnamefont {F.}~\bibnamefont {Casola}}, \bibinfo {author} {\bibfnamefont {T.}~\bibnamefont {Taniguchi}}, \bibinfo {author} {\bibfnamefont {K.}~\bibnamefont {Watanabe}}, \bibinfo {author} {\bibfnamefont {M.~M.}\ \bibnamefont
  {Fogler}}, \bibinfo {author} {\bibfnamefont {P.}~\bibnamefont {Kim}}, \bibinfo {author} {\bibfnamefont {A.}~\bibnamefont {Yacoby}},\ and\ \bibinfo {author} {\bibfnamefont {R.~L.}\ \bibnamefont {Walsworth}},\ }\bibfield  {title} {\bibinfo {title} {Imaging viscous flow of the {{Dirac}} fluid in graphene},\ }\href {https://doi.org/10.1038/s41586-020-2507-2} {\bibfield  {journal} {\bibinfo  {journal} {Nature}\ }\textbf {\bibinfo {volume} {583}},\ \bibinfo {pages} {537} (\bibinfo {year} {2020})},\ \Eprint {https://arxiv.org/abs/1905.10791} {arXiv:1905.10791} \BibitemShut {NoStop}%
\bibitem [{\citenamefont {Sulpizio}\ \emph {et~al.}(2019)\citenamefont {Sulpizio}, \citenamefont {Ella}, \citenamefont {Rozen}, \citenamefont {Birkbeck}, \citenamefont {Perello}, \citenamefont {Dutta}, \citenamefont {{Ben-Shalom}}, \citenamefont {Taniguchi}, \citenamefont {Watanabe}, \citenamefont {Holder}, \citenamefont {Queiroz}, \citenamefont {Principi}, \citenamefont {Stern}, \citenamefont {Scaffidi}, \citenamefont {Geim},\ and\ \citenamefont {Ilani}}]{Sulpizio-Ilani:2019}%
  \BibitemOpen
  \bibfield  {author} {\bibinfo {author} {\bibfnamefont {J.~A.}\ \bibnamefont {Sulpizio}}, \bibinfo {author} {\bibfnamefont {L.}~\bibnamefont {Ella}}, \bibinfo {author} {\bibfnamefont {A.}~\bibnamefont {Rozen}}, \bibinfo {author} {\bibfnamefont {J.}~\bibnamefont {Birkbeck}}, \bibinfo {author} {\bibfnamefont {D.~J.}\ \bibnamefont {Perello}}, \bibinfo {author} {\bibfnamefont {D.}~\bibnamefont {Dutta}}, \bibinfo {author} {\bibfnamefont {M.}~\bibnamefont {{Ben-Shalom}}}, \bibinfo {author} {\bibfnamefont {T.}~\bibnamefont {Taniguchi}}, \bibinfo {author} {\bibfnamefont {K.}~\bibnamefont {Watanabe}}, \bibinfo {author} {\bibfnamefont {T.}~\bibnamefont {Holder}}, \bibinfo {author} {\bibfnamefont {R.}~\bibnamefont {Queiroz}}, \bibinfo {author} {\bibfnamefont {A.}~\bibnamefont {Principi}}, \bibinfo {author} {\bibfnamefont {A.}~\bibnamefont {Stern}}, \bibinfo {author} {\bibfnamefont {T.}~\bibnamefont {Scaffidi}}, \bibinfo {author} {\bibfnamefont {A.~K.}\ \bibnamefont {Geim}},\ and\ \bibinfo {author} {\bibfnamefont
  {S.}~\bibnamefont {Ilani}},\ }\bibfield  {title} {\bibinfo {title} {Visualizing {{Poiseuille}} flow of hydrodynamic electrons},\ }\href {https://doi.org/10.1038/s41586-019-1788-9} {\bibfield  {journal} {\bibinfo  {journal} {Nature}\ }\textbf {\bibinfo {volume} {576}},\ \bibinfo {pages} {75} (\bibinfo {year} {2019})},\ \Eprint {https://arxiv.org/abs/1905.11662} {arXiv:1905.11662} \BibitemShut {NoStop}%
\bibitem [{\citenamefont {Samaddar}\ \emph {et~al.}(2021)\citenamefont {Samaddar}, \citenamefont {Strasdas}, \citenamefont {Jan{\ss}en}, \citenamefont {Just}, \citenamefont {Johnsen}, \citenamefont {Wang}, \citenamefont {Uzlu}, \citenamefont {Li}, \citenamefont {Neumaier}, \citenamefont {Liebmann},\ and\ \citenamefont {Morgenstern}}]{Samaddar-Morgenstern:2021}%
  \BibitemOpen
  \bibfield  {author} {\bibinfo {author} {\bibfnamefont {S.}~\bibnamefont {Samaddar}}, \bibinfo {author} {\bibfnamefont {J.}~\bibnamefont {Strasdas}}, \bibinfo {author} {\bibfnamefont {K.}~\bibnamefont {Jan{\ss}en}}, \bibinfo {author} {\bibfnamefont {S.}~\bibnamefont {Just}}, \bibinfo {author} {\bibfnamefont {T.}~\bibnamefont {Johnsen}}, \bibinfo {author} {\bibfnamefont {Z.}~\bibnamefont {Wang}}, \bibinfo {author} {\bibfnamefont {B.}~\bibnamefont {Uzlu}}, \bibinfo {author} {\bibfnamefont {S.}~\bibnamefont {Li}}, \bibinfo {author} {\bibfnamefont {D.}~\bibnamefont {Neumaier}}, \bibinfo {author} {\bibfnamefont {M.}~\bibnamefont {Liebmann}},\ and\ \bibinfo {author} {\bibfnamefont {M.}~\bibnamefont {Morgenstern}},\ }\bibfield  {title} {\bibinfo {title} {Evidence for {{Local Spots}} of {{Viscous Electron Flow}} in {{Graphene}} at {{Moderate Mobility}}},\ }\href {https://doi.org/10.1021/acs.nanolett.1c01145} {\bibfield  {journal} {\bibinfo  {journal} {Nano Lett.}\ }\textbf {\bibinfo {volume} {21}},\ \bibinfo {pages}
  {9365} (\bibinfo {year} {2021})},\ \Eprint {https://arxiv.org/abs/2103.11466} {arXiv:2103.11466} \BibitemShut {NoStop}%
\bibitem [{\citenamefont {Kumar}\ \emph {et~al.}(2022)\citenamefont {Kumar}, \citenamefont {Birkbeck}, \citenamefont {Sulpizio}, \citenamefont {Perello}, \citenamefont {Taniguchi}, \citenamefont {Watanabe}, \citenamefont {Reuven}, \citenamefont {Scaffidi}, \citenamefont {Stern}, \citenamefont {Geim},\ and\ \citenamefont {Ilani}}]{Kumar-Ilani:2021}%
  \BibitemOpen
  \bibfield  {author} {\bibinfo {author} {\bibfnamefont {C.}~\bibnamefont {Kumar}}, \bibinfo {author} {\bibfnamefont {J.}~\bibnamefont {Birkbeck}}, \bibinfo {author} {\bibfnamefont {J.~A.}\ \bibnamefont {Sulpizio}}, \bibinfo {author} {\bibfnamefont {D.}~\bibnamefont {Perello}}, \bibinfo {author} {\bibfnamefont {T.}~\bibnamefont {Taniguchi}}, \bibinfo {author} {\bibfnamefont {K.}~\bibnamefont {Watanabe}}, \bibinfo {author} {\bibfnamefont {O.}~\bibnamefont {Reuven}}, \bibinfo {author} {\bibfnamefont {T.}~\bibnamefont {Scaffidi}}, \bibinfo {author} {\bibfnamefont {A.}~\bibnamefont {Stern}}, \bibinfo {author} {\bibfnamefont {A.~K.}\ \bibnamefont {Geim}},\ and\ \bibinfo {author} {\bibfnamefont {S.}~\bibnamefont {Ilani}},\ }\bibfield  {title} {\bibinfo {title} {Imaging hydrodynamic electrons flowing without {{Landauer}}--{{Sharvin}} resistance},\ }\href {https://doi.org/10.1038/s41586-022-05002-7} {\bibfield  {journal} {\bibinfo  {journal} {Nature}\ }\textbf {\bibinfo {volume} {609}},\ \bibinfo {pages} {276}
  (\bibinfo {year} {2022})},\ \Eprint {https://arxiv.org/abs/2111.06412} {arXiv:2111.06412} \BibitemShut {NoStop}%
\bibitem [{\citenamefont {Jenkins}\ \emph {et~al.}(2022)\citenamefont {Jenkins}, \citenamefont {Baumann}, \citenamefont {Zhou}, \citenamefont {Meynell}, \citenamefont {Daipeng}, \citenamefont {Watanabe}, \citenamefont {Taniguchi}, \citenamefont {Lucas}, \citenamefont {Young},\ and\ \citenamefont {Bleszynski~Jayich}}]{Jenkins-BleszynskiJayich-ImagingBreakdownOhmic-2022}%
  \BibitemOpen
  \bibfield  {author} {\bibinfo {author} {\bibfnamefont {A.}~\bibnamefont {Jenkins}}, \bibinfo {author} {\bibfnamefont {S.}~\bibnamefont {Baumann}}, \bibinfo {author} {\bibfnamefont {H.}~\bibnamefont {Zhou}}, \bibinfo {author} {\bibfnamefont {S.~A.}\ \bibnamefont {Meynell}}, \bibinfo {author} {\bibfnamefont {Y.}~\bibnamefont {Daipeng}}, \bibinfo {author} {\bibfnamefont {K.}~\bibnamefont {Watanabe}}, \bibinfo {author} {\bibfnamefont {T.}~\bibnamefont {Taniguchi}}, \bibinfo {author} {\bibfnamefont {A.}~\bibnamefont {Lucas}}, \bibinfo {author} {\bibfnamefont {A.~F.}\ \bibnamefont {Young}},\ and\ \bibinfo {author} {\bibfnamefont {A.~C.}\ \bibnamefont {Bleszynski~Jayich}},\ }\bibfield  {title} {\bibinfo {title} {Imaging the {{Breakdown}} of {{Ohmic Transport}} in {{Graphene}}},\ }\href {https://doi.org/10.1103/PhysRevLett.129.087701} {\bibfield  {journal} {\bibinfo  {journal} {Phys. Rev. Lett.}\ }\textbf {\bibinfo {volume} {129}},\ \bibinfo {pages} {087701} (\bibinfo {year} {2022})},\ \Eprint
  {https://arxiv.org/abs/2002.05065} {arXiv:2002.05065 [cond-mat.mes-hall]} \BibitemShut {NoStop}%
\bibitem [{\citenamefont {Krebs}\ \emph {et~al.}(2023)\citenamefont {Krebs}, \citenamefont {Behn}, \citenamefont {Li}, \citenamefont {Smith}, \citenamefont {Watanabe}, \citenamefont {Taniguchi}, \citenamefont {Levchenko},\ and\ \citenamefont {Brar}}]{Krebs-Brar-ImagingBreakingElectrostatic-2023}%
  \BibitemOpen
  \bibfield  {author} {\bibinfo {author} {\bibfnamefont {Z.~J.}\ \bibnamefont {Krebs}}, \bibinfo {author} {\bibfnamefont {W.~A.}\ \bibnamefont {Behn}}, \bibinfo {author} {\bibfnamefont {S.}~\bibnamefont {Li}}, \bibinfo {author} {\bibfnamefont {K.~J.}\ \bibnamefont {Smith}}, \bibinfo {author} {\bibfnamefont {K.}~\bibnamefont {Watanabe}}, \bibinfo {author} {\bibfnamefont {T.}~\bibnamefont {Taniguchi}}, \bibinfo {author} {\bibfnamefont {A.}~\bibnamefont {Levchenko}},\ and\ \bibinfo {author} {\bibfnamefont {V.~W.}\ \bibnamefont {Brar}},\ }\bibfield  {title} {\bibinfo {title} {Imaging the breaking of electrostatic dams in graphene for ballistic and viscous fluids},\ }\href {https://doi.org/10.1126/science.abm6073} {\bibfield  {journal} {\bibinfo  {journal} {Science}\ }\textbf {\bibinfo {volume} {379}},\ \bibinfo {pages} {671} (\bibinfo {year} {2023})},\ \Eprint {https://arxiv.org/abs/2106.07212} {arXiv:2106.07212} \BibitemShut {NoStop}%
\bibitem [{\citenamefont {Gooth}\ \emph {et~al.}(2018)\citenamefont {Gooth}, \citenamefont {Menges}, \citenamefont {Kumar}, \citenamefont {S{\"u}{$\beta$}}, \citenamefont {Shekhar}, \citenamefont {Sun}, \citenamefont {Drechsler}, \citenamefont {Zierold}, \citenamefont {Felser},\ and\ \citenamefont {Gotsmann}}]{Gooth-Felser:2018}%
  \BibitemOpen
  \bibfield  {author} {\bibinfo {author} {\bibfnamefont {J.}~\bibnamefont {Gooth}}, \bibinfo {author} {\bibfnamefont {F.}~\bibnamefont {Menges}}, \bibinfo {author} {\bibfnamefont {N.}~\bibnamefont {Kumar}}, \bibinfo {author} {\bibfnamefont {V.}~\bibnamefont {S{\"u}{$\beta$}}}, \bibinfo {author} {\bibfnamefont {C.}~\bibnamefont {Shekhar}}, \bibinfo {author} {\bibfnamefont {Y.}~\bibnamefont {Sun}}, \bibinfo {author} {\bibfnamefont {U.}~\bibnamefont {Drechsler}}, \bibinfo {author} {\bibfnamefont {R.}~\bibnamefont {Zierold}}, \bibinfo {author} {\bibfnamefont {C.}~\bibnamefont {Felser}},\ and\ \bibinfo {author} {\bibfnamefont {B.}~\bibnamefont {Gotsmann}},\ }\bibfield  {title} {\bibinfo {title} {Thermal and electrical signatures of a hydrodynamic electron fluid in tungsten diphosphide},\ }\href {https://doi.org/10.1038/s41467-018-06688-y} {\bibfield  {journal} {\bibinfo  {journal} {Nat. Commun.}\ }\textbf {\bibinfo {volume} {9}},\ \bibinfo {pages} {4093} (\bibinfo {year} {2018})},\ \Eprint
  {https://arxiv.org/abs/1706.05925} {arXiv:1706.05925} \BibitemShut {NoStop}%
\bibitem [{\citenamefont {Jaoui}\ \emph {et~al.}(2018)\citenamefont {Jaoui}, \citenamefont {Fauqu{\'e}}, \citenamefont {Rischau}, \citenamefont {Subedi}, \citenamefont {Fu}, \citenamefont {Gooth}, \citenamefont {Kumar}, \citenamefont {S{\"u}{\ss}}, \citenamefont {Maslov}, \citenamefont {Felser},\ and\ \citenamefont {Behnia}}]{Jaoui-Behnia-DepartureWiedemannFranz-2018}%
  \BibitemOpen
  \bibfield  {author} {\bibinfo {author} {\bibfnamefont {A.}~\bibnamefont {Jaoui}}, \bibinfo {author} {\bibfnamefont {B.}~\bibnamefont {Fauqu{\'e}}}, \bibinfo {author} {\bibfnamefont {C.~W.}\ \bibnamefont {Rischau}}, \bibinfo {author} {\bibfnamefont {A.}~\bibnamefont {Subedi}}, \bibinfo {author} {\bibfnamefont {C.}~\bibnamefont {Fu}}, \bibinfo {author} {\bibfnamefont {J.}~\bibnamefont {Gooth}}, \bibinfo {author} {\bibfnamefont {N.}~\bibnamefont {Kumar}}, \bibinfo {author} {\bibfnamefont {V.}~\bibnamefont {S{\"u}{\ss}}}, \bibinfo {author} {\bibfnamefont {D.~L.}\ \bibnamefont {Maslov}}, \bibinfo {author} {\bibfnamefont {C.}~\bibnamefont {Felser}},\ and\ \bibinfo {author} {\bibfnamefont {K.}~\bibnamefont {Behnia}},\ }\bibfield  {title} {\bibinfo {title} {Departure from the {{Wiedemann}}--{{Franz}} law in {{WP}}{$_2$} driven by mismatch in {{T-square}} resistivity prefactors},\ }\href {https://doi.org/10.1038/s41535-018-0136-x} {\bibfield  {journal} {\bibinfo  {journal} {npj Quant Mater}\ }\textbf {\bibinfo
  {volume} {3}},\ \bibinfo {pages} {64} (\bibinfo {year} {2018})},\ \Eprint {https://arxiv.org/abs/1806.04094} {arXiv:1806.04094} \BibitemShut {NoStop}%
\bibitem [{\citenamefont {Link}\ \emph {et~al.}(2018)\citenamefont {Link}, \citenamefont {Narozhny}, \citenamefont {Kiselev},\ and\ \citenamefont {Schmalian}}]{Link-Schmalian:2017}%
  \BibitemOpen
  \bibfield  {author} {\bibinfo {author} {\bibfnamefont {J.~M.}\ \bibnamefont {Link}}, \bibinfo {author} {\bibfnamefont {B.~N.}\ \bibnamefont {Narozhny}}, \bibinfo {author} {\bibfnamefont {E.~I.}\ \bibnamefont {Kiselev}},\ and\ \bibinfo {author} {\bibfnamefont {J.}~\bibnamefont {Schmalian}},\ }\bibfield  {title} {\bibinfo {title} {Out-of-{{Bounds Hydrodynamics}} in {{Anisotropic Dirac Fluids}}},\ }\href {https://doi.org/10.1103/PhysRevLett.120.196801} {\bibfield  {journal} {\bibinfo  {journal} {Phys. Rev. Lett.}\ }\textbf {\bibinfo {volume} {120}},\ \bibinfo {pages} {196801} (\bibinfo {year} {2018})},\ \Eprint {https://arxiv.org/abs/1708.02759} {arXiv:1708.02759} \BibitemShut {NoStop}%
\bibitem [{\citenamefont {Rao}\ and\ \citenamefont {Bradlyn}(2020)}]{Rao-Bradlyn:2019}%
  \BibitemOpen
  \bibfield  {author} {\bibinfo {author} {\bibfnamefont {P.}~\bibnamefont {Rao}}\ and\ \bibinfo {author} {\bibfnamefont {B.}~\bibnamefont {Bradlyn}},\ }\bibfield  {title} {\bibinfo {title} {Hall {{Viscosity}} in {{Quantum Systems}} with {{Discrete Symmetry}}: {{Point Group}} and {{Lattice Anisotropy}}},\ }\href {https://doi.org/10.1103/PhysRevX.10.021005} {\bibfield  {journal} {\bibinfo  {journal} {Phys. Rev. X}\ }\textbf {\bibinfo {volume} {10}},\ \bibinfo {pages} {021005} (\bibinfo {year} {2020})},\ \Eprint {https://arxiv.org/abs/1910.10727} {arXiv:1910.10727} \BibitemShut {NoStop}%
\bibitem [{\citenamefont {Offertaler}\ and\ \citenamefont {Bradlyn}(2019)}]{Offertaler-Bradlyn:2019}%
  \BibitemOpen
  \bibfield  {author} {\bibinfo {author} {\bibfnamefont {B.}~\bibnamefont {Offertaler}}\ and\ \bibinfo {author} {\bibfnamefont {B.}~\bibnamefont {Bradlyn}},\ }\bibfield  {title} {\bibinfo {title} {Viscoelastic response of quantum {{Hall}} fluids in a tilted field},\ }\href {https://doi.org/10.1103/PhysRevB.99.035427} {\bibfield  {journal} {\bibinfo  {journal} {Phys. Rev. B}\ }\textbf {\bibinfo {volume} {99}},\ \bibinfo {pages} {035427} (\bibinfo {year} {2019})},\ \Eprint {https://arxiv.org/abs/1811.08443} {arXiv:1811.08443} \BibitemShut {NoStop}%
\bibitem [{\citenamefont {Cook}\ and\ \citenamefont {Lucas}(2019)}]{Cook-Lucas:2019}%
  \BibitemOpen
  \bibfield  {author} {\bibinfo {author} {\bibfnamefont {C.~Q.}\ \bibnamefont {Cook}}\ and\ \bibinfo {author} {\bibfnamefont {A.}~\bibnamefont {Lucas}},\ }\bibfield  {title} {\bibinfo {title} {Electron hydrodynamics with a polygonal {{Fermi}} surface},\ }\href {https://doi.org/10.1103/PhysRevB.99.235148} {\bibfield  {journal} {\bibinfo  {journal} {Phys. Rev. B}\ }\textbf {\bibinfo {volume} {99}},\ \bibinfo {pages} {235148} (\bibinfo {year} {2019})},\ \Eprint {https://arxiv.org/abs/1903.05652} {arXiv:1903.05652} \BibitemShut {NoStop}%
\bibitem [{\citenamefont {{Pe{\~n}a-Benitez}}\ \emph {et~al.}(2019)\citenamefont {{Pe{\~n}a-Benitez}}, \citenamefont {Saha},\ and\ \citenamefont {Sur{\'o}wka}}]{Pena-Benitez-Surowka:2019}%
  \BibitemOpen
  \bibfield  {author} {\bibinfo {author} {\bibfnamefont {F.}~\bibnamefont {{Pe{\~n}a-Benitez}}}, \bibinfo {author} {\bibfnamefont {K.}~\bibnamefont {Saha}},\ and\ \bibinfo {author} {\bibfnamefont {P.}~\bibnamefont {Sur{\'o}wka}},\ }\bibfield  {title} {\bibinfo {title} {Berry curvature and {{Hall}} viscosities in an anisotropic {{Dirac}} semimetal},\ }\href {https://doi.org/10.1103/PhysRevB.99.045141} {\bibfield  {journal} {\bibinfo  {journal} {Phys. Rev. B}\ }\textbf {\bibinfo {volume} {99}},\ \bibinfo {pages} {045141} (\bibinfo {year} {2019})},\ \Eprint {https://arxiv.org/abs/1805.09827} {arXiv:1805.09827} \BibitemShut {NoStop}%
\bibitem [{\citenamefont {Varnavides}\ \emph {et~al.}(2020)\citenamefont {Varnavides}, \citenamefont {Jermyn}, \citenamefont {Anikeeva}, \citenamefont {Felser},\ and\ \citenamefont {Narang}}]{Varnavides-Narang:2020}%
  \BibitemOpen
  \bibfield  {author} {\bibinfo {author} {\bibfnamefont {G.}~\bibnamefont {Varnavides}}, \bibinfo {author} {\bibfnamefont {A.~S.}\ \bibnamefont {Jermyn}}, \bibinfo {author} {\bibfnamefont {P.}~\bibnamefont {Anikeeva}}, \bibinfo {author} {\bibfnamefont {C.}~\bibnamefont {Felser}},\ and\ \bibinfo {author} {\bibfnamefont {P.}~\bibnamefont {Narang}},\ }\bibfield  {title} {\bibinfo {title} {Electron hydrodynamics in anisotropic materials},\ }\href {https://doi.org/10.1038/s41467-020-18553-y} {\bibfield  {journal} {\bibinfo  {journal} {Nat. Commun.}\ }\textbf {\bibinfo {volume} {11}},\ \bibinfo {pages} {4710} (\bibinfo {year} {2020})},\ \Eprint {https://arxiv.org/abs/2002.08976} {arXiv:2002.08976} \BibitemShut {NoStop}%
\bibitem [{\citenamefont {Robredo}\ \emph {et~al.}(2021)\citenamefont {Robredo}, \citenamefont {Rao}, \citenamefont {{de Juan}}, \citenamefont {Bergara}, \citenamefont {Ma{\~n}es}, \citenamefont {Cortijo}, \citenamefont {Vergniory},\ and\ \citenamefont {Bradlyn}}]{Robredo-Bradlyn:2021}%
  \BibitemOpen
  \bibfield  {author} {\bibinfo {author} {\bibfnamefont {I.}~\bibnamefont {Robredo}}, \bibinfo {author} {\bibfnamefont {P.}~\bibnamefont {Rao}}, \bibinfo {author} {\bibfnamefont {F.}~\bibnamefont {{de Juan}}}, \bibinfo {author} {\bibfnamefont {A.}~\bibnamefont {Bergara}}, \bibinfo {author} {\bibfnamefont {J.~L.}\ \bibnamefont {Ma{\~n}es}}, \bibinfo {author} {\bibfnamefont {A.}~\bibnamefont {Cortijo}}, \bibinfo {author} {\bibfnamefont {M.~G.}\ \bibnamefont {Vergniory}},\ and\ \bibinfo {author} {\bibfnamefont {B.}~\bibnamefont {Bradlyn}},\ }\bibfield  {title} {\bibinfo {title} {Cubic {{Hall}} viscosity in three-dimensional topological semimetals},\ }\href {https://doi.org/10.1103/PhysRevResearch.3.L032068} {\bibfield  {journal} {\bibinfo  {journal} {Phys. Rev. Res.}\ }\textbf {\bibinfo {volume} {3}},\ \bibinfo {pages} {L032068} (\bibinfo {year} {2021})},\ \Eprint {https://arxiv.org/abs/2102.02226} {arXiv:2102.02226} \BibitemShut {NoStop}%
\bibitem [{\citenamefont {Huang}\ and\ \citenamefont {Lucas}(2022)}]{Huang-Lucas:2022}%
  \BibitemOpen
  \bibfield  {author} {\bibinfo {author} {\bibfnamefont {X.}~\bibnamefont {Huang}}\ and\ \bibinfo {author} {\bibfnamefont {A.}~\bibnamefont {Lucas}},\ }\bibfield  {title} {\bibinfo {title} {Hydrodynamic effective field theories with discrete rotational symmetry},\ }\href {https://doi.org/10.1007/JHEP03(2022)082} {\bibfield  {journal} {\bibinfo  {journal} {J. High Energy Phys.}\ }\textbf {\bibinfo {volume} {2022}}\bibfield  {number} {\bibinfo  {number} { (3)},\ \bibinfo {pages} {82}},\ }\Eprint {https://arxiv.org/abs/2201.03565} {arXiv:2201.03565} \BibitemShut {NoStop}%
\bibitem [{\citenamefont {Herasymchuk}\ \emph {et~al.}(2024)\citenamefont {Herasymchuk}, \citenamefont {Gorbar},\ and\ \citenamefont {Sukhachov}}]{Herasymchuk-Sukhachov-ViscoelasticResponseAnisotropic-2024}%
  \BibitemOpen
  \bibfield  {author} {\bibinfo {author} {\bibfnamefont {A.~A.}\ \bibnamefont {Herasymchuk}}, \bibinfo {author} {\bibfnamefont {E.~V.}\ \bibnamefont {Gorbar}},\ and\ \bibinfo {author} {\bibfnamefont {P.~O.}\ \bibnamefont {Sukhachov}},\ }\bibfield  {title} {\bibinfo {title} {Viscoelastic response and anisotropic hydrodynamics in {{Weyl}} semimetals},\ }\href {https://doi.org/10.1103/PhysRevB.110.035133} {\bibfield  {journal} {\bibinfo  {journal} {Phys. Rev. B}\ }\textbf {\bibinfo {volume} {110}},\ \bibinfo {pages} {035133} (\bibinfo {year} {2024})},\ \Eprint {https://arxiv.org/abs/2402.19304} {arXiv:2402.19304 [cond-mat.mes-hall]} \BibitemShut {NoStop}%
\bibitem [{\citenamefont {{\v S}mejkal}\ \emph {et~al.}(2020)\citenamefont {{\v S}mejkal}, \citenamefont {{Gonz{\'a}lez-Hern{\'a}ndez}}, \citenamefont {Jungwirth},\ and\ \citenamefont {Sinova}}]{Smejkal-Sinova:2020}%
  \BibitemOpen
  \bibfield  {author} {\bibinfo {author} {\bibfnamefont {L.}~\bibnamefont {{\v S}mejkal}}, \bibinfo {author} {\bibfnamefont {R.}~\bibnamefont {{Gonz{\'a}lez-Hern{\'a}ndez}}}, \bibinfo {author} {\bibfnamefont {T.}~\bibnamefont {Jungwirth}},\ and\ \bibinfo {author} {\bibfnamefont {J.}~\bibnamefont {Sinova}},\ }\bibfield  {title} {\bibinfo {title} {Crystal time-reversal symmetry breaking and spontaneous {{Hall}} effect in collinear antiferromagnets},\ }\href {https://doi.org/10.1126/sciadv.aaz8809} {\bibfield  {journal} {\bibinfo  {journal} {Sci. Adv.}\ }\textbf {\bibinfo {volume} {6}},\ \bibinfo {pages} {eaaz8809} (\bibinfo {year} {2020})},\ \Eprint {https://arxiv.org/abs/1901.00445} {arXiv:1901.00445} \BibitemShut {NoStop}%
\bibitem [{\citenamefont {{\v S}mejkal}\ \emph {et~al.}(2022{\natexlab{a}})\citenamefont {{\v S}mejkal}, \citenamefont {Sinova},\ and\ \citenamefont {Jungwirth}}]{Smejkal-Jungwirth:2022}%
  \BibitemOpen
  \bibfield  {author} {\bibinfo {author} {\bibfnamefont {L.}~\bibnamefont {{\v S}mejkal}}, \bibinfo {author} {\bibfnamefont {J.}~\bibnamefont {Sinova}},\ and\ \bibinfo {author} {\bibfnamefont {T.}~\bibnamefont {Jungwirth}},\ }\bibfield  {title} {\bibinfo {title} {Emerging {{Research Landscape}} of {{Altermagnetism}}},\ }\href {https://doi.org/10.1103/PhysRevX.12.040501} {\bibfield  {journal} {\bibinfo  {journal} {Phys. Rev. X}\ }\textbf {\bibinfo {volume} {12}},\ \bibinfo {pages} {040501} (\bibinfo {year} {2022}{\natexlab{a}})},\ \Eprint {https://arxiv.org/abs/2204.10844} {arXiv:2204.10844} \BibitemShut {NoStop}%
\bibitem [{\citenamefont {{\v S}mejkal}\ \emph {et~al.}(2022{\natexlab{b}})\citenamefont {{\v S}mejkal}, \citenamefont {Sinova},\ and\ \citenamefont {Jungwirth}}]{Smejkal-Jungwirth-ConventionalFerromagnetismAntiferromagnetism-2022}%
  \BibitemOpen
  \bibfield  {author} {\bibinfo {author} {\bibfnamefont {L.}~\bibnamefont {{\v S}mejkal}}, \bibinfo {author} {\bibfnamefont {J.}~\bibnamefont {Sinova}},\ and\ \bibinfo {author} {\bibfnamefont {T.}~\bibnamefont {Jungwirth}},\ }\bibfield  {title} {\bibinfo {title} {Beyond {{Conventional Ferromagnetism}} and {{Antiferromagnetism}}: {{A Phase}} with {{Nonrelativistic Spin}} and {{Crystal Rotation Symmetry}}},\ }\href {https://doi.org/10.1103/PhysRevX.12.031042} {\bibfield  {journal} {\bibinfo  {journal} {Phys. Rev. X}\ }\textbf {\bibinfo {volume} {12}},\ \bibinfo {pages} {031042} (\bibinfo {year} {2022}{\natexlab{b}})},\ \Eprint {https://arxiv.org/abs/2105.05820} {arXiv:2105.05820} \BibitemShut {NoStop}%
\bibitem [{\citenamefont {Noda}\ \emph {et~al.}(2016)\citenamefont {Noda}, \citenamefont {Ohno},\ and\ \citenamefont {Nakamura}}]{Noda-Nakamura-MomentumdependentBandSpin-2016}%
  \BibitemOpen
  \bibfield  {author} {\bibinfo {author} {\bibfnamefont {Y.}~\bibnamefont {Noda}}, \bibinfo {author} {\bibfnamefont {K.}~\bibnamefont {Ohno}},\ and\ \bibinfo {author} {\bibfnamefont {S.}~\bibnamefont {Nakamura}},\ }\bibfield  {title} {\bibinfo {title} {Momentum-dependent band spin splitting in semiconducting {{MnO}}{$_2$} : A density functional calculation},\ }\href {https://doi.org/10.1039/C5CP07806G} {\bibfield  {journal} {\bibinfo  {journal} {Phys. Chem. Chem. Phys.}\ }\textbf {\bibinfo {volume} {18}},\ \bibinfo {pages} {13294} (\bibinfo {year} {2016})}\BibitemShut {NoStop}%
\bibitem [{\citenamefont {Hayami}\ \emph {et~al.}(2019)\citenamefont {Hayami}, \citenamefont {Yanagi},\ and\ \citenamefont {Kusunose}}]{Hayami-Kusunose-MomentumDependentSpinSplitting-2019}%
  \BibitemOpen
  \bibfield  {author} {\bibinfo {author} {\bibfnamefont {S.}~\bibnamefont {Hayami}}, \bibinfo {author} {\bibfnamefont {Y.}~\bibnamefont {Yanagi}},\ and\ \bibinfo {author} {\bibfnamefont {H.}~\bibnamefont {Kusunose}},\ }\bibfield  {title} {\bibinfo {title} {Momentum-{{Dependent Spin Splitting}} by {{Collinear Antiferromagnetic Ordering}}},\ }\href {https://doi.org/10.7566/JPSJ.88.123702} {\bibfield  {journal} {\bibinfo  {journal} {J. Phys. Soc. Jpn.}\ }\textbf {\bibinfo {volume} {88}},\ \bibinfo {pages} {123702} (\bibinfo {year} {2019})},\ \Eprint {https://arxiv.org/abs/1908.08680} {arXiv:1908.08680 [cond-mat]} \BibitemShut {NoStop}%
\bibitem [{\citenamefont {Ahn}\ \emph {et~al.}(2019)\citenamefont {Ahn}, \citenamefont {Hariki}, \citenamefont {Lee},\ and\ \citenamefont {Kune{\v s}}}]{Ahn-Kunes:2019}%
  \BibitemOpen
  \bibfield  {author} {\bibinfo {author} {\bibfnamefont {K.-H.}\ \bibnamefont {Ahn}}, \bibinfo {author} {\bibfnamefont {A.}~\bibnamefont {Hariki}}, \bibinfo {author} {\bibfnamefont {K.-W.}\ \bibnamefont {Lee}},\ and\ \bibinfo {author} {\bibfnamefont {J.}~\bibnamefont {Kune{\v s}}},\ }\bibfield  {title} {\bibinfo {title} {Antiferromagnetism in {{RuO}}{$_2$} as d-wave {{Pomeranchuk}} instability},\ }\href {https://doi.org/10.1103/PhysRevB.99.184432} {\bibfield  {journal} {\bibinfo  {journal} {Phys. Rev. B}\ }\textbf {\bibinfo {volume} {99}},\ \bibinfo {pages} {184432} (\bibinfo {year} {2019})},\ \Eprint {https://arxiv.org/abs/1902.04436} {arXiv:1902.04436} \BibitemShut {NoStop}%
\bibitem [{\citenamefont {Yuan}\ \emph {et~al.}(2020)\citenamefont {Yuan}, \citenamefont {Wang}, \citenamefont {Luo}, \citenamefont {Rashba},\ and\ \citenamefont {Zunger}}]{Yuan-Zunger:2020}%
  \BibitemOpen
  \bibfield  {author} {\bibinfo {author} {\bibfnamefont {L.-D.}\ \bibnamefont {Yuan}}, \bibinfo {author} {\bibfnamefont {Z.}~\bibnamefont {Wang}}, \bibinfo {author} {\bibfnamefont {J.-W.}\ \bibnamefont {Luo}}, \bibinfo {author} {\bibfnamefont {E.~I.}\ \bibnamefont {Rashba}},\ and\ \bibinfo {author} {\bibfnamefont {A.}~\bibnamefont {Zunger}},\ }\bibfield  {title} {\bibinfo {title} {Giant momentum-dependent spin splitting in centrosymmetric low- {$\mathds{Z}$} antiferromagnets},\ }\href {https://doi.org/10.1103/PhysRevB.102.014422} {\bibfield  {journal} {\bibinfo  {journal} {Phys. Rev. B}\ }\textbf {\bibinfo {volume} {102}},\ \bibinfo {pages} {014422} (\bibinfo {year} {2020})},\ \Eprint {https://arxiv.org/abs/1912.12689} {arXiv:1912.12689} \BibitemShut {NoStop}%
\bibitem [{\citenamefont {Yuan}\ \emph {et~al.}(2021)\citenamefont {Yuan}, \citenamefont {Wang}, \citenamefont {Luo},\ and\ \citenamefont {Zunger}}]{Yuan-Zunger-PredictionLowZCollinear-2021}%
  \BibitemOpen
  \bibfield  {author} {\bibinfo {author} {\bibfnamefont {L.-D.}\ \bibnamefont {Yuan}}, \bibinfo {author} {\bibfnamefont {Z.}~\bibnamefont {Wang}}, \bibinfo {author} {\bibfnamefont {J.-W.}\ \bibnamefont {Luo}},\ and\ \bibinfo {author} {\bibfnamefont {A.}~\bibnamefont {Zunger}},\ }\bibfield  {title} {\bibinfo {title} {Prediction of low-{{Z}} collinear and noncollinear antiferromagnetic compounds having momentum-dependent spin splitting even without spin-orbit coupling},\ }\href {https://doi.org/10.1103/PhysRevMaterials.5.014409} {\bibfield  {journal} {\bibinfo  {journal} {Phys. Rev. Materials}\ }\textbf {\bibinfo {volume} {5}},\ \bibinfo {pages} {014409} (\bibinfo {year} {2021})},\ \Eprint {https://arxiv.org/abs/2008.08532} {arXiv:2008.08532 [cond-mat.mtrl-sci]} \BibitemShut {NoStop}%
\bibitem [{\citenamefont {Ma}\ \emph {et~al.}(2021)\citenamefont {Ma}, \citenamefont {Hu}, \citenamefont {Li}, \citenamefont {Liu}, \citenamefont {Yao}, \citenamefont {Jia},\ and\ \citenamefont {Liu}}]{Ma-Liu-MultifunctionalAntiferromagneticMaterials-2021}%
  \BibitemOpen
  \bibfield  {author} {\bibinfo {author} {\bibfnamefont {H.-Y.}\ \bibnamefont {Ma}}, \bibinfo {author} {\bibfnamefont {M.}~\bibnamefont {Hu}}, \bibinfo {author} {\bibfnamefont {N.}~\bibnamefont {Li}}, \bibinfo {author} {\bibfnamefont {J.}~\bibnamefont {Liu}}, \bibinfo {author} {\bibfnamefont {W.}~\bibnamefont {Yao}}, \bibinfo {author} {\bibfnamefont {J.-F.}\ \bibnamefont {Jia}},\ and\ \bibinfo {author} {\bibfnamefont {J.}~\bibnamefont {Liu}},\ }\bibfield  {title} {\bibinfo {title} {Multifunctional antiferromagnetic materials with giant piezomagnetism and noncollinear spin current},\ }\href {https://doi.org/10.1038/s41467-021-23127-7} {\bibfield  {journal} {\bibinfo  {journal} {Nat Commun}\ }\textbf {\bibinfo {volume} {12}},\ \bibinfo {pages} {2846} (\bibinfo {year} {2021})},\ \Eprint {https://arxiv.org/abs/2104.00561} {arXiv:2104.00561 [cond-mat.mtrl-sci]} \BibitemShut {NoStop}%
\bibitem [{\citenamefont {Mazin}(2022)}]{Mazin:2022}%
  \BibitemOpen
  \bibfield  {author} {\bibinfo {author} {\bibfnamefont {I.}~\bibnamefont {Mazin}},\ }\bibfield  {title} {\bibinfo {title} {Editorial: {{Altermagnetism}}---{{A New Punch Line}} of {{Fundamental Magnetism}}},\ }\href {https://doi.org/10.1103/PhysRevX.12.040002} {\bibfield  {journal} {\bibinfo  {journal} {Phys. Rev. X}\ }\textbf {\bibinfo {volume} {12}},\ \bibinfo {pages} {040002} (\bibinfo {year} {2022})}\BibitemShut {NoStop}%
\bibitem [{\citenamefont {{Gonz{\'a}lez-Hern{\'a}ndez}}\ \emph {et~al.}(2021)\citenamefont {{Gonz{\'a}lez-Hern{\'a}ndez}}, \citenamefont {{\v S}mejkal}, \citenamefont {V{\'y}born{\'y}}, \citenamefont {Yahagi}, \citenamefont {Sinova}, \citenamefont {Jungwirth},\ and\ \citenamefont {{\v Z}elezn{\'y}}}]{Gonzalez-Hernandez-Zelezny:2021}%
  \BibitemOpen
  \bibfield  {author} {\bibinfo {author} {\bibfnamefont {R.}~\bibnamefont {{Gonz{\'a}lez-Hern{\'a}ndez}}}, \bibinfo {author} {\bibfnamefont {L.}~\bibnamefont {{\v S}mejkal}}, \bibinfo {author} {\bibfnamefont {K.}~\bibnamefont {V{\'y}born{\'y}}}, \bibinfo {author} {\bibfnamefont {Y.}~\bibnamefont {Yahagi}}, \bibinfo {author} {\bibfnamefont {J.}~\bibnamefont {Sinova}}, \bibinfo {author} {\bibfnamefont {T.}~\bibnamefont {Jungwirth}},\ and\ \bibinfo {author} {\bibfnamefont {J.}~\bibnamefont {{\v Z}elezn{\'y}}},\ }\bibfield  {title} {\bibinfo {title} {Efficient {{Electrical Spin Splitter Based}} on {{Nonrelativistic Collinear Antiferromagnetism}}},\ }\href {https://doi.org/10.1103/PhysRevLett.126.127701} {\bibfield  {journal} {\bibinfo  {journal} {Phys. Rev. Lett.}\ }\textbf {\bibinfo {volume} {126}},\ \bibinfo {pages} {127701} (\bibinfo {year} {2021})},\ \Eprint {https://arxiv.org/abs/2002.07073} {arXiv:2002.07073} \BibitemShut {NoStop}%
\bibitem [{\citenamefont {Karube}\ \emph {et~al.}(2022)\citenamefont {Karube}, \citenamefont {Tanaka}, \citenamefont {Sugawara}, \citenamefont {Kadoguchi}, \citenamefont {Kohda},\ and\ \citenamefont {Nitta}}]{Karube-Nitta:2022}%
  \BibitemOpen
  \bibfield  {author} {\bibinfo {author} {\bibfnamefont {S.}~\bibnamefont {Karube}}, \bibinfo {author} {\bibfnamefont {T.}~\bibnamefont {Tanaka}}, \bibinfo {author} {\bibfnamefont {D.}~\bibnamefont {Sugawara}}, \bibinfo {author} {\bibfnamefont {N.}~\bibnamefont {Kadoguchi}}, \bibinfo {author} {\bibfnamefont {M.}~\bibnamefont {Kohda}},\ and\ \bibinfo {author} {\bibfnamefont {J.}~\bibnamefont {Nitta}},\ }\bibfield  {title} {\bibinfo {title} {Observation of {{Spin-Splitter Torque}} in {{Collinear Antiferromagnetic RuO}}{$_2$}},\ }\href {https://doi.org/10.1103/PhysRevLett.129.137201} {\bibfield  {journal} {\bibinfo  {journal} {Phys. Rev. Lett.}\ }\textbf {\bibinfo {volume} {129}},\ \bibinfo {pages} {137201} (\bibinfo {year} {2022})},\ \Eprint {https://arxiv.org/abs/2111.07487} {arXiv:2111.07487} \BibitemShut {NoStop}%
\bibitem [{\citenamefont {Bai}\ \emph {et~al.}(2022)\citenamefont {Bai}, \citenamefont {Han}, \citenamefont {Feng}, \citenamefont {Zhou}, \citenamefont {Su}, \citenamefont {Wang}, \citenamefont {Liao}, \citenamefont {Zhu}, \citenamefont {Chen}, \citenamefont {Pan}, \citenamefont {Fan},\ and\ \citenamefont {Song}}]{Bai-Song:2022}%
  \BibitemOpen
  \bibfield  {author} {\bibinfo {author} {\bibfnamefont {H.}~\bibnamefont {Bai}}, \bibinfo {author} {\bibfnamefont {L.}~\bibnamefont {Han}}, \bibinfo {author} {\bibfnamefont {X.~Y.}\ \bibnamefont {Feng}}, \bibinfo {author} {\bibfnamefont {Y.~J.}\ \bibnamefont {Zhou}}, \bibinfo {author} {\bibfnamefont {R.~X.}\ \bibnamefont {Su}}, \bibinfo {author} {\bibfnamefont {Q.}~\bibnamefont {Wang}}, \bibinfo {author} {\bibfnamefont {L.~Y.}\ \bibnamefont {Liao}}, \bibinfo {author} {\bibfnamefont {W.~X.}\ \bibnamefont {Zhu}}, \bibinfo {author} {\bibfnamefont {X.~Z.}\ \bibnamefont {Chen}}, \bibinfo {author} {\bibfnamefont {F.}~\bibnamefont {Pan}}, \bibinfo {author} {\bibfnamefont {X.~L.}\ \bibnamefont {Fan}},\ and\ \bibinfo {author} {\bibfnamefont {C.}~\bibnamefont {Song}},\ }\bibfield  {title} {\bibinfo {title} {Observation of {{Spin Splitting Torque}} in a {{Collinear Antiferromagnet RuO}}{$_2$}},\ }\href {https://doi.org/10.1103/PhysRevLett.128.197202} {\bibfield  {journal} {\bibinfo  {journal} {Phys. Rev. Lett.}\
  }\textbf {\bibinfo {volume} {128}},\ \bibinfo {pages} {197202} (\bibinfo {year} {2022})},\ \Eprint {https://arxiv.org/abs/2109.05933} {arXiv:2109.05933} \BibitemShut {NoStop}%
\bibitem [{\citenamefont {Bose}\ \emph {et~al.}(2022)\citenamefont {Bose}, \citenamefont {Schreiber}, \citenamefont {Jain}, \citenamefont {Shao}, \citenamefont {Nair}, \citenamefont {Sun}, \citenamefont {Zhang}, \citenamefont {Muller}, \citenamefont {Tsymbal}, \citenamefont {Schlom},\ and\ \citenamefont {Ralph}}]{Bose-Ralph:2022}%
  \BibitemOpen
  \bibfield  {author} {\bibinfo {author} {\bibfnamefont {A.}~\bibnamefont {Bose}}, \bibinfo {author} {\bibfnamefont {N.~J.}\ \bibnamefont {Schreiber}}, \bibinfo {author} {\bibfnamefont {R.}~\bibnamefont {Jain}}, \bibinfo {author} {\bibfnamefont {D.-F.}\ \bibnamefont {Shao}}, \bibinfo {author} {\bibfnamefont {H.~P.}\ \bibnamefont {Nair}}, \bibinfo {author} {\bibfnamefont {J.}~\bibnamefont {Sun}}, \bibinfo {author} {\bibfnamefont {X.~S.}\ \bibnamefont {Zhang}}, \bibinfo {author} {\bibfnamefont {D.~A.}\ \bibnamefont {Muller}}, \bibinfo {author} {\bibfnamefont {E.~Y.}\ \bibnamefont {Tsymbal}}, \bibinfo {author} {\bibfnamefont {D.~G.}\ \bibnamefont {Schlom}},\ and\ \bibinfo {author} {\bibfnamefont {D.~C.}\ \bibnamefont {Ralph}},\ }\bibfield  {title} {\bibinfo {title} {Tilted spin current generated by the collinear antiferromagnet ruthenium dioxide},\ }\href {https://doi.org/10.1038/s41928-022-00744-8} {\bibfield  {journal} {\bibinfo  {journal} {Nat. Electron.}\ }\textbf {\bibinfo {volume} {5}},\ \bibinfo {pages}
  {267} (\bibinfo {year} {2022})},\ \Eprint {https://arxiv.org/abs/2108.09150} {arXiv:2108.09150} \BibitemShut {NoStop}%
\bibitem [{\citenamefont {{\v S}mejkal}\ \emph {et~al.}(2022{\natexlab{c}})\citenamefont {{\v S}mejkal}, \citenamefont {Hellenes}, \citenamefont {{Gonz{\'a}lez-Hern{\'a}ndez}}, \citenamefont {Sinova},\ and\ \citenamefont {Jungwirth}}]{Smejkal-Jungwirth-GiantTunnelingMagnetoresistance-2022}%
  \BibitemOpen
  \bibfield  {author} {\bibinfo {author} {\bibfnamefont {L.}~\bibnamefont {{\v S}mejkal}}, \bibinfo {author} {\bibfnamefont {A.~B.}\ \bibnamefont {Hellenes}}, \bibinfo {author} {\bibfnamefont {R.}~\bibnamefont {{Gonz{\'a}lez-Hern{\'a}ndez}}}, \bibinfo {author} {\bibfnamefont {J.}~\bibnamefont {Sinova}},\ and\ \bibinfo {author} {\bibfnamefont {T.}~\bibnamefont {Jungwirth}},\ }\bibfield  {title} {\bibinfo {title} {Giant and {{Tunneling Magnetoresistance}} in {{Unconventional Collinear Antiferromagnets}} with {{Nonrelativistic Spin-Momentum Coupling}}},\ }\href {https://doi.org/10.1103/PhysRevX.12.011028} {\bibfield  {journal} {\bibinfo  {journal} {Phys. Rev. X}\ }\textbf {\bibinfo {volume} {12}},\ \bibinfo {pages} {011028} (\bibinfo {year} {2022}{\natexlab{c}})},\ \Eprint {https://arxiv.org/abs/2103.12664} {arXiv:2103.12664 [cond-mat.mes-hall]} \BibitemShut {NoStop}%
\bibitem [{\citenamefont {Samanta}\ \emph {et~al.}(2024)\citenamefont {Samanta}, \citenamefont {Jiang}, \citenamefont {Paudel}, \citenamefont {Shao},\ and\ \citenamefont {Tsymbal}}]{Samanta-Tsymbal-TunnelingMagnetoresistanceMagnetic-2023}%
  \BibitemOpen
  \bibfield  {author} {\bibinfo {author} {\bibfnamefont {K.}~\bibnamefont {Samanta}}, \bibinfo {author} {\bibfnamefont {Y.-Y.}\ \bibnamefont {Jiang}}, \bibinfo {author} {\bibfnamefont {T.~R.}\ \bibnamefont {Paudel}}, \bibinfo {author} {\bibfnamefont {D.-F.}\ \bibnamefont {Shao}},\ and\ \bibinfo {author} {\bibfnamefont {E.~Y.}\ \bibnamefont {Tsymbal}},\ }\bibfield  {title} {\bibinfo {title} {Tunneling magnetoresistance in magnetic tunnel junctions with a single ferromagnetic electrode},\ }\href {https://doi.org/10.1103/PhysRevB.109.174407} {\bibfield  {journal} {\bibinfo  {journal} {Phys. Rev. B}\ }\textbf {\bibinfo {volume} {109}},\ \bibinfo {pages} {174407} (\bibinfo {year} {2024})},\ \Eprint {https://arxiv.org/abs/2310.02139} {arXiv:2310.02139 [cond-mat, physics:physics]} \BibitemShut {NoStop}%
\bibitem [{\citenamefont {Cui}\ \emph {et~al.}(2023)\citenamefont {Cui}, \citenamefont {Zhu}, \citenamefont {Yao}, \citenamefont {Cui},\ and\ \citenamefont {Yang}}]{Cui-Yang-GiantSpinHallTunneling-2023}%
  \BibitemOpen
  \bibfield  {author} {\bibinfo {author} {\bibfnamefont {Q.}~\bibnamefont {Cui}}, \bibinfo {author} {\bibfnamefont {Y.}~\bibnamefont {Zhu}}, \bibinfo {author} {\bibfnamefont {X.}~\bibnamefont {Yao}}, \bibinfo {author} {\bibfnamefont {P.}~\bibnamefont {Cui}},\ and\ \bibinfo {author} {\bibfnamefont {H.}~\bibnamefont {Yang}},\ }\bibfield  {title} {\bibinfo {title} {Giant spin-{{Hall}} and tunneling magnetoresistance effects based on a two-dimensional nonrelativistic antiferromagnetic metal},\ }\href {https://doi.org/10.1103/PhysRevB.108.024410} {\bibfield  {journal} {\bibinfo  {journal} {Phys. Rev. B}\ }\textbf {\bibinfo {volume} {108}},\ \bibinfo {pages} {024410} (\bibinfo {year} {2023})}\BibitemShut {NoStop}%
\bibitem [{\citenamefont {{\=O}ik{\'e}}\ \emph {et~al.}(2024)\citenamefont {{\=O}ik{\'e}}, \citenamefont {Shinada},\ and\ \citenamefont {Peters}}]{Oike-Peters-NonlinearMagnetoelectricEffect-2024}%
  \BibitemOpen
  \bibfield  {author} {\bibinfo {author} {\bibfnamefont {J.}~\bibnamefont {{\=O}ik{\'e}}}, \bibinfo {author} {\bibfnamefont {K.}~\bibnamefont {Shinada}},\ and\ \bibinfo {author} {\bibfnamefont {R.}~\bibnamefont {Peters}},\ }\bibfield  {title} {\bibinfo {title} {Nonlinear magnetoelectric effect under magnetic octupole order: {{Application}} to a d -wave altermagnet and a pyrochlore lattice with all-in/all-out magnetic order},\ }\href {https://doi.org/10.1103/PhysRevB.110.184407} {\bibfield  {journal} {\bibinfo  {journal} {Phys. Rev. B}\ }\textbf {\bibinfo {volume} {110}},\ \bibinfo {pages} {184407} (\bibinfo {year} {2024})}\BibitemShut {NoStop}%
\bibitem [{\citenamefont {{\v S}mejkal}(2024)}]{Smejkal-AltermagneticMultiferroicsAltermagnetoelectric-2024}%
  \BibitemOpen
  \bibfield  {author} {\bibinfo {author} {\bibfnamefont {L.}~\bibnamefont {{\v S}mejkal}},\ }\href {https://doi.org/10.48550/arXiv.2411.19928} {\bibinfo {title} {Altermagnetic multiferroics and altermagnetoelectric effect}} (\bibinfo {year} {2024}),\ \Eprint {https://arxiv.org/abs/2411.19928} {arXiv:2411.19928 [cond-mat]} \BibitemShut {NoStop}%
\bibitem [{\citenamefont {Sun}\ \emph {et~al.}(2024)\citenamefont {Sun}, \citenamefont {Wang}, \citenamefont {Yang}, \citenamefont {Liu}, \citenamefont {Wang}, \citenamefont {Huang},\ and\ \citenamefont {Cheng}}]{Sun-Cheng-RobustMagnetoelectricCoupling-2024}%
  \BibitemOpen
  \bibfield  {author} {\bibinfo {author} {\bibfnamefont {W.}~\bibnamefont {Sun}}, \bibinfo {author} {\bibfnamefont {W.}~\bibnamefont {Wang}}, \bibinfo {author} {\bibfnamefont {C.}~\bibnamefont {Yang}}, \bibinfo {author} {\bibfnamefont {Y.}~\bibnamefont {Liu}}, \bibinfo {author} {\bibfnamefont {X.}~\bibnamefont {Wang}}, \bibinfo {author} {\bibfnamefont {S.}~\bibnamefont {Huang}},\ and\ \bibinfo {author} {\bibfnamefont {Z.}~\bibnamefont {Cheng}},\ }\href {https://doi.org/10.48550/arXiv.2412.05970} {\bibinfo {title} {Robust magnetoelectric coupling in altermagnetic-ferroelectric type-{{III}} multiferroics}} (\bibinfo {year} {2024}),\ \Eprint {https://arxiv.org/abs/2412.05970} {arXiv:2412.05970 [cond-mat]} \BibitemShut {NoStop}%
\bibitem [{\citenamefont {Lin}\ \emph {et~al.}(2024)\citenamefont {Lin}, \citenamefont {Zhang}, \citenamefont {Lu},\ and\ \citenamefont {Xie}}]{Lin-Xie-CoulombDragAltermagnets-2024}%
  \BibitemOpen
  \bibfield  {author} {\bibinfo {author} {\bibfnamefont {H.-J.}\ \bibnamefont {Lin}}, \bibinfo {author} {\bibfnamefont {S.-B.}\ \bibnamefont {Zhang}}, \bibinfo {author} {\bibfnamefont {H.-Z.}\ \bibnamefont {Lu}},\ and\ \bibinfo {author} {\bibfnamefont {X.~C.}\ \bibnamefont {Xie}},\ }\href {https://doi.org/10.48550/arXiv.2412.13927} {\bibinfo {title} {Coulomb {{Drag}} in {{Altermagnets}}}} (\bibinfo {year} {2024}),\ \Eprint {https://arxiv.org/abs/2412.13927} {arXiv:2412.13927 [cond-mat]} \BibitemShut {NoStop}%
\bibitem [{\citenamefont {Sun}\ \emph {et~al.}(2023)\citenamefont {Sun}, \citenamefont {Brataas},\ and\ \citenamefont {Linder}}]{Sun-Linder:2023}%
  \BibitemOpen
  \bibfield  {author} {\bibinfo {author} {\bibfnamefont {C.}~\bibnamefont {Sun}}, \bibinfo {author} {\bibfnamefont {A.}~\bibnamefont {Brataas}},\ and\ \bibinfo {author} {\bibfnamefont {J.}~\bibnamefont {Linder}},\ }\bibfield  {title} {\bibinfo {title} {Andreev reflection in altermagnets},\ }\href {https://doi.org/10.1103/PhysRevB.108.054511} {\bibfield  {journal} {\bibinfo  {journal} {Phys. Rev. B}\ }\textbf {\bibinfo {volume} {108}},\ \bibinfo {pages} {054511} (\bibinfo {year} {2023})},\ \Eprint {https://arxiv.org/abs/2303.14236} {arXiv:2303.14236} \BibitemShut {NoStop}%
\bibitem [{\citenamefont {Papaj}(2023)}]{Papaj:2023}%
  \BibitemOpen
  \bibfield  {author} {\bibinfo {author} {\bibfnamefont {M.}~\bibnamefont {Papaj}},\ }\bibfield  {title} {\bibinfo {title} {Andreev reflection at the altermagnet-superconductor interface},\ }\href {https://doi.org/10.1103/PhysRevB.108.L060508} {\bibfield  {journal} {\bibinfo  {journal} {Phys. Rev. B}\ }\textbf {\bibinfo {volume} {108}},\ \bibinfo {pages} {L060508} (\bibinfo {year} {2023})},\ \Eprint {https://arxiv.org/abs/2305.03856} {arXiv:2305.03856} \BibitemShut {NoStop}%
\bibitem [{\citenamefont {Beenakker}\ and\ \citenamefont {Vakhtel}(2023)}]{Beenakker-Vakhtel:2023}%
  \BibitemOpen
  \bibfield  {author} {\bibinfo {author} {\bibfnamefont {C.~W.~J.}\ \bibnamefont {Beenakker}}\ and\ \bibinfo {author} {\bibfnamefont {T.}~\bibnamefont {Vakhtel}},\ }\bibfield  {title} {\bibinfo {title} {Phase-shifted {{Andreev}} levels in an altermagnet {{Josephson}} junction},\ }\href {https://doi.org/10.1103/PhysRevB.108.075425} {\bibfield  {journal} {\bibinfo  {journal} {Phys. Rev. B}\ }\textbf {\bibinfo {volume} {108}},\ \bibinfo {pages} {075425} (\bibinfo {year} {2023})},\ \Eprint {https://arxiv.org/abs/2306.16300} {arXiv:2306.16300} \BibitemShut {NoStop}%
\bibitem [{\citenamefont {Nagae}\ \emph {et~al.}(2025)\citenamefont {Nagae}, \citenamefont {Schnyder},\ and\ \citenamefont {Ikegaya}}]{Nagae-Ikegaya-SpinpolarizedSpecularAndreev-2024}%
  \BibitemOpen
  \bibfield  {author} {\bibinfo {author} {\bibfnamefont {Y.}~\bibnamefont {Nagae}}, \bibinfo {author} {\bibfnamefont {A.~P.}\ \bibnamefont {Schnyder}},\ and\ \bibinfo {author} {\bibfnamefont {S.}~\bibnamefont {Ikegaya}},\ }\bibfield  {title} {\bibinfo {title} {{Spin-polarized specular Andreev reflections in altermagnets}},\ }\href {https://doi.org/10.1103/PhysRevB.111.L100507} {\bibfield  {journal} {\bibinfo  {journal} {Phys. Rev. B}\ }\textbf {\bibinfo {volume} {111}},\ \bibinfo {pages} {L100507} (\bibinfo {year} {2025})},\ \Eprint {https://arxiv.org/abs/2403.07117} {arXiv:2403.07117} \BibitemShut {NoStop}%
\bibitem [{\citenamefont {Das}\ and\ \citenamefont {Soori}(2024)}]{Das-Soori-CrossedAndreevReflection-2024}%
  \BibitemOpen
  \bibfield  {author} {\bibinfo {author} {\bibfnamefont {S.}~\bibnamefont {Das}}\ and\ \bibinfo {author} {\bibfnamefont {A.}~\bibnamefont {Soori}},\ }\bibfield  {title} {\bibinfo {title} {Crossed {{Andreev}} reflection in altermagnets},\ }\href {https://doi.org/10.1103/PhysRevB.109.245424} {\bibfield  {journal} {\bibinfo  {journal} {Phys. Rev. B}\ }\textbf {\bibinfo {volume} {109}},\ \bibinfo {pages} {245424} (\bibinfo {year} {2024})},\ \Eprint {https://arxiv.org/abs/2402.08263} {arXiv:2402.08263 [cond-mat.mes-hall]} \BibitemShut {NoStop}%
\bibitem [{\citenamefont {Ouassou}\ \emph {et~al.}(2023)\citenamefont {Ouassou}, \citenamefont {Brataas},\ and\ \citenamefont {Linder}}]{Ouassou-Linder:2023}%
  \BibitemOpen
  \bibfield  {author} {\bibinfo {author} {\bibfnamefont {J.~A.}\ \bibnamefont {Ouassou}}, \bibinfo {author} {\bibfnamefont {A.}~\bibnamefont {Brataas}},\ and\ \bibinfo {author} {\bibfnamefont {J.}~\bibnamefont {Linder}},\ }\bibfield  {title} {\bibinfo {title} {Dc {{Josephson Effect}} in {{Altermagnets}}},\ }\href {https://doi.org/10.1103/PhysRevLett.131.076003} {\bibfield  {journal} {\bibinfo  {journal} {Phys. Rev. Lett.}\ }\textbf {\bibinfo {volume} {131}},\ \bibinfo {pages} {076003} (\bibinfo {year} {2023})},\ \Eprint {https://arxiv.org/abs/2301.03603} {arXiv:2301.03603} \BibitemShut {NoStop}%
\bibitem [{\citenamefont {Zhang}\ \emph {et~al.}(2024)\citenamefont {Zhang}, \citenamefont {Hu},\ and\ \citenamefont {Neupert}}]{Zhang-Neupert-FinitemomentumCooperPairing-2024}%
  \BibitemOpen
  \bibfield  {author} {\bibinfo {author} {\bibfnamefont {S.-B.}\ \bibnamefont {Zhang}}, \bibinfo {author} {\bibfnamefont {L.-H.}\ \bibnamefont {Hu}},\ and\ \bibinfo {author} {\bibfnamefont {T.}~\bibnamefont {Neupert}},\ }\bibfield  {title} {\bibinfo {title} {Finite-momentum {{Cooper}} pairing in proximitized altermagnets},\ }\href {https://doi.org/10.1038/s41467-024-45951-3} {\bibfield  {journal} {\bibinfo  {journal} {Nat Commun}\ }\textbf {\bibinfo {volume} {15}},\ \bibinfo {pages} {1801} (\bibinfo {year} {2024})},\ \Eprint {https://arxiv.org/abs/2302.13185} {arXiv:2302.13185} \BibitemShut {NoStop}%
\bibitem [{\citenamefont {Sun}\ \emph {et~al.}(2025)\citenamefont {Sun}, \citenamefont {Zhang}, \citenamefont {Li},\ and\ \citenamefont {Trauzettel}}]{Sun-Trauzettel-TunableSecondHarmornic-2025}%
  \BibitemOpen
  \bibfield  {author} {\bibinfo {author} {\bibfnamefont {H.-P.}\ \bibnamefont {Sun}}, \bibinfo {author} {\bibfnamefont {S.-B.}\ \bibnamefont {Zhang}}, \bibinfo {author} {\bibfnamefont {C.-A.}\ \bibnamefont {Li}},\ and\ \bibinfo {author} {\bibfnamefont {B.}~\bibnamefont {Trauzettel}},\ }\bibfield  {title} {\bibinfo {title} {Tunable second harmornic in altermagnetic {{Josephson}} junctions},\ }\href {https://doi.org/10.1103/PhysRevB.111.165406} {\bibfield  {journal} {\bibinfo  {journal} {Phys. Rev. B}\ }\textbf {\bibinfo {volume} {111}},\ \bibinfo {pages} {165406} (\bibinfo {year} {2025})},\ \Eprint {https://arxiv.org/abs/2407.19413} {arXiv:2407.19413 [cond-mat]} \BibitemShut {NoStop}%
\bibitem [{\citenamefont {Giil}\ \emph {et~al.}(2024)\citenamefont {Giil}, \citenamefont {Brekke}, \citenamefont {Linder},\ and\ \citenamefont {Brataas}}]{Giil-Brataas:2024hat}%
  \BibitemOpen
  \bibfield  {author} {\bibinfo {author} {\bibfnamefont {H.~G.}\ \bibnamefont {Giil}}, \bibinfo {author} {\bibfnamefont {B.}~\bibnamefont {Brekke}}, \bibinfo {author} {\bibfnamefont {J.}~\bibnamefont {Linder}},\ and\ \bibinfo {author} {\bibfnamefont {A.}~\bibnamefont {Brataas}},\ }\bibfield  {title} {\bibinfo {title} {Quasiclassical theory of superconducting spin-splitter effects and spin-filtering via altermagnets},\ }\href {https://doi.org/10.1103/PhysRevB.110.L140506} {\bibfield  {journal} {\bibinfo  {journal} {Phys. Rev. B}\ }\textbf {\bibinfo {volume} {110}},\ \bibinfo {pages} {L140506} (\bibinfo {year} {2024})},\ \Eprint {https://arxiv.org/abs/2403.04851} {arXiv:2403.04851} \BibitemShut {NoStop}%
\bibitem [{\citenamefont {Zyuzin}(2024)}]{Zyuzin-MagnetoelectricEffectSuperconductors-2024}%
  \BibitemOpen
  \bibfield  {author} {\bibinfo {author} {\bibfnamefont {A.~A.}\ \bibnamefont {Zyuzin}},\ }\bibfield  {title} {\bibinfo {title} {Magnetoelectric effect in superconductors with $d$-wave magnetization},\ }\href {https://doi.org/10.1103/PhysRevB.109.L220505} {\bibfield  {journal} {\bibinfo  {journal} {Phys. Rev. B}\ }\textbf {\bibinfo {volume} {109}},\ \bibinfo {pages} {L220505} (\bibinfo {year} {2024})},\ \Eprint {https://arxiv.org/abs/2402.15459} {arXiv:2402.15459} \BibitemShut {NoStop}%
\bibitem [{\citenamefont {Hu}\ \emph {et~al.}(2025)\citenamefont {Hu}, \citenamefont {Matsyshyn},\ and\ \citenamefont {Song}}]{Hu-Song-NonlinearSuperconductingMagnetoelectric-2024}%
  \BibitemOpen
  \bibfield  {author} {\bibinfo {author} {\bibfnamefont {J.-X.}\ \bibnamefont {Hu}}, \bibinfo {author} {\bibfnamefont {O.}~\bibnamefont {Matsyshyn}},\ and\ \bibinfo {author} {\bibfnamefont {J.~C.~W.}\ \bibnamefont {Song}},\ }\bibfield  {title} {\bibinfo {title} {Nonlinear superconducting magnetoelectric effect},\ }\href {https://doi.org/10.1103/PhysRevLett.134.026001} {\bibfield  {journal} {\bibinfo  {journal} {Phys. Rev. Lett.}\ }\textbf {\bibinfo {volume} {134}},\ \bibinfo {pages} {026001} (\bibinfo {year} {2025})},\ \Eprint {https://arxiv.org/abs/2404.18616} {arXiv:2404.18616} \BibitemShut {NoStop}%
\bibitem [{\citenamefont {Kokkeler}\ \emph {et~al.}(2025)\citenamefont {Kokkeler}, \citenamefont {Tokatly},\ and\ \citenamefont {Bergeret}}]{Kokkeler-Bergeret-QuantumTransportTheory-2024}%
  \BibitemOpen
  \bibfield  {author} {\bibinfo {author} {\bibfnamefont {T.}~\bibnamefont {Kokkeler}}, \bibinfo {author} {\bibfnamefont {I.}~\bibnamefont {Tokatly}},\ and\ \bibinfo {author} {\bibfnamefont {F.~S.}\ \bibnamefont {Bergeret}},\ }\bibfield  {title} {\bibinfo {title} {{Quantum transport theory for unconventional magnets: Interplay of altermagnetism and p-wave magnetism with superconductivity}},\ }\href {https://doi.org/10.21468/SciPostPhys.18.6.178} {\bibfield  {journal} {\bibinfo  {journal} {SciPost Phys.}\ }\textbf {\bibinfo {volume} {18}},\ \bibinfo {pages} {178} (\bibinfo {year} {2025})},\ \Eprint {https://arxiv.org/abs/2412.10236} {arXiv:2412.10236} \BibitemShut {NoStop}%
\bibitem [{\citenamefont {Sukhachov}\ \emph {et~al.}(2024)\citenamefont {Sukhachov}, \citenamefont {Hodt},\ and\ \citenamefont {Linder}}]{Sukhachov-Linder-ThermoelectricEffectAltermagnetSuperconductor-2024}%
  \BibitemOpen
  \bibfield  {author} {\bibinfo {author} {\bibfnamefont {P.~O.}\ \bibnamefont {Sukhachov}}, \bibinfo {author} {\bibfnamefont {E.~W.}\ \bibnamefont {Hodt}},\ and\ \bibinfo {author} {\bibfnamefont {J.}~\bibnamefont {Linder}},\ }\bibfield  {title} {\bibinfo {title} {Thermoelectric {{Effect}} in {{Altermagnet-Superconductor Junctions}}},\ }\href {https://doi.org/10.1103/PhysRevB.110.094508} {\bibfield  {journal} {\bibinfo  {journal} {Phys. Rev. B}\ }\textbf {\bibinfo {volume} {110}},\ \bibinfo {pages} {094508} (\bibinfo {year} {2024})},\ \Eprint {https://arxiv.org/abs/2404.10038} {arXiv:2404.10038 [cond-mat]} \BibitemShut {NoStop}%
\bibitem [{\citenamefont {Bai}\ \emph {et~al.}(2024)\citenamefont {Bai}, \citenamefont {Feng}, \citenamefont {Liu}, \citenamefont {{\v S}mejkal}, \citenamefont {Mokrousov},\ and\ \citenamefont {Yao}}]{Bai-Yao-AltermagnetismExploringNew-2024}%
  \BibitemOpen
  \bibfield  {author} {\bibinfo {author} {\bibfnamefont {L.}~\bibnamefont {Bai}}, \bibinfo {author} {\bibfnamefont {W.}~\bibnamefont {Feng}}, \bibinfo {author} {\bibfnamefont {S.}~\bibnamefont {Liu}}, \bibinfo {author} {\bibfnamefont {L.}~\bibnamefont {{\v S}mejkal}}, \bibinfo {author} {\bibfnamefont {Y.}~\bibnamefont {Mokrousov}},\ and\ \bibinfo {author} {\bibfnamefont {Y.}~\bibnamefont {Yao}},\ }\bibfield  {title} {\bibinfo {title} {Altermagnetism: {{Exploring New Frontiers}} in {{Magnetism}} and {{Spintronics}}},\ }\href {https://doi.org/10.1002/adfm.202409327} {\bibfield  {journal} {\bibinfo  {journal} {Adv Funct Materials}\ ,\ \bibinfo {pages} {2409327}} (\bibinfo {year} {2024})},\ \Eprint {https://arxiv.org/abs/2406.02123} {arXiv:2406.02123 [cond-mat]} \BibitemShut {NoStop}%
\bibitem [{\citenamefont {Krempask{\'y}}\ \emph {et~al.}(2024)\citenamefont {Krempask{\'y}}, \citenamefont {{\v S}mejkal}, \citenamefont {D'Souza}, \citenamefont {Hajlaoui}, \citenamefont {Springholz}, \citenamefont {Uhl{\'i}{\v r}ov{\'a}}, \citenamefont {Alarab}, \citenamefont {Constantinou}, \citenamefont {Strocov}, \citenamefont {Usanov}, \citenamefont {Pudelko}, \citenamefont {{Gonz{\'a}lez-Hern{\'a}ndez}}, \citenamefont {Birk~Hellenes}, \citenamefont {Jansa}, \citenamefont {Reichlov{\'a}}, \citenamefont {{\v S}ob{\'a}{\v n}}, \citenamefont {Gonzalez~Betancourt}, \citenamefont {Wadley}, \citenamefont {Sinova}, \citenamefont {Kriegner}, \citenamefont {Min{\'a}r}, \citenamefont {Dil},\ and\ \citenamefont {Jungwirth}}]{Krempasky-Jungwirth:2024}%
  \BibitemOpen
  \bibfield  {author} {\bibinfo {author} {\bibfnamefont {J.}~\bibnamefont {Krempask{\'y}}}, \bibinfo {author} {\bibfnamefont {L.}~\bibnamefont {{\v S}mejkal}}, \bibinfo {author} {\bibfnamefont {S.~W.}\ \bibnamefont {D'Souza}}, \bibinfo {author} {\bibfnamefont {M.}~\bibnamefont {Hajlaoui}}, \bibinfo {author} {\bibfnamefont {G.}~\bibnamefont {Springholz}}, \bibinfo {author} {\bibfnamefont {K.}~\bibnamefont {Uhl{\'i}{\v r}ov{\'a}}}, \bibinfo {author} {\bibfnamefont {F.}~\bibnamefont {Alarab}}, \bibinfo {author} {\bibfnamefont {P.~C.}\ \bibnamefont {Constantinou}}, \bibinfo {author} {\bibfnamefont {V.}~\bibnamefont {Strocov}}, \bibinfo {author} {\bibfnamefont {D.}~\bibnamefont {Usanov}}, \bibinfo {author} {\bibfnamefont {W.~R.}\ \bibnamefont {Pudelko}}, \bibinfo {author} {\bibfnamefont {R.}~\bibnamefont {{Gonz{\'a}lez-Hern{\'a}ndez}}}, \bibinfo {author} {\bibfnamefont {A.}~\bibnamefont {Birk~Hellenes}}, \bibinfo {author} {\bibfnamefont {Z.}~\bibnamefont {Jansa}}, \bibinfo {author} {\bibfnamefont {H.}~\bibnamefont
  {Reichlov{\'a}}}, \bibinfo {author} {\bibfnamefont {Z.}~\bibnamefont {{\v S}ob{\'a}{\v n}}}, \bibinfo {author} {\bibfnamefont {R.~D.}\ \bibnamefont {Gonzalez~Betancourt}}, \bibinfo {author} {\bibfnamefont {P.}~\bibnamefont {Wadley}}, \bibinfo {author} {\bibfnamefont {J.}~\bibnamefont {Sinova}}, \bibinfo {author} {\bibfnamefont {D.}~\bibnamefont {Kriegner}}, \bibinfo {author} {\bibfnamefont {J.}~\bibnamefont {Min{\'a}r}}, \bibinfo {author} {\bibfnamefont {J.~H.}\ \bibnamefont {Dil}},\ and\ \bibinfo {author} {\bibfnamefont {T.}~\bibnamefont {Jungwirth}},\ }\bibfield  {title} {\bibinfo {title} {Altermagnetic lifting of {{Kramers}} spin degeneracy},\ }\href {https://doi.org/10.1038/s41586-023-06907-7} {\bibfield  {journal} {\bibinfo  {journal} {Nature}\ }\textbf {\bibinfo {volume} {626}},\ \bibinfo {pages} {517} (\bibinfo {year} {2024})},\ \Eprint {https://arxiv.org/abs/2308.10681} {arXiv:2308.10681 [physics.app-ph]} \BibitemShut {NoStop}%
\bibitem [{\citenamefont {Osumi}\ \emph {et~al.}(2024)\citenamefont {Osumi}, \citenamefont {Souma}, \citenamefont {Aoyama}, \citenamefont {Yamauchi}, \citenamefont {Honma}, \citenamefont {Nakayama}, \citenamefont {Takahashi}, \citenamefont {Ohgushi},\ and\ \citenamefont {Sato}}]{Osumi-Sato-ObservationGiantBand-2024}%
  \BibitemOpen
  \bibfield  {author} {\bibinfo {author} {\bibfnamefont {T.}~\bibnamefont {Osumi}}, \bibinfo {author} {\bibfnamefont {S.}~\bibnamefont {Souma}}, \bibinfo {author} {\bibfnamefont {T.}~\bibnamefont {Aoyama}}, \bibinfo {author} {\bibfnamefont {K.}~\bibnamefont {Yamauchi}}, \bibinfo {author} {\bibfnamefont {A.}~\bibnamefont {Honma}}, \bibinfo {author} {\bibfnamefont {K.}~\bibnamefont {Nakayama}}, \bibinfo {author} {\bibfnamefont {T.}~\bibnamefont {Takahashi}}, \bibinfo {author} {\bibfnamefont {K.}~\bibnamefont {Ohgushi}},\ and\ \bibinfo {author} {\bibfnamefont {T.}~\bibnamefont {Sato}},\ }\bibfield  {title} {\bibinfo {title} {Observation of a giant band splitting in altermagnetic {{MnTe}}},\ }\href {https://doi.org/10.1103/PhysRevB.109.115102} {\bibfield  {journal} {\bibinfo  {journal} {Phys. Rev. B}\ }\textbf {\bibinfo {volume} {109}},\ \bibinfo {pages} {115102} (\bibinfo {year} {2024})}\BibitemShut {NoStop}%
\bibitem [{\citenamefont {Reimers}\ \emph {et~al.}(2024)\citenamefont {Reimers}, \citenamefont {Odenbreit}, \citenamefont {{\v S}mejkal}, \citenamefont {Strocov}, \citenamefont {Constantinou}, \citenamefont {Hellenes}, \citenamefont {Jaeschke~Ubiergo}, \citenamefont {Campos}, \citenamefont {Bharadwaj}, \citenamefont {Chakraborty}, \citenamefont {Denneulin}, \citenamefont {Shi}, \citenamefont {{Dunin-Borkowski}}, \citenamefont {Das}, \citenamefont {Kl{\"a}ui}, \citenamefont {Sinova},\ and\ \citenamefont {Jourdan}}]{Reimers-Jourdan:2023}%
  \BibitemOpen
  \bibfield  {author} {\bibinfo {author} {\bibfnamefont {S.}~\bibnamefont {Reimers}}, \bibinfo {author} {\bibfnamefont {L.}~\bibnamefont {Odenbreit}}, \bibinfo {author} {\bibfnamefont {L.}~\bibnamefont {{\v S}mejkal}}, \bibinfo {author} {\bibfnamefont {V.~N.}\ \bibnamefont {Strocov}}, \bibinfo {author} {\bibfnamefont {P.}~\bibnamefont {Constantinou}}, \bibinfo {author} {\bibfnamefont {A.~B.}\ \bibnamefont {Hellenes}}, \bibinfo {author} {\bibfnamefont {R.}~\bibnamefont {Jaeschke~Ubiergo}}, \bibinfo {author} {\bibfnamefont {W.~H.}\ \bibnamefont {Campos}}, \bibinfo {author} {\bibfnamefont {V.~K.}\ \bibnamefont {Bharadwaj}}, \bibinfo {author} {\bibfnamefont {A.}~\bibnamefont {Chakraborty}}, \bibinfo {author} {\bibfnamefont {T.}~\bibnamefont {Denneulin}}, \bibinfo {author} {\bibfnamefont {W.}~\bibnamefont {Shi}}, \bibinfo {author} {\bibfnamefont {R.~E.}\ \bibnamefont {{Dunin-Borkowski}}}, \bibinfo {author} {\bibfnamefont {S.}~\bibnamefont {Das}}, \bibinfo {author} {\bibfnamefont {M.}~\bibnamefont {Kl{\"a}ui}},
  \bibinfo {author} {\bibfnamefont {J.}~\bibnamefont {Sinova}},\ and\ \bibinfo {author} {\bibfnamefont {M.}~\bibnamefont {Jourdan}},\ }\bibfield  {title} {\bibinfo {title} {Direct observation of altermagnetic band splitting in {{CrSb}} thin films},\ }\href {https://doi.org/10.1038/s41467-024-46476-5} {\bibfield  {journal} {\bibinfo  {journal} {Nat. Commun.}\ }\textbf {\bibinfo {volume} {15}},\ \bibinfo {pages} {2116} (\bibinfo {year} {2024})},\ \Eprint {https://arxiv.org/abs/2310.17280} {arXiv:2310.17280} \BibitemShut {NoStop}%
\bibitem [{\citenamefont {Zeng}\ \emph {et~al.}(2024)\citenamefont {Zeng}, \citenamefont {Zhu}, \citenamefont {Zhu}, \citenamefont {Liu}, \citenamefont {Ma}, \citenamefont {Hao}, \citenamefont {Liu}, \citenamefont {Qu}, \citenamefont {Yang}, \citenamefont {Jiang}, \citenamefont {Yamagami}, \citenamefont {Arita}, \citenamefont {Zhang}, \citenamefont {Shao}, \citenamefont {Dai}, \citenamefont {Shimada}, \citenamefont {Liu}, \citenamefont {Ye}, \citenamefont {Huang}, \citenamefont {Liu},\ and\ \citenamefont {Liu}}]{Zeng-Liu-ObservationSpinSplitting-2024}%
  \BibitemOpen
  \bibfield  {author} {\bibinfo {author} {\bibfnamefont {M.}~\bibnamefont {Zeng}}, \bibinfo {author} {\bibfnamefont {M.-Y.}\ \bibnamefont {Zhu}}, \bibinfo {author} {\bibfnamefont {Y.-P.}\ \bibnamefont {Zhu}}, \bibinfo {author} {\bibfnamefont {X.-R.}\ \bibnamefont {Liu}}, \bibinfo {author} {\bibfnamefont {X.-M.}\ \bibnamefont {Ma}}, \bibinfo {author} {\bibfnamefont {Y.-J.}\ \bibnamefont {Hao}}, \bibinfo {author} {\bibfnamefont {P.}~\bibnamefont {Liu}}, \bibinfo {author} {\bibfnamefont {G.}~\bibnamefont {Qu}}, \bibinfo {author} {\bibfnamefont {Y.}~\bibnamefont {Yang}}, \bibinfo {author} {\bibfnamefont {Z.}~\bibnamefont {Jiang}}, \bibinfo {author} {\bibfnamefont {K.}~\bibnamefont {Yamagami}}, \bibinfo {author} {\bibfnamefont {M.}~\bibnamefont {Arita}}, \bibinfo {author} {\bibfnamefont {X.}~\bibnamefont {Zhang}}, \bibinfo {author} {\bibfnamefont {T.-H.}\ \bibnamefont {Shao}}, \bibinfo {author} {\bibfnamefont {Y.}~\bibnamefont {Dai}}, \bibinfo {author} {\bibfnamefont {K.}~\bibnamefont {Shimada}}, \bibinfo {author}
  {\bibfnamefont {Z.}~\bibnamefont {Liu}}, \bibinfo {author} {\bibfnamefont {M.}~\bibnamefont {Ye}}, \bibinfo {author} {\bibfnamefont {Y.}~\bibnamefont {Huang}}, \bibinfo {author} {\bibfnamefont {Q.}~\bibnamefont {Liu}},\ and\ \bibinfo {author} {\bibfnamefont {C.}~\bibnamefont {Liu}},\ }\bibfield  {title} {\bibinfo {title} {Observation of {{Spin Splitting}} in {{Room}}-{{Temperature Metallic Antiferromagnet CrSb}}},\ }\href {https://doi.org/10.1002/advs.202406529} {\bibfield  {journal} {\bibinfo  {journal} {Advanced Science}\ ,\ \bibinfo {pages} {2406529}} (\bibinfo {year} {2024})},\ \Eprint {https://arxiv.org/abs/2405.12679} {arXiv:2405.12679 [cond-mat]} \BibitemShut {NoStop}%
\bibitem [{\citenamefont {Ding}\ \emph {et~al.}(2024)\citenamefont {Ding}, \citenamefont {Jiang}, \citenamefont {Chen}, \citenamefont {Tao}, \citenamefont {Liu}, \citenamefont {Li}, \citenamefont {Liu}, \citenamefont {Sun}, \citenamefont {Cheng}, \citenamefont {Liu}, \citenamefont {Yang}, \citenamefont {Zhang}, \citenamefont {Deng}, \citenamefont {Jing}, \citenamefont {Huang}, \citenamefont {Shi}, \citenamefont {Ye}, \citenamefont {Qiao}, \citenamefont {Wang}, \citenamefont {Guo}, \citenamefont {Feng},\ and\ \citenamefont {Shen}}]{Ding-Shen-LargeBandsplittingWave-2024}%
  \BibitemOpen
  \bibfield  {author} {\bibinfo {author} {\bibfnamefont {J.}~\bibnamefont {Ding}}, \bibinfo {author} {\bibfnamefont {Z.}~\bibnamefont {Jiang}}, \bibinfo {author} {\bibfnamefont {X.}~\bibnamefont {Chen}}, \bibinfo {author} {\bibfnamefont {Z.}~\bibnamefont {Tao}}, \bibinfo {author} {\bibfnamefont {Z.}~\bibnamefont {Liu}}, \bibinfo {author} {\bibfnamefont {T.}~\bibnamefont {Li}}, \bibinfo {author} {\bibfnamefont {J.}~\bibnamefont {Liu}}, \bibinfo {author} {\bibfnamefont {J.}~\bibnamefont {Sun}}, \bibinfo {author} {\bibfnamefont {J.}~\bibnamefont {Cheng}}, \bibinfo {author} {\bibfnamefont {J.}~\bibnamefont {Liu}}, \bibinfo {author} {\bibfnamefont {Y.}~\bibnamefont {Yang}}, \bibinfo {author} {\bibfnamefont {R.}~\bibnamefont {Zhang}}, \bibinfo {author} {\bibfnamefont {L.}~\bibnamefont {Deng}}, \bibinfo {author} {\bibfnamefont {W.}~\bibnamefont {Jing}}, \bibinfo {author} {\bibfnamefont {Y.}~\bibnamefont {Huang}}, \bibinfo {author} {\bibfnamefont {Y.}~\bibnamefont {Shi}}, \bibinfo {author} {\bibfnamefont
  {M.}~\bibnamefont {Ye}}, \bibinfo {author} {\bibfnamefont {S.}~\bibnamefont {Qiao}}, \bibinfo {author} {\bibfnamefont {Y.}~\bibnamefont {Wang}}, \bibinfo {author} {\bibfnamefont {Y.}~\bibnamefont {Guo}}, \bibinfo {author} {\bibfnamefont {D.}~\bibnamefont {Feng}},\ and\ \bibinfo {author} {\bibfnamefont {D.}~\bibnamefont {Shen}},\ }\bibfield  {title} {\bibinfo {title} {Large band splitting in $g$-wave altermagnet crsb},\ }\href {https://doi.org/10.1103/PhysRevLett.133.206401} {\bibfield  {journal} {\bibinfo  {journal} {Phys. Rev. Lett.}\ }\textbf {\bibinfo {volume} {133}},\ \bibinfo {pages} {206401} (\bibinfo {year} {2024})},\ \Eprint {https://arxiv.org/abs/2405.12687} {arXiv:2405.12687} \BibitemShut {NoStop}%
\bibitem [{\citenamefont {Jeong}\ \emph {et~al.}(2024)\citenamefont {Jeong}, \citenamefont {Choi}, \citenamefont {Nair}, \citenamefont {Buiarelli}, \citenamefont {Pourbahari}, \citenamefont {Oh}, \citenamefont {Bassim}, \citenamefont {Seo}, \citenamefont {Choi}, \citenamefont {Fernandes}, \citenamefont {Birol}, \citenamefont {Zhao}, \citenamefont {Lee},\ and\ \citenamefont {Jalan}}]{Jeong-Jalan-AltermagneticPolarMetallic-2024}%
  \BibitemOpen
  \bibfield  {author} {\bibinfo {author} {\bibfnamefont {S.~G.}\ \bibnamefont {Jeong}}, \bibinfo {author} {\bibfnamefont {I.~H.}\ \bibnamefont {Choi}}, \bibinfo {author} {\bibfnamefont {S.}~\bibnamefont {Nair}}, \bibinfo {author} {\bibfnamefont {L.}~\bibnamefont {Buiarelli}}, \bibinfo {author} {\bibfnamefont {B.}~\bibnamefont {Pourbahari}}, \bibinfo {author} {\bibfnamefont {J.~Y.}\ \bibnamefont {Oh}}, \bibinfo {author} {\bibfnamefont {N.}~\bibnamefont {Bassim}}, \bibinfo {author} {\bibfnamefont {A.}~\bibnamefont {Seo}}, \bibinfo {author} {\bibfnamefont {W.~S.}\ \bibnamefont {Choi}}, \bibinfo {author} {\bibfnamefont {R.~M.}\ \bibnamefont {Fernandes}}, \bibinfo {author} {\bibfnamefont {T.}~\bibnamefont {Birol}}, \bibinfo {author} {\bibfnamefont {L.}~\bibnamefont {Zhao}}, \bibinfo {author} {\bibfnamefont {J.~S.}\ \bibnamefont {Lee}},\ and\ \bibinfo {author} {\bibfnamefont {B.}~\bibnamefont {Jalan}},\ }\href@noop {} {\bibinfo {title} {Altermagnetic {{Polar Metallic}} phase in {{Ultra-Thin Epitaxially-Strained
  RuO2 Films}}}} (\bibinfo {year} {2024}),\ \Eprint {https://arxiv.org/abs/2405.05838} {arXiv:2405.05838 [cond-mat]} \BibitemShut {NoStop}%
\bibitem [{\citenamefont {Betancourt}\ \emph {et~al.}(2024)\citenamefont {Betancourt}, \citenamefont {Zub{\'a}{\v c}}, \citenamefont {Geishendorf}, \citenamefont {Ritzinger}, \citenamefont {R{\r u}{\v z}i{\v c}kov{\'a}}, \citenamefont {Kotte}, \citenamefont {{\v Z}elezn{\'y}}, \citenamefont {Olejn{\'i}k}, \citenamefont {Springholz}, \citenamefont {B{\"u}chner}, \citenamefont {Thomas}, \citenamefont {V{\'y}born{\'y}}, \citenamefont {Jungwirth}, \citenamefont {Reichlov{\'a}},\ and\ \citenamefont {Kriegner}}]{Betancourt2024}%
  \BibitemOpen
  \bibfield  {author} {\bibinfo {author} {\bibfnamefont {R.~D.~G.}\ \bibnamefont {Betancourt}}, \bibinfo {author} {\bibfnamefont {J.}~\bibnamefont {Zub{\'a}{\v c}}}, \bibinfo {author} {\bibfnamefont {K.}~\bibnamefont {Geishendorf}}, \bibinfo {author} {\bibfnamefont {P.}~\bibnamefont {Ritzinger}}, \bibinfo {author} {\bibfnamefont {B.}~\bibnamefont {R{\r u}{\v z}i{\v c}kov{\'a}}}, \bibinfo {author} {\bibfnamefont {T.}~\bibnamefont {Kotte}}, \bibinfo {author} {\bibfnamefont {J.}~\bibnamefont {{\v Z}elezn{\'y}}}, \bibinfo {author} {\bibfnamefont {K.}~\bibnamefont {Olejn{\'i}k}}, \bibinfo {author} {\bibfnamefont {G.}~\bibnamefont {Springholz}}, \bibinfo {author} {\bibfnamefont {B.}~\bibnamefont {B{\"u}chner}}, \bibinfo {author} {\bibfnamefont {A.}~\bibnamefont {Thomas}}, \bibinfo {author} {\bibfnamefont {K.}~\bibnamefont {V{\'y}born{\'y}}}, \bibinfo {author} {\bibfnamefont {T.}~\bibnamefont {Jungwirth}}, \bibinfo {author} {\bibfnamefont {H.}~\bibnamefont {Reichlov{\'a}}},\ and\ \bibinfo {author} {\bibfnamefont
  {D.}~\bibnamefont {Kriegner}},\ }\bibfield  {title} {\bibinfo {title} {Anisotropic magnetoresistance in altermagnetic {{MnTe}}},\ }\href {https://doi.org/10.1038/s44306-024-00046-z} {\bibfield  {journal} {\bibinfo  {journal} {npj Spintronics}\ }\textbf {\bibinfo {volume} {2}},\ \bibinfo {pages} {45} (\bibinfo {year} {2024})},\ \Eprint {https://arxiv.org/abs/2404.16516} {arXiv:2404.16516} \BibitemShut {NoStop}%
\bibitem [{\citenamefont {Weber}\ \emph {et~al.}(2024)\citenamefont {Weber}, \citenamefont {Wust}, \citenamefont {Haag}, \citenamefont {Akashdeep}, \citenamefont {Leckron}, \citenamefont {Schmitt}, \citenamefont {Ramos}, \citenamefont {Kikkawa}, \citenamefont {Saitoh}, \citenamefont {Kl{\"a}ui}, \citenamefont {{\v S}mejkal}, \citenamefont {Sinova}, \citenamefont {Aeschlimann}, \citenamefont {Jakob}, \citenamefont {Stadtm{\"u}ller},\ and\ \citenamefont {Schneider}}]{Weber-Schneider-AllOpticalExcitation-2024}%
  \BibitemOpen
  \bibfield  {author} {\bibinfo {author} {\bibfnamefont {M.}~\bibnamefont {Weber}}, \bibinfo {author} {\bibfnamefont {S.}~\bibnamefont {Wust}}, \bibinfo {author} {\bibfnamefont {L.}~\bibnamefont {Haag}}, \bibinfo {author} {\bibfnamefont {A.}~\bibnamefont {Akashdeep}}, \bibinfo {author} {\bibfnamefont {K.}~\bibnamefont {Leckron}}, \bibinfo {author} {\bibfnamefont {C.}~\bibnamefont {Schmitt}}, \bibinfo {author} {\bibfnamefont {R.}~\bibnamefont {Ramos}}, \bibinfo {author} {\bibfnamefont {T.}~\bibnamefont {Kikkawa}}, \bibinfo {author} {\bibfnamefont {E.}~\bibnamefont {Saitoh}}, \bibinfo {author} {\bibfnamefont {M.}~\bibnamefont {Kl{\"a}ui}}, \bibinfo {author} {\bibfnamefont {L.}~\bibnamefont {{\v S}mejkal}}, \bibinfo {author} {\bibfnamefont {J.}~\bibnamefont {Sinova}}, \bibinfo {author} {\bibfnamefont {M.}~\bibnamefont {Aeschlimann}}, \bibinfo {author} {\bibfnamefont {G.}~\bibnamefont {Jakob}}, \bibinfo {author} {\bibfnamefont {B.}~\bibnamefont {Stadtm{\"u}ller}},\ and\ \bibinfo {author} {\bibfnamefont {H.~C.}\
  \bibnamefont {Schneider}},\ }\href {http://arxiv.org/abs/2408.05187} {\bibinfo {title} {All optical excitation of spin polarization in d-wave altermagnets}} (\bibinfo {year} {2024}),\ \Eprint {https://arxiv.org/abs/2408.05187} {arXiv:2408.05187 [cond-mat]} \BibitemShut {NoStop}%
\bibitem [{\citenamefont {Li}\ \emph {et~al.}(2024)\citenamefont {Li}, \citenamefont {Hu}, \citenamefont {Li}, \citenamefont {Wang}, \citenamefont {Chen}, \citenamefont {Thiagarajan}, \citenamefont {Leandersson}, \citenamefont {Polley}, \citenamefont {Kim}, \citenamefont {Liu}, \citenamefont {Fulga}, \citenamefont {Vergniory}, \citenamefont {Janson}, \citenamefont {Tjernberg},\ and\ \citenamefont {van~den Brink}}]{Li-Brink-TopologicalWeylAltermagnetism-2024}%
  \BibitemOpen
  \bibfield  {author} {\bibinfo {author} {\bibfnamefont {C.}~\bibnamefont {Li}}, \bibinfo {author} {\bibfnamefont {M.}~\bibnamefont {Hu}}, \bibinfo {author} {\bibfnamefont {Z.}~\bibnamefont {Li}}, \bibinfo {author} {\bibfnamefont {Y.}~\bibnamefont {Wang}}, \bibinfo {author} {\bibfnamefont {W.}~\bibnamefont {Chen}}, \bibinfo {author} {\bibfnamefont {B.}~\bibnamefont {Thiagarajan}}, \bibinfo {author} {\bibfnamefont {M.}~\bibnamefont {Leandersson}}, \bibinfo {author} {\bibfnamefont {C.}~\bibnamefont {Polley}}, \bibinfo {author} {\bibfnamefont {T.}~\bibnamefont {Kim}}, \bibinfo {author} {\bibfnamefont {H.}~\bibnamefont {Liu}}, \bibinfo {author} {\bibfnamefont {C.}~\bibnamefont {Fulga}}, \bibinfo {author} {\bibfnamefont {M.~G.}\ \bibnamefont {Vergniory}}, \bibinfo {author} {\bibfnamefont {O.}~\bibnamefont {Janson}}, \bibinfo {author} {\bibfnamefont {O.}~\bibnamefont {Tjernberg}},\ and\ \bibinfo {author} {\bibfnamefont {J.}~\bibnamefont {van~den Brink}},\ }\href {https://doi.org/10.48550/arXiv.2405.14777} {\bibinfo
  {title} {Topological {{Weyl Altermagnetism}} in {{CrSb}}}} (\bibinfo {year} {2024}),\ \Eprint {https://arxiv.org/abs/2405.14777} {arXiv:2405.14777 [cond-mat]} \BibitemShut {NoStop}%
\bibitem [{\citenamefont {Regmi}\ \emph {et~al.}(2025)\citenamefont {Regmi}, \citenamefont {Bhandari}, \citenamefont {Thapa}, \citenamefont {Hao}, \citenamefont {Sharma}, \citenamefont {McKenzie}, \citenamefont {Chen}, \citenamefont {Nayak}, \citenamefont {El~Gazzah}, \citenamefont {M{\'a}rkus}, \citenamefont {Forr{\'o}}, \citenamefont {Liu}, \citenamefont {Cao}, \citenamefont {Mitchell}, \citenamefont {Mazin},\ and\ \citenamefont {Ghimire}}]{Regmi-Ghimire-AltermagnetismLayeredIntercalated-2024}%
  \BibitemOpen
  \bibfield  {author} {\bibinfo {author} {\bibfnamefont {R.~B.}\ \bibnamefont {Regmi}}, \bibinfo {author} {\bibfnamefont {H.}~\bibnamefont {Bhandari}}, \bibinfo {author} {\bibfnamefont {B.}~\bibnamefont {Thapa}}, \bibinfo {author} {\bibfnamefont {Y.}~\bibnamefont {Hao}}, \bibinfo {author} {\bibfnamefont {N.}~\bibnamefont {Sharma}}, \bibinfo {author} {\bibfnamefont {J.}~\bibnamefont {McKenzie}}, \bibinfo {author} {\bibfnamefont {X.}~\bibnamefont {Chen}}, \bibinfo {author} {\bibfnamefont {A.}~\bibnamefont {Nayak}}, \bibinfo {author} {\bibfnamefont {M.}~\bibnamefont {El~Gazzah}}, \bibinfo {author} {\bibfnamefont {B.~G.}\ \bibnamefont {M{\'a}rkus}}, \bibinfo {author} {\bibfnamefont {L.}~\bibnamefont {Forr{\'o}}}, \bibinfo {author} {\bibfnamefont {X.}~\bibnamefont {Liu}}, \bibinfo {author} {\bibfnamefont {H.}~\bibnamefont {Cao}}, \bibinfo {author} {\bibfnamefont {J.~F.}\ \bibnamefont {Mitchell}}, \bibinfo {author} {\bibfnamefont {I.~I.}\ \bibnamefont {Mazin}},\ and\ \bibinfo {author} {\bibfnamefont {N.~J.}\
  \bibnamefont {Ghimire}},\ }\bibfield  {title} {\bibinfo {title} {Electronic structure of a layered altermagnetic compound {$\text{Co}\text{Nb}_4\text{Se}_8$}},\ }\href {https://doi.org/10.1038/s41467-025-58642-4} {\bibfield  {journal} {\bibinfo  {journal} {Nature Communications}\ }\textbf {\bibinfo {volume} {16}},\ \bibinfo {pages} {4399} (\bibinfo {year} {2025})},\ \Eprint {https://arxiv.org/abs/2503.16670} {arXiv:2503.16670} \BibitemShut {NoStop}%
\bibitem [{\citenamefont {Dale}\ \emph {et~al.}(2024)\citenamefont {Dale}, \citenamefont {Ashour}, \citenamefont {Vila}, \citenamefont {Regmi}, \citenamefont {Fox}, \citenamefont {Johnson}, \citenamefont {Fedorov}, \citenamefont {Stibor}, \citenamefont {Ghimire},\ and\ \citenamefont {Griffin}}]{Dale-Griffin-NonrelativisticSpinSplitting-2024}%
  \BibitemOpen
  \bibfield  {author} {\bibinfo {author} {\bibfnamefont {N.}~\bibnamefont {Dale}}, \bibinfo {author} {\bibfnamefont {O.~A.}\ \bibnamefont {Ashour}}, \bibinfo {author} {\bibfnamefont {M.}~\bibnamefont {Vila}}, \bibinfo {author} {\bibfnamefont {R.~B.}\ \bibnamefont {Regmi}}, \bibinfo {author} {\bibfnamefont {J.}~\bibnamefont {Fox}}, \bibinfo {author} {\bibfnamefont {C.~W.}\ \bibnamefont {Johnson}}, \bibinfo {author} {\bibfnamefont {A.}~\bibnamefont {Fedorov}}, \bibinfo {author} {\bibfnamefont {A.}~\bibnamefont {Stibor}}, \bibinfo {author} {\bibfnamefont {N.~J.}\ \bibnamefont {Ghimire}},\ and\ \bibinfo {author} {\bibfnamefont {S.~M.}\ \bibnamefont {Griffin}},\ }\href {https://doi.org/10.48550/arXiv.2411.18761} {\bibinfo {title} {Non-relativistic spin splitting above and below the {{Fermi}} level in a {$g$}-wave altermagnet}} (\bibinfo {year} {2024}),\ \Eprint {https://arxiv.org/abs/2411.18761} {arXiv:2411.18761 [cond-mat]} \BibitemShut {NoStop}%
\bibitem [{\citenamefont {Zhang}\ \emph {et~al.}(2025)\citenamefont {Zhang}, \citenamefont {Cheng}, \citenamefont {Yin}, \citenamefont {Liu}, \citenamefont {Deng}, \citenamefont {Qiao}, \citenamefont {Shi}, \citenamefont {Zhang}, \citenamefont {Lin}, \citenamefont {Liu}, \citenamefont {Ye}, \citenamefont {Huang}, \citenamefont {Meng}, \citenamefont {Zhang}, \citenamefont {Okuda}, \citenamefont {Shimada}, \citenamefont {Cui}, \citenamefont {Zhao}, \citenamefont {Cao}, \citenamefont {Qiao}, \citenamefont {Liu},\ and\ \citenamefont {Chen}}]{Zhang-Chen-CrystalsymmetrypairedSpinValley-2025}%
  \BibitemOpen
  \bibfield  {author} {\bibinfo {author} {\bibfnamefont {F.}~\bibnamefont {Zhang}}, \bibinfo {author} {\bibfnamefont {X.}~\bibnamefont {Cheng}}, \bibinfo {author} {\bibfnamefont {Z.}~\bibnamefont {Yin}}, \bibinfo {author} {\bibfnamefont {C.}~\bibnamefont {Liu}}, \bibinfo {author} {\bibfnamefont {L.}~\bibnamefont {Deng}}, \bibinfo {author} {\bibfnamefont {Y.}~\bibnamefont {Qiao}}, \bibinfo {author} {\bibfnamefont {Z.}~\bibnamefont {Shi}}, \bibinfo {author} {\bibfnamefont {S.}~\bibnamefont {Zhang}}, \bibinfo {author} {\bibfnamefont {J.}~\bibnamefont {Lin}}, \bibinfo {author} {\bibfnamefont {Z.}~\bibnamefont {Liu}}, \bibinfo {author} {\bibfnamefont {M.}~\bibnamefont {Ye}}, \bibinfo {author} {\bibfnamefont {Y.}~\bibnamefont {Huang}}, \bibinfo {author} {\bibfnamefont {X.}~\bibnamefont {Meng}}, \bibinfo {author} {\bibfnamefont {C.}~\bibnamefont {Zhang}}, \bibinfo {author} {\bibfnamefont {T.}~\bibnamefont {Okuda}}, \bibinfo {author} {\bibfnamefont {K.}~\bibnamefont {Shimada}}, \bibinfo {author} {\bibfnamefont
  {S.}~\bibnamefont {Cui}}, \bibinfo {author} {\bibfnamefont {Y.}~\bibnamefont {Zhao}}, \bibinfo {author} {\bibfnamefont {G.-H.}\ \bibnamefont {Cao}}, \bibinfo {author} {\bibfnamefont {S.}~\bibnamefont {Qiao}}, \bibinfo {author} {\bibfnamefont {J.}~\bibnamefont {Liu}},\ and\ \bibinfo {author} {\bibfnamefont {C.}~\bibnamefont {Chen}},\ }\bibfield  {title} {\bibinfo {title} {Crystal-symmetry-paired spin--valley locking in a layered room-temperature metallic altermagnet candidate},\ }\bibfield  {journal} {\bibinfo  {journal} {Nat. Phys.}\ }\href {https://doi.org/10.1038/s41567-025-02864-2} {10.1038/s41567-025-02864-2} (\bibinfo {year} {2025})\BibitemShut {NoStop}%
\bibitem [{\citenamefont {Jiang}\ \emph {et~al.}(2025)\citenamefont {Jiang}, \citenamefont {Hu}, \citenamefont {Bai}, \citenamefont {Song}, \citenamefont {Mu}, \citenamefont {Qu}, \citenamefont {Li}, \citenamefont {Zhu}, \citenamefont {Pi}, \citenamefont {Wei}, \citenamefont {Sun}, \citenamefont {Huang}, \citenamefont {Zheng}, \citenamefont {Peng}, \citenamefont {He}, \citenamefont {Li}, \citenamefont {Luo}, \citenamefont {Li}, \citenamefont {Chen}, \citenamefont {Li}, \citenamefont {Weng},\ and\ \citenamefont {Qian}}]{Jiang-Qian-MetallicRoomtemperatureDwave-2025}%
  \BibitemOpen
  \bibfield  {author} {\bibinfo {author} {\bibfnamefont {B.}~\bibnamefont {Jiang}}, \bibinfo {author} {\bibfnamefont {M.}~\bibnamefont {Hu}}, \bibinfo {author} {\bibfnamefont {J.}~\bibnamefont {Bai}}, \bibinfo {author} {\bibfnamefont {Z.}~\bibnamefont {Song}}, \bibinfo {author} {\bibfnamefont {C.}~\bibnamefont {Mu}}, \bibinfo {author} {\bibfnamefont {G.}~\bibnamefont {Qu}}, \bibinfo {author} {\bibfnamefont {W.}~\bibnamefont {Li}}, \bibinfo {author} {\bibfnamefont {W.}~\bibnamefont {Zhu}}, \bibinfo {author} {\bibfnamefont {H.}~\bibnamefont {Pi}}, \bibinfo {author} {\bibfnamefont {Z.}~\bibnamefont {Wei}}, \bibinfo {author} {\bibfnamefont {Y.-J.}\ \bibnamefont {Sun}}, \bibinfo {author} {\bibfnamefont {Y.}~\bibnamefont {Huang}}, \bibinfo {author} {\bibfnamefont {X.}~\bibnamefont {Zheng}}, \bibinfo {author} {\bibfnamefont {Y.}~\bibnamefont {Peng}}, \bibinfo {author} {\bibfnamefont {L.}~\bibnamefont {He}}, \bibinfo {author} {\bibfnamefont {S.}~\bibnamefont {Li}}, \bibinfo {author} {\bibfnamefont {J.}~\bibnamefont
  {Luo}}, \bibinfo {author} {\bibfnamefont {Z.}~\bibnamefont {Li}}, \bibinfo {author} {\bibfnamefont {G.}~\bibnamefont {Chen}}, \bibinfo {author} {\bibfnamefont {H.}~\bibnamefont {Li}}, \bibinfo {author} {\bibfnamefont {H.}~\bibnamefont {Weng}},\ and\ \bibinfo {author} {\bibfnamefont {T.}~\bibnamefont {Qian}},\ }\bibfield  {title} {\bibinfo {title} {A metallic room-temperature d-wave altermagnet},\ }\bibfield  {journal} {\bibinfo  {journal} {Nat. Phys.}\ }\href {https://doi.org/10.1038/s41567-025-02822-y} {10.1038/s41567-025-02822-y} (\bibinfo {year} {2025})\BibitemShut {NoStop}%
\bibitem [{\citenamefont {Sakhya}\ \emph {et~al.}(2025)\citenamefont {Sakhya}, \citenamefont {Mondal}, \citenamefont {Sprague}, \citenamefont {Regmi}, \citenamefont {Kumay}, \citenamefont {Sheokand}, \citenamefont {Mazin}, \citenamefont {Ghimire},\ and\ \citenamefont {Neupane}}]{Sakhya-Neupane-ElectronicStructureLayered-2025}%
  \BibitemOpen
  \bibfield  {author} {\bibinfo {author} {\bibfnamefont {A.~P.}\ \bibnamefont {Sakhya}}, \bibinfo {author} {\bibfnamefont {M.~I.}\ \bibnamefont {Mondal}}, \bibinfo {author} {\bibfnamefont {M.}~\bibnamefont {Sprague}}, \bibinfo {author} {\bibfnamefont {R.~B.}\ \bibnamefont {Regmi}}, \bibinfo {author} {\bibfnamefont {A.~K.}\ \bibnamefont {Kumay}}, \bibinfo {author} {\bibfnamefont {H.}~\bibnamefont {Sheokand}}, \bibinfo {author} {\bibfnamefont {I.~I.}\ \bibnamefont {Mazin}}, \bibinfo {author} {\bibfnamefont {N.~J.}\ \bibnamefont {Ghimire}},\ and\ \bibinfo {author} {\bibfnamefont {M.}~\bibnamefont {Neupane}},\ }\href {https://arxiv.org/abs/2503.16670} {\bibinfo {title} {Electronic structure of a layered altermagnetic compound conb4se8}} (\bibinfo {year} {2025}),\ \Eprint {https://arxiv.org/abs/2503.16670} {arXiv:2503.16670 [cond-mat.mes-hall]} \BibitemShut {NoStop}%
\bibitem [{\citenamefont {Song}\ \emph {et~al.}(2025)\citenamefont {Song}, \citenamefont {Bai}, \citenamefont {Zhou}, \citenamefont {Han}, \citenamefont {Reichlova}, \citenamefont {Dil}, \citenamefont {Liu}, \citenamefont {Chen},\ and\ \citenamefont {Pan}}]{Song-Pan-AltermagnetsNewClass-2025}%
  \BibitemOpen
  \bibfield  {author} {\bibinfo {author} {\bibfnamefont {C.}~\bibnamefont {Song}}, \bibinfo {author} {\bibfnamefont {H.}~\bibnamefont {Bai}}, \bibinfo {author} {\bibfnamefont {Z.}~\bibnamefont {Zhou}}, \bibinfo {author} {\bibfnamefont {L.}~\bibnamefont {Han}}, \bibinfo {author} {\bibfnamefont {H.}~\bibnamefont {Reichlova}}, \bibinfo {author} {\bibfnamefont {J.~H.}\ \bibnamefont {Dil}}, \bibinfo {author} {\bibfnamefont {J.}~\bibnamefont {Liu}}, \bibinfo {author} {\bibfnamefont {X.}~\bibnamefont {Chen}},\ and\ \bibinfo {author} {\bibfnamefont {F.}~\bibnamefont {Pan}},\ }\bibfield  {title} {\bibinfo {title} {Altermagnets as a new class of functional materials},\ }\bibfield  {journal} {\bibinfo  {journal} {Nat Rev Mater}\ }\href {https://doi.org/10.1038/s41578-025-00779-1} {10.1038/s41578-025-00779-1} (\bibinfo {year} {2025})\BibitemShut {NoStop}%
\bibitem [{Note1()}]{Note1}%
  \BibitemOpen
  \bibinfo {note} {The spin-dependent effective chemical potential can be achieved by applying a magnetic field via the Zeeman term. In 2D systems, orbital effects can be ignored if the field is in the plane of the material.}\BibitemShut {Stop}%
\bibitem [{\citenamefont {Bradlyn}\ \emph {et~al.}(2012)\citenamefont {Bradlyn}, \citenamefont {Goldstein},\ and\ \citenamefont {Read}}]{Bradlyn-Read:2012}%
  \BibitemOpen
  \bibfield  {author} {\bibinfo {author} {\bibfnamefont {B.}~\bibnamefont {Bradlyn}}, \bibinfo {author} {\bibfnamefont {M.}~\bibnamefont {Goldstein}},\ and\ \bibinfo {author} {\bibfnamefont {N.}~\bibnamefont {Read}},\ }\bibfield  {title} {\bibinfo {title} {Kubo formulas for viscosity: {{Hall}} viscosity, {{Ward}} identities, and the relation with conductivity},\ }\href {https://doi.org/10.1103/PhysRevB.86.245309} {\bibfield  {journal} {\bibinfo  {journal} {Phys. Rev. B}\ }\textbf {\bibinfo {volume} {86}},\ \bibinfo {pages} {245309} (\bibinfo {year} {2012})},\ \Eprint {https://arxiv.org/abs/1207.7021} {arXiv:1207.7021} \BibitemShut {NoStop}%
\bibitem [{\citenamefont {Bateman}\ and\ \citenamefont {Erd\'elyi}(1953)}]{Bateman-Erdelyi:book-t1}%
  \BibitemOpen
  \bibfield  {author} {\bibinfo {author} {\bibfnamefont {H.}~\bibnamefont {Bateman}}\ and\ \bibinfo {author} {\bibfnamefont {A.}~\bibnamefont {Erd\'elyi}},\ }\href {https://resolver.caltech.edu/CaltechAUTHORS:20140123-104529738} {\emph {\bibinfo {title} {Higher {{Transcendental Functions}}, Vol. 1}}}\ (\bibinfo  {publisher} {McGraw-Hill},\ \bibinfo {address} {New York},\ \bibinfo {year} {1953})\BibitemShut {NoStop}%
\bibitem [{\citenamefont {Callaway}(1959)}]{Callaway-ModelLatticeThermal-1959}%
  \BibitemOpen
  \bibfield  {author} {\bibinfo {author} {\bibfnamefont {J.}~\bibnamefont {Callaway}},\ }\bibfield  {title} {\bibinfo {title} {Model for {{Lattice Thermal Conductivity}} at {{Low Temperatures}}},\ }\href {https://doi.org/10.1103/PhysRev.113.1046} {\bibfield  {journal} {\bibinfo  {journal} {Phys. Rev.}\ }\textbf {\bibinfo {volume} {113}},\ \bibinfo {pages} {1046} (\bibinfo {year} {1959})}\BibitemShut {NoStop}%
\bibitem [{\citenamefont {Gantmakher}\ and\ \citenamefont {Levinson}(1987)}]{Gantmakher-Levinson-CarrierScatteringMetals-1987}%
  \BibitemOpen
  \bibfield  {author} {\bibinfo {author} {\bibfnamefont {V.~F.}\ \bibnamefont {Gantmakher}}\ and\ \bibinfo {author} {\bibfnamefont {Y.~B.}\ \bibnamefont {Levinson}},\ }\href {https://www.elsevier.com/books/carrier-scattering-in-metals-and-semiconductors/gantmakher/978-0-444-87025-4} {\emph {\bibinfo {title} {Carrier {{Scattering}} in {{Metals}} and {{Semiconductors}}}}},\ Modern Problems in Condensed Matter Sciences\ (\bibinfo  {publisher} {North-Holland},\ \bibinfo {address} {Amsterdam},\ \bibinfo {year} {1987})\BibitemShut {NoStop}%
\bibitem [{\citenamefont {Gurzhi}\ \emph {et~al.}(2006)\citenamefont {Gurzhi}, \citenamefont {Kalinenko}, \citenamefont {Kopeliovich}, \citenamefont {Pyshkin},\ and\ \citenamefont {Yanovsky}}]{Gurzhi-Yanovsky:2006}%
  \BibitemOpen
  \bibfield  {author} {\bibinfo {author} {\bibfnamefont {R.~N.}\ \bibnamefont {Gurzhi}}, \bibinfo {author} {\bibfnamefont {A.~N.}\ \bibnamefont {Kalinenko}}, \bibinfo {author} {\bibfnamefont {A.~I.}\ \bibnamefont {Kopeliovich}}, \bibinfo {author} {\bibfnamefont {P.~V.}\ \bibnamefont {Pyshkin}},\ and\ \bibinfo {author} {\bibfnamefont {A.~V.}\ \bibnamefont {Yanovsky}},\ }\bibfield  {title} {\bibinfo {title} {Dynamics of a spin-polarized electron liquid: {{Spin}} oscillations with a low decay},\ }\href {https://doi.org/10.1103/PhysRevB.73.153204} {\bibfield  {journal} {\bibinfo  {journal} {Phys. Rev. B}\ }\textbf {\bibinfo {volume} {73}},\ \bibinfo {pages} {153204} (\bibinfo {year} {2006})},\ \Eprint {https://arxiv.org/abs/1109.1872} {arXiv:1109.1872} \BibitemShut {NoStop}%
\bibitem [{\citenamefont {D'Amico}\ and\ \citenamefont {Vignale}(2000)}]{DAmico-Vignale-TheorySpinCoulomb-2000}%
  \BibitemOpen
  \bibfield  {author} {\bibinfo {author} {\bibfnamefont {I.}~\bibnamefont {D'Amico}}\ and\ \bibinfo {author} {\bibfnamefont {G.}~\bibnamefont {Vignale}},\ }\bibfield  {title} {\bibinfo {title} {Theory of spin {{Coulomb}} drag in spin-polarized transport},\ }\href {https://doi.org/10.1103/PhysRevB.62.4853} {\bibfield  {journal} {\bibinfo  {journal} {Phys. Rev. B}\ }\textbf {\bibinfo {volume} {62}},\ \bibinfo {pages} {4853} (\bibinfo {year} {2000})},\ \Eprint {https://arxiv.org/abs/cond-mat/9912412} {arXiv:cond-mat/9912412} \BibitemShut {NoStop}%
\bibitem [{\citenamefont {Matsuo}\ \emph {et~al.}(2017)\citenamefont {Matsuo}, \citenamefont {Ohnuma},\ and\ \citenamefont {Maekawa}}]{Matsuo:2017}%
  \BibitemOpen
  \bibfield  {author} {\bibinfo {author} {\bibfnamefont {M.}~\bibnamefont {Matsuo}}, \bibinfo {author} {\bibfnamefont {Y.}~\bibnamefont {Ohnuma}},\ and\ \bibinfo {author} {\bibfnamefont {S.}~\bibnamefont {Maekawa}},\ }\bibfield  {title} {\bibinfo {title} {Theory of spin hydrodynamic generation},\ }\href {https://doi.org/10.1103/PhysRevB.96.020401} {\bibfield  {journal} {\bibinfo  {journal} {Phys. Rev. B}\ }\textbf {\bibinfo {volume} {96}},\ \bibinfo {pages} {020401} (\bibinfo {year} {2017})},\ \Eprint {https://arxiv.org/abs/1706.06521} {arxiv:1706.06521} \BibitemShut {NoStop}%
\bibitem [{\citenamefont {Hodt}\ \emph {et~al.}(2024)\citenamefont {Hodt}, \citenamefont {Sukhachov},\ and\ \citenamefont {Linder}}]{Hodt-Linder-InterfaceinducedMagnetizationAltermagnets-2024}%
  \BibitemOpen
  \bibfield  {author} {\bibinfo {author} {\bibfnamefont {E.~W.}\ \bibnamefont {Hodt}}, \bibinfo {author} {\bibfnamefont {P.}~\bibnamefont {Sukhachov}},\ and\ \bibinfo {author} {\bibfnamefont {J.}~\bibnamefont {Linder}},\ }\bibfield  {title} {\bibinfo {title} {Interface-induced magnetization in altermagnets and antiferromagnets},\ }\href {https://doi.org/10.1103/PhysRevB.110.054446} {\bibfield  {journal} {\bibinfo  {journal} {Phys. Rev. B}\ }\textbf {\bibinfo {volume} {110}},\ \bibinfo {pages} {054446} (\bibinfo {year} {2024})},\ \Eprint {https://arxiv.org/abs/2406.07603} {arXiv:2406.07603 [cond-mat]} \BibitemShut {NoStop}%
\bibitem [{\citenamefont {Landau}\ and\ \citenamefont {Lifshitz}(2013)}]{Landau:t6-2013}%
  \BibitemOpen
  \bibfield  {author} {\bibinfo {author} {\bibfnamefont {L.~D.}\ \bibnamefont {Landau}}\ and\ \bibinfo {author} {\bibfnamefont {E.~M.}\ \bibnamefont {Lifshitz}},\ }\href {https://www.elsevier.com/books/fluid-mechanics/landau/978-0-08-057073-0} {\emph {\bibinfo {title} {Fluid {{Mechanics}}}}}\ (\bibinfo  {publisher} {Butterworth-Heinemann},\ \bibinfo {address} {Oxford},\ \bibinfo {year} {2013})\BibitemShut {NoStop}%
\bibitem [{\citenamefont {Levitov}\ and\ \citenamefont {Falkovich}(2016)}]{Levitov-Falkovich:2016}%
  \BibitemOpen
  \bibfield  {author} {\bibinfo {author} {\bibfnamefont {L.}~\bibnamefont {Levitov}}\ and\ \bibinfo {author} {\bibfnamefont {G.}~\bibnamefont {Falkovich}},\ }\bibfield  {title} {\bibinfo {title} {Electron viscosity, current vortices and negative nonlocal resistance in graphene},\ }\href {https://doi.org/10.1038/nphys3667} {\bibfield  {journal} {\bibinfo  {journal} {Nat. Phys.}\ }\textbf {\bibinfo {volume} {12}},\ \bibinfo {pages} {672} (\bibinfo {year} {2016})},\ \Eprint {https://arxiv.org/abs/1508.00836} {arXiv:1508.00836} \BibitemShut {NoStop}%
\bibitem [{\citenamefont {Falkovich}\ and\ \citenamefont {Levitov}(2017)}]{Falkovich-Levitov:2017}%
  \BibitemOpen
  \bibfield  {author} {\bibinfo {author} {\bibfnamefont {G.}~\bibnamefont {Falkovich}}\ and\ \bibinfo {author} {\bibfnamefont {L.}~\bibnamefont {Levitov}},\ }\bibfield  {title} {\bibinfo {title} {Linking {{Spatial Distributions}} of {{Potential}} and {{Current}} in {{Viscous Electronics}}},\ }\href {https://doi.org/10.1103/PhysRevLett.119.066601} {\bibfield  {journal} {\bibinfo  {journal} {Phys. Rev. Lett.}\ }\textbf {\bibinfo {volume} {119}},\ \bibinfo {pages} {066601} (\bibinfo {year} {2017})},\ \Eprint {https://arxiv.org/abs/1607.00986} {arXiv:1607.00986} \BibitemShut {NoStop}%
\bibitem [{\citenamefont {Danz}\ and\ \citenamefont {Narozhny}(2020)}]{Danz-Narozhny:2019}%
  \BibitemOpen
  \bibfield  {author} {\bibinfo {author} {\bibfnamefont {S.}~\bibnamefont {Danz}}\ and\ \bibinfo {author} {\bibfnamefont {B.~N.}\ \bibnamefont {Narozhny}},\ }\bibfield  {title} {\bibinfo {title} {Vorticity of viscous electronic flow in graphene},\ }\href {https://doi.org/10.1088/2053-1583/ab7bfa} {\bibfield  {journal} {\bibinfo  {journal} {2D Mater.}\ }\textbf {\bibinfo {volume} {7}},\ \bibinfo {pages} {035001} (\bibinfo {year} {2020})},\ \Eprint {https://arxiv.org/abs/1910.14473} {arXiv:1910.14473} \BibitemShut {NoStop}%
\bibitem [{\citenamefont {Torre}\ \emph {et~al.}(2015)\citenamefont {Torre}, \citenamefont {Tomadin}, \citenamefont {Geim},\ and\ \citenamefont {Polini}}]{Torre-Polini:2015}%
  \BibitemOpen
  \bibfield  {author} {\bibinfo {author} {\bibfnamefont {I.}~\bibnamefont {Torre}}, \bibinfo {author} {\bibfnamefont {A.}~\bibnamefont {Tomadin}}, \bibinfo {author} {\bibfnamefont {A.~K.}\ \bibnamefont {Geim}},\ and\ \bibinfo {author} {\bibfnamefont {M.}~\bibnamefont {Polini}},\ }\bibfield  {title} {\bibinfo {title} {Nonlocal transport and the hydrodynamic shear viscosity in graphene},\ }\href {https://doi.org/10.1103/PhysRevB.92.165433} {\bibfield  {journal} {\bibinfo  {journal} {Phys. Rev. B}\ }\textbf {\bibinfo {volume} {92}},\ \bibinfo {pages} {165433} (\bibinfo {year} {2015})},\ \Eprint {https://arxiv.org/abs/1508.00363} {arXiv:1508.00363} \BibitemShut {NoStop}%
\bibitem [{\citenamefont {Bychkov}\ and\ \citenamefont {Rashba}(1984)}]{Bychkov-Rashba-OscillatoryEffectsMagnetic-1984}%
  \BibitemOpen
  \bibfield  {author} {\bibinfo {author} {\bibfnamefont {Y.~A.}\ \bibnamefont {Bychkov}}\ and\ \bibinfo {author} {\bibfnamefont {E.~I.}\ \bibnamefont {Rashba}},\ }\bibfield  {title} {\bibinfo {title} {Oscillatory effects and the magnetic susceptibility of carriers in inversion layers},\ }\href {https://doi.org/10.1088/0022-3719/17/33/015} {\bibfield  {journal} {\bibinfo  {journal} {J. Phys. C: Solid State Phys.}\ }\textbf {\bibinfo {volume} {17}},\ \bibinfo {pages} {6039} (\bibinfo {year} {1984})}\BibitemShut {NoStop}%
\bibitem [{\citenamefont {Maze}\ \emph {et~al.}(2008)\citenamefont {Maze}, \citenamefont {Stanwix}, \citenamefont {Hodges}, \citenamefont {Hong}, \citenamefont {Taylor}, \citenamefont {Cappellaro}, \citenamefont {Jiang}, \citenamefont {Dutt}, \citenamefont {Togan}, \citenamefont {Zibrov}, \citenamefont {Yacoby}, \citenamefont {Walsworth},\ and\ \citenamefont {Lukin}}]{Maze-Lukin-NanoscaleMagneticSensing-2008}%
  \BibitemOpen
  \bibfield  {author} {\bibinfo {author} {\bibfnamefont {J.~R.}\ \bibnamefont {Maze}}, \bibinfo {author} {\bibfnamefont {P.~L.}\ \bibnamefont {Stanwix}}, \bibinfo {author} {\bibfnamefont {J.~S.}\ \bibnamefont {Hodges}}, \bibinfo {author} {\bibfnamefont {S.}~\bibnamefont {Hong}}, \bibinfo {author} {\bibfnamefont {J.~M.}\ \bibnamefont {Taylor}}, \bibinfo {author} {\bibfnamefont {P.}~\bibnamefont {Cappellaro}}, \bibinfo {author} {\bibfnamefont {L.}~\bibnamefont {Jiang}}, \bibinfo {author} {\bibfnamefont {M.~V.~G.}\ \bibnamefont {Dutt}}, \bibinfo {author} {\bibfnamefont {E.}~\bibnamefont {Togan}}, \bibinfo {author} {\bibfnamefont {A.~S.}\ \bibnamefont {Zibrov}}, \bibinfo {author} {\bibfnamefont {A.}~\bibnamefont {Yacoby}}, \bibinfo {author} {\bibfnamefont {R.~L.}\ \bibnamefont {Walsworth}},\ and\ \bibinfo {author} {\bibfnamefont {M.~D.}\ \bibnamefont {Lukin}},\ }\bibfield  {title} {\bibinfo {title} {Nanoscale magnetic sensing with an individual electronic spin in diamond},\ }\href
  {https://doi.org/10.1038/nature07279} {\bibfield  {journal} {\bibinfo  {journal} {Nature}\ }\textbf {\bibinfo {volume} {455}},\ \bibinfo {pages} {644} (\bibinfo {year} {2008})}\BibitemShut {NoStop}%
\bibitem [{\citenamefont {Levine}\ \emph {et~al.}(2019)\citenamefont {Levine}, \citenamefont {Turner}, \citenamefont {Kehayias}, \citenamefont {Hart}, \citenamefont {Langellier}, \citenamefont {Trubko}, \citenamefont {Glenn}, \citenamefont {Fu},\ and\ \citenamefont {Walsworth}}]{Levine-Walsworth-PrinciplesTechniquesQuantum-2019}%
  \BibitemOpen
  \bibfield  {author} {\bibinfo {author} {\bibfnamefont {E.~V.}\ \bibnamefont {Levine}}, \bibinfo {author} {\bibfnamefont {M.~J.}\ \bibnamefont {Turner}}, \bibinfo {author} {\bibfnamefont {P.}~\bibnamefont {Kehayias}}, \bibinfo {author} {\bibfnamefont {C.~A.}\ \bibnamefont {Hart}}, \bibinfo {author} {\bibfnamefont {N.}~\bibnamefont {Langellier}}, \bibinfo {author} {\bibfnamefont {R.}~\bibnamefont {Trubko}}, \bibinfo {author} {\bibfnamefont {D.~R.}\ \bibnamefont {Glenn}}, \bibinfo {author} {\bibfnamefont {R.~R.}\ \bibnamefont {Fu}},\ and\ \bibinfo {author} {\bibfnamefont {R.~L.}\ \bibnamefont {Walsworth}},\ }\bibfield  {title} {\bibinfo {title} {Principles and techniques of the quantum diamond microscope},\ }\href {https://doi.org/10.1515/nanoph-2019-0209} {\bibfield  {journal} {\bibinfo  {journal} {Nanophotonics}\ }\textbf {\bibinfo {volume} {8}},\ \bibinfo {pages} {1945} (\bibinfo {year} {2019})}\BibitemShut {NoStop}%
\bibitem [{\citenamefont {Barry}\ \emph {et~al.}(2020)\citenamefont {Barry}, \citenamefont {Schloss}, \citenamefont {Bauch}, \citenamefont {Turner}, \citenamefont {Hart}, \citenamefont {Pham},\ and\ \citenamefont {Walsworth}}]{Barry-Walsworth:2019-QSM}%
  \BibitemOpen
  \bibfield  {author} {\bibinfo {author} {\bibfnamefont {J.~F.}\ \bibnamefont {Barry}}, \bibinfo {author} {\bibfnamefont {J.~M.}\ \bibnamefont {Schloss}}, \bibinfo {author} {\bibfnamefont {E.}~\bibnamefont {Bauch}}, \bibinfo {author} {\bibfnamefont {M.~J.}\ \bibnamefont {Turner}}, \bibinfo {author} {\bibfnamefont {C.~A.}\ \bibnamefont {Hart}}, \bibinfo {author} {\bibfnamefont {L.~M.}\ \bibnamefont {Pham}},\ and\ \bibinfo {author} {\bibfnamefont {R.~L.}\ \bibnamefont {Walsworth}},\ }\bibfield  {title} {\bibinfo {title} {Sensitivity optimization for {{NV-diamond}} magnetometry},\ }\href {https://doi.org/10.1103/RevModPhys.92.015004} {\bibfield  {journal} {\bibinfo  {journal} {Rev. Mod. Phys.}\ }\textbf {\bibinfo {volume} {92}},\ \bibinfo {pages} {015004} (\bibinfo {year} {2020})},\ \Eprint {https://arxiv.org/abs/1903.08176v2} {arXiv:1903.08176v2} \BibitemShut {NoStop}%
\bibitem [{\citenamefont {Marchiori}\ \emph {et~al.}(2021)\citenamefont {Marchiori}, \citenamefont {Ceccarelli}, \citenamefont {Rossi}, \citenamefont {Lorenzelli}, \citenamefont {Degen},\ and\ \citenamefont {Poggio}}]{Marchiori-Poggio-NanoscaleMagneticField-2021}%
  \BibitemOpen
  \bibfield  {author} {\bibinfo {author} {\bibfnamefont {E.}~\bibnamefont {Marchiori}}, \bibinfo {author} {\bibfnamefont {L.}~\bibnamefont {Ceccarelli}}, \bibinfo {author} {\bibfnamefont {N.}~\bibnamefont {Rossi}}, \bibinfo {author} {\bibfnamefont {L.}~\bibnamefont {Lorenzelli}}, \bibinfo {author} {\bibfnamefont {C.~L.}\ \bibnamefont {Degen}},\ and\ \bibinfo {author} {\bibfnamefont {M.}~\bibnamefont {Poggio}},\ }\bibfield  {title} {\bibinfo {title} {Nanoscale magnetic field imaging for {{2D}} materials},\ }\href {https://doi.org/10.1038/s42254-021-00380-9} {\bibfield  {journal} {\bibinfo  {journal} {Nature Reviews Physics}\ }\textbf {\bibinfo {volume} {4}},\ \bibinfo {pages} {49} (\bibinfo {year} {2021})}\BibitemShut {NoStop}%
\bibitem [{\citenamefont {Rovny}\ \emph {et~al.}(2024)\citenamefont {Rovny}, \citenamefont {Gopalakrishnan}, \citenamefont {Jayich}, \citenamefont {Maletinsky}, \citenamefont {Demler},\ and\ \citenamefont {de~Leon}}]{Rovny-Leon-NewOpportunitiesCondensed-2024}%
  \BibitemOpen
  \bibfield  {author} {\bibinfo {author} {\bibfnamefont {J.}~\bibnamefont {Rovny}}, \bibinfo {author} {\bibfnamefont {S.}~\bibnamefont {Gopalakrishnan}}, \bibinfo {author} {\bibfnamefont {A.~C.~B.}\ \bibnamefont {Jayich}}, \bibinfo {author} {\bibfnamefont {P.}~\bibnamefont {Maletinsky}}, \bibinfo {author} {\bibfnamefont {E.}~\bibnamefont {Demler}},\ and\ \bibinfo {author} {\bibfnamefont {N.~P.}\ \bibnamefont {de~Leon}},\ }\href {https://doi.org/10.48550/arXiv.2403.13710} {\bibinfo {title} {New opportunities in condensed matter physics for nanoscale quantum sensors}} (\bibinfo {year} {2024}),\ \Eprint {https://arxiv.org/abs/2403.13710} {arXiv:2403.13710 [cond-mat]} \BibitemShut {NoStop}%
\bibitem [{\citenamefont {{Aharon-Steinberg}}\ \emph {et~al.}(2022)\citenamefont {{Aharon-Steinberg}}, \citenamefont {V{\"o}lkl}, \citenamefont {Kaplan}, \citenamefont {Pariari}, \citenamefont {Roy}, \citenamefont {Holder}, \citenamefont {Wolf}, \citenamefont {Meltzer}, \citenamefont {Myasoedov}, \citenamefont {Huber}, \citenamefont {Yan}, \citenamefont {Falkovich}, \citenamefont {Levitov}, \citenamefont {H{\"u}cker},\ and\ \citenamefont {Zeldov}}]{Aharon-Steinberg2022}%
  \BibitemOpen
  \bibfield  {author} {\bibinfo {author} {\bibfnamefont {A.}~\bibnamefont {{Aharon-Steinberg}}}, \bibinfo {author} {\bibfnamefont {T.}~\bibnamefont {V{\"o}lkl}}, \bibinfo {author} {\bibfnamefont {A.}~\bibnamefont {Kaplan}}, \bibinfo {author} {\bibfnamefont {A.~K.}\ \bibnamefont {Pariari}}, \bibinfo {author} {\bibfnamefont {I.}~\bibnamefont {Roy}}, \bibinfo {author} {\bibfnamefont {T.}~\bibnamefont {Holder}}, \bibinfo {author} {\bibfnamefont {Y.}~\bibnamefont {Wolf}}, \bibinfo {author} {\bibfnamefont {A.~Y.}\ \bibnamefont {Meltzer}}, \bibinfo {author} {\bibfnamefont {Y.}~\bibnamefont {Myasoedov}}, \bibinfo {author} {\bibfnamefont {M.~E.}\ \bibnamefont {Huber}}, \bibinfo {author} {\bibfnamefont {B.}~\bibnamefont {Yan}}, \bibinfo {author} {\bibfnamefont {G.}~\bibnamefont {Falkovich}}, \bibinfo {author} {\bibfnamefont {L.~S.}\ \bibnamefont {Levitov}}, \bibinfo {author} {\bibfnamefont {M.}~\bibnamefont {H{\"u}cker}},\ and\ \bibinfo {author} {\bibfnamefont {E.}~\bibnamefont {Zeldov}},\ }\bibfield  {title} {\bibinfo
  {title} {Direct observation of vortices in an electron fluid},\ }\href {https://doi.org/10.1038/s41586-022-04794-y} {\bibfield  {journal} {\bibinfo  {journal} {Nature}\ }\textbf {\bibinfo {volume} {607}},\ \bibinfo {pages} {74} (\bibinfo {year} {2022})},\ \Eprint {https://arxiv.org/abs/2202.02798} {arXiv:2202.02798} \BibitemShut {NoStop}%
\bibitem [{\citenamefont {Baddorf}(2007)}]{Baddorf-ScanningTunnelingPotentiometry-2007}%
  \BibitemOpen
  \bibfield  {author} {\bibinfo {author} {\bibfnamefont {A.~P.}\ \bibnamefont {Baddorf}},\ }\bibfield  {title} {\bibinfo {title} {Scanning {{Tunneling Potentiometry}}: {{The Power}} of {{STM}} applied to {{Electrical Transport}}},\ }in\ \href {https://doi.org/10.1007/978-0-387-28668-6_2} {\emph {\bibinfo {booktitle} {Scanning {{Probe Microscopy}}}}},\ \bibinfo {editor} {edited by\ \bibinfo {editor} {\bibfnamefont {S.}~\bibnamefont {Kalinin}}\ and\ \bibinfo {editor} {\bibfnamefont {A.}~\bibnamefont {Gruverman}}}\ (\bibinfo  {publisher} {Springer New York},\ \bibinfo {address} {New York, NY},\ \bibinfo {year} {2007})\ pp.\ \bibinfo {pages} {11--30}\BibitemShut {NoStop}%
\bibitem [{\citenamefont {Johnson}\ and\ \citenamefont {Silsbee}()}]{Johnson-Silsbee-InterfacialChargespinCoupling-1985}%
  \BibitemOpen
  \bibfield  {author} {\bibinfo {author} {\bibfnamefont {M.}~\bibnamefont {Johnson}}\ and\ \bibinfo {author} {\bibfnamefont {R.~H.}\ \bibnamefont {Silsbee}},\ }\bibfield  {title} {\bibinfo {title} {Interfacial charge-spin coupling: {{Injection}} and detection of spin magnetization in metals},\ }\href {https://doi.org/10.1103/PhysRevLett.55.1790} {\bibfield  {journal} {\bibinfo  {journal} {Physical Review Letters}\ }\textbf {\bibinfo {volume} {55}},\ \bibinfo {pages} {1790}}\BibitemShut {NoStop}%
\bibitem [{\citenamefont {Jedema}\ \emph {et~al.}()\citenamefont {Jedema}, \citenamefont {Filip},\ and\ \citenamefont {Van~Wees}}]{Jedema-VanWees-ElectricalSpinInjection-2001}%
  \BibitemOpen
  \bibfield  {author} {\bibinfo {author} {\bibfnamefont {F.~J.}\ \bibnamefont {Jedema}}, \bibinfo {author} {\bibfnamefont {A.~T.}\ \bibnamefont {Filip}},\ and\ \bibinfo {author} {\bibfnamefont {B.~J.}\ \bibnamefont {Van~Wees}},\ }\bibfield  {title} {\bibinfo {title} {Electrical spin injection and accumulation at room temperature in an all-metal mesoscopic spin valve},\ }\href {https://doi.org/10.1038/35066533} {\bibfield  {journal} {\bibinfo  {journal} {Nature}\ }\textbf {\bibinfo {volume} {410}},\ \bibinfo {pages} {345}}\BibitemShut {NoStop}%
\bibitem [{\citenamefont {Rai}\ \emph {et~al.}(2025)\citenamefont {Rai}, \citenamefont {Patra}, \citenamefont {Bera}, \citenamefont {Deb}, \citenamefont {Mondal}, \citenamefont {Mahadevan},\ and\ \citenamefont {Kumar}}]{Rai-Kumar-DirectionDependentConductionPolarity-2025}%
  \BibitemOpen
  \bibfield  {author} {\bibinfo {author} {\bibfnamefont {B.}~\bibnamefont {Rai}}, \bibinfo {author} {\bibfnamefont {K.}~\bibnamefont {Patra}}, \bibinfo {author} {\bibfnamefont {S.}~\bibnamefont {Bera}}, \bibinfo {author} {\bibfnamefont {K.}~\bibnamefont {Deb}}, \bibinfo {author} {\bibfnamefont {M.}~\bibnamefont {Mondal}}, \bibinfo {author} {\bibfnamefont {P.}~\bibnamefont {Mahadevan}},\ and\ \bibinfo {author} {\bibfnamefont {N.}~\bibnamefont {Kumar}},\ }\href {https://doi.org/10.48550/arXiv.2502.02231} {\bibinfo {title} {Direction-{{Dependent Conduction Polarity}} in {{Altermagnetic CrSb}}}} (\bibinfo {year} {2025}),\ \Eprint {https://arxiv.org/abs/2502.02231} {arXiv:2502.02231 [cond-mat]} \BibitemShut {NoStop}%
\bibitem [{\citenamefont {Herasymchuk}\ \emph {et~al.}(2025)\citenamefont {Herasymchuk}, \citenamefont {Hallberg}, \citenamefont {Hodt}, \citenamefont {Linder}, \citenamefont {Gorbar},\ and\ \citenamefont {Sukhachov}}]{Herasymchuk-Sukhachov-ElectricSpinCurrent-2025}%
  \BibitemOpen
  \bibfield  {author} {\bibinfo {author} {\bibfnamefont {A.~A.}\ \bibnamefont {Herasymchuk}}, \bibinfo {author} {\bibfnamefont {K.~B.}\ \bibnamefont {Hallberg}}, \bibinfo {author} {\bibfnamefont {E.~W.}\ \bibnamefont {Hodt}}, \bibinfo {author} {\bibfnamefont {J.}~\bibnamefont {Linder}}, \bibinfo {author} {\bibfnamefont {E.~V.}\ \bibnamefont {Gorbar}},\ and\ \bibinfo {author} {\bibfnamefont {P.}~\bibnamefont {Sukhachov}},\ }\href {https://doi.org/10.48550/arXiv.2507.08072} {\bibinfo {title} {Electric and spin current vortices in altermagnets}} (\bibinfo {year} {2025}),\ \Eprint {https://arxiv.org/abs/2507.08072} {arXiv:2507.08072 [cond-mat]} \BibitemShut {NoStop}%
\bibitem [{\citenamefont {{Wolfram Research, Inc.}}(2024)}]{Wolfram}%
  \BibitemOpen
  \bibfield  {author} {\bibinfo {author} {\bibnamefont {{Wolfram Research, Inc.}}},\ }\href {https://www.wolfram.com/mathematica} {\bibinfo {title} {Mathematica}} (\bibinfo {year} {2024})\BibitemShut {NoStop}%
\end{thebibliography}%

\end{document}